\documentclass[final,3p,fleqn,11pt]{elsarticle}

\usepackage{amssymb}
\usepackage{multicol}
\usepackage[svgnames]{xcolor}

\usepackage{booktabs}
\usepackage{array}
\newcommand{\PreserveBackslash}[1]{\let\temp=\\#1\let\\=\temp}
\newcolumntype{C}[1]{>{\PreserveBackslash\centering}p{#1}}
\newcolumntype{R}[1]{>{\PreserveBackslash\raggedleft}p{#1}}
\newcolumntype{L}[1]{>{\PreserveBackslash\raggedright}p{#1}}

\usepackage{amsmath}
\usepackage{amsfonts}
\usepackage{amsthm}
\usepackage{mathtools}
\usepackage{subfig}
\usepackage{lineno}
\usepackage[hidelinks]{hyperref}
\usepackage{textcomp}
\usepackage{algorithm,algpseudocode,algorithmicx}
\usepackage{pifont}
\usepackage{tabularx}

\usepackage{physics}
\usepackage{MnSymbol}

\usepackage{float}
\usepackage{array,multirow}
\usepackage{color, colortbl}
\usepackage{tikz}
\usetikzlibrary{patterns}
\usetikzlibrary{spy}
\usetikzlibrary{spy,decorations.fractals,shapes.misc}

\usepackage{pgfplots}
\pgfplotsset{compat=1.16}
\newsavebox\Axis

\definecolor{lightgray}{gray}{0.80}

\usetikzlibrary{decorations.pathreplacing}
\usepackage[listings,theorems,skins,breakable]{tcolorbox}
\tcbset{skin=enhanced}
\newtcolorbox{lbracebox}[1][Word]{%
   frame hidden,enlarge left by=2cm,width=\linewidth-2cm,%
  overlay unbroken = {\draw [decorate,decoration={brace,amplitude=10pt},]%
                     (frame.south west)-- (frame.north west)
                    node [black,midway,left,xshift=-.6cm] {#1};},% 
}

\newcommand{\Bezier}{B\'ezier~}

\definecolor{grey1}{rgb}{0.5, 0.5, 0.5}
\definecolor{green1}{rgb}{0.4660, 0.6740, 0.1880} 
\definecolor{blue1}{rgb}{0.26, 0.41, 0.88} 
\definecolor{red1}{rgb}{0.8600, 0.0800, 0.2400}
\definecolor{yellow1}{rgb}{0.93, 0.75, 0.125}
\definecolor{purple1}{rgb}{0.4940, 0.1840, 0.5560}
\definecolor{lightblue1}{rgb}{0.3010, 0.7450, 0.9330}
\definecolor{bordeaux1}{rgb}{0.6350, 0.0780, 0.1840}
\definecolor{brown1}{rgb}{0.65, 0.16, 0.16}
\definecolor{pink1}{rgb}{1.0, 0.08, 0.58}

\definecolor{burntorange}{rgb}{0.8702,0.4791176,0}

\usepackage{color, soul}
\definecolor{blue2}{RGB}{125, 249, 255}

\DeclareRobustCommand{\reviewerI}[1]{{\sethlcolor{pink}\hl{#1}}}
\DeclareRobustCommand{\reviewerII}[1]{{\sethlcolor{yellow}\hl{#1}}}

\DeclareRobustCommand{\changed}[1]{{\sethlcolor{blue2}\hl{#1}}}

\soulregister\reviewerI{1}
\soulregister\reviewerII{1}
\soulregister\changed{1}

\soulregister\ref7
\soulregister\pageref7
\soulregister\eqref7
\soulregister\cite7

\makeatletter
\renewcommand*\env@matrix[1][*\c@MaxMatrixCols c]{%
  \hskip -\arraycolsep
  \let\@ifnextchar\new@ifnextchar
  \array{#1}}
\makeatother

\theoremstyle{plain}
\newtheorem{theorem}{Theorem}[section]

\newtheorem{remark}[theorem]{Remark}
\newtheorem{example}[theorem]{Example}

\theoremstyle{definition}

\newcommand{\vect}[1]{\boldsymbol{#1}} 									% a vector
\newcommand{\mat}[1]{\mathbf{#1}} 											% a vector
\newcommand{\eigenvec}{U}             % symbol for eigenvector
\newcommand{\laplace}{\Delta}         % symbol for laplacian
\newcommand{\noMode}{n}               % mode number
\newcommand{\domain}{\Omega}														% physical domain
\newcommand{\boundary}{\Gamma}													% physical domain boundary
\newcommand{\pardomain}{\hat{\Omega}}										% parametric domain
\newcommand{\ndofs}{N}																	% number of basis functions (ndofs)
\newcommand{\trialf}{B}																	% trial func
\newcommand{\testf}{\tilde{B}}													% proposed test func
\newcommand{\dual}{\bar{\trialf}}												% exact dual B-spline
\newcommand{\adual}{\hat{\trialf}}                      % approx. dual 

\newcommand{\trialfv}{\mat{b}}													% vector of trial func
\newcommand{\testfv}{\mat{\tilde{b}}}									% vector of proposed test func
\newcommand{\adualv}{\mat{\hat{b}}}										% vector of approximate dual B-spline
\newcommand{\dualv}{\mat{\bar{b}}}											% vector of exact dual 

\newcommand{\map}{\Phi}                                    % mapping operator
\newcommand{\detJ}{C}                                   % determinant of Jacobian

\newcommand{\invmassM}{\hat{\mathbf{G}}_r^{-1}}
\newcommand{\invmassMf}{\hat{\mathbf{G}}^{-1}}
\newcommand{\massMf}{\hat{\mathbf{G}}}
\newcommand{\massM}{\hat{\mathbf{G}}_r}

\newcommand{\knotvect}{\boldsymbol{\Xi}}               % knot vector

\newcommand{\sspace}[2]{\mathbb{S}^{#1}_{#2}}					% univariate B-spline
\newcommand{\Span}[1]{\text{span} \left( #1 \right)}					% span of spline space
\allowdisplaybreaks

\modulolinenumbers[5]

\begin{document}

\begin{frontmatter}

\title{Towards higher-order accurate mass lumping in explicit isogeometric analysis for structural dynamics}

\author[address1]{Thi-Hoa Nguyen}
\ead{nguyen@mechanik.tu-darmstadt.de}

\author[address1]{Ren\'e R. Hiemstra}
\ead{hiemstra@mechanik.tu-darmstadt.de}

\author[address1]{Sascha Eisentr\"ager}
\ead{eisentraeger@mechanik.tu-darmstadt.de}

\author[address1]{Dominik Schillinger\corref{cor1}}
\ead{schillinger@mechanik.tu-darmstadt.de}

\cortext[cor1]{Corresponding author}

\address[address1]{Institute for Mechanics, Computational Mechanics Group, Technical University of Darmstadt, Germany\\[1.26cm]
\begin{large} 
Dedicated to Thomas J.\ R.\ Hughes on the occasion of his 80th birthday.
\end{large}}

\begin{abstract}
We present a mass lumping approach based on an isogeometric Petrov-Galerkin method that preserves higher-order spatial accuracy in explicit dynamics calculations irrespective of the polynomial degree of the spline approximation. To discretize the test function space, our method uses an approximate dual basis, whose functions are smooth, have local support and satisfy approximate bi-orthogonality with respect to a trial space of B-splines. The resulting mass matrix is ``close'' to the identity matrix. Specifically, a lumped version of this mass matrix preserves all relevant polynomials when utilized in a Galerkin projection. Consequently, the mass matrix can be lumped (via row-sum lumping) without compromising spatial accuracy in explicit dynamics calculations. We also address the imposition of Dirichlet boundary conditions and the preservation of approximate bi-orthogonality under geometric mappings. In addition, we establish a link between the exact dual and approximate dual basis functions via an iterative algorithm that improves the approximate dual basis towards exact bi-orthogonality. We demonstrate the performance of our higher-order accurate mass lumping approach via convergence studies and spectral analyses of discretized beam, plate and shell models.
\end{abstract}

\begin{highlights}
	\item We present a mass lumping approach based on an isogeometric Petrov-Galerkin method.
	\item It is based on smooth, locally supported, approximate dual functions to B-splines.
	\item The resulting mass matrix can be lumped without compromising the spatial accuracy in explicit dynamics calculations.
	\item We demonstrate its performance via beam, plate and shell benchmarks.
\end{highlights}


\end{frontmatter}

\newpage
\tableofcontents
\newpage

% \linenumbers

\section{Introduction}

I first arrived at the University of Texas at Austin in spring 2011, where I joined the Hughes research group as a visiting PhD student. I remember one afternoon in the old JP's Java, where Tom used to take us for a drink. Someone was talking about a leading figure in computational mechanics in his sixties who was thinking about moving to Harvard. I blurted out: ``Why would they hire him? He will probably retire in two years..." Being the gentleman he is, Tom looked at me smiling and said: ``Dominik, you know, \textit{I} moved to Texas in my sixties..." and I completed for myself: and instead of retiring you (once more) revolutionized the world of computational mechanics by inventing IGA. Fast forward one and a half years - I had graduated in the meantime and was now a postdoc in his group - I told Tom that I was about to accept a job offer from a German aircraft engines manufacturer. To my surprise, he immediately told me that I couldn't do that. Combining rational arguments and subtle challenging, he then convinced me of the following plan: I would stay in Texas for six months more and submit a few applications for open professorship positions in the U.S., and in case I did not secure an academic job by then, I could start a career designing aircraft engines. The bottomline: by late 2013, Tom and his unique environment had transformed me from a PhD student into an aspiring assistant professor at the University of Minnesota. 
Happy Birthday, Tom, and thank you for all you have done and achieved for the field of computational mechanics, for our community and for so many of us colleagues and friends. (Dominik Schillinger)\\

Isogeometric analysis (IGA) was initiated in 2005 by Tom and his students \cite{Hughes:05.1,Cottrell:09.1} with the goal of bridging the gap between computer aided geometric design (CAD) and finite element analysis (FEA). The core idea of IGA is to use the same \textit{smooth} and \textit{higher-order} spline basis functions for the representation of both geometry in CAD and the approximation of physics-based field solutions in FEA \cite{Haberleitner:17.1}. 
IGA enables finite element approximations that are higher-order smooth, a feature that cannot be achieved in \reviewerII{FEA based on nodal basis functions} \cite{Hughes:17.1}. 
A key property of higher-order IGA, already discussed in one of the first articles \cite{Cottrell:06.1}, is its well-behaved discrete spectrum of eigenfrequencies and eigenmodes. The associated potential of IGA for higher-order accurate explicit dynamics calculations, however, lies largely idle to this day. 
One reason is that no practical methodology exists yet that reconciles higher-order accuracy and mass lumping. In this paper, we will focus on a new Petrov-Galerkin method that enables a higher-order accurate mass lumping that could help close this gap.

\subsection{Explicit dynamics and mass lumping in FEA}

Finite element discretizations of transient problems in structural mechanics generate semidiscrete systems of coupled second-order ordinary differential equations of the following type \cite{hartmann2015mass}:
\begin{align}
	\mat{M} \, \vect{\ddot{u}}^n \; = \; \mat{F}^n_{\text{ext}} \; - \; \int_{\Omega} \boldsymbol{B}(\vect{x}^n)^T \; \boldsymbol{\sigma}^n \; \mathrm{d}\Omega \; ,
\label{eq1}
\end{align}
where $\mat{M}$ is the mass matrix, $\vect{u}$ the displacement vector, $\mat{F}_{\text{ext}}$ the vector of external forces, $\boldsymbol{B}$ the discrete gradient operator, $\boldsymbol{\sigma}$ the Cauchy stress, and $\Omega$ the problem domain. The superscript $n$ is the time index that indicates that a variable is evaluated at time instant $t^n$.

In many applications such as virtual testing of vehicle crashworthiness or the simulation-based design of metal forming processes, the internal force vector of the discrete system \eqref{eq1} involves shell elements, nonlinear material models, and contact with friction. For shell problems, most established finite element formulations are based on shear-deformable Reissner-Mindlin-type models \cite{Bischoff:04.1}, which allow the consistent evaluation of the integral in \eqref{eq1} with standard nodal basis functions. 
Due to their cost effectiveness and robustness against locking, state-of-the-art commercial software packages such as LS-Dyna or Pam-Crash mostly rely on lowest-order linear basis functions with reduced quadrature and hourglass control, which require very fine meshes to achieve accurate solutions on complex geometries. 

For the resulting large and ill-conditioned systems, however, iterative equation solvers required by implicit methods do not converge well and are thus prohibitively expensive \cite{hartmann2015mass}.

As a result, explicit dynamics based on explicit second-order accurate time integration methods such as the central difference method or variations thereof \cite{hughes_finite_2003}, has 
established itself 
as the key technology for efficiently solving \eqref{eq1} \cite{belytschko1984explicit,benson1992computational}. Given a lumping scheme to diagonalize the mass matrix, its inversion in \eqref{eq1} becomes trivial and the dominating cost of the explicit multistep procedure shifts entirely to the evaluation of the internal force vector. In addition, a diagonal mass matrix significantly reduces memory and facilitates parallel computing, as global system matrices and the associate assembly or inversion procedures do never explicitly occur. An alternative pathway to explicit dynamics with diagonal mass matrices in FEA is the use of \reviewerII{so-called Gauss-Lobatto} Lagrange basis functions with nodes at the Gauss-Lobatto points in conjunction with Gauss-Lobatto quadrature \cite{ArticleFried1975,Canuto:07.1,Willberg:12.1, Schillinger2014}. This is the only technique established so far that obtains higher-order spatial accuracy in explicit dynamics with a diagonal mass matrix \cite{ArticleDuczek2019b,ArticleDuczek2019a}.
\reviewerII{We note that within the last decade, there has been a new increase in research and development in higher-order explicit structural analysis with $C^0$-continuous finite elements, both in academia, see e.g. }\cite{sprague2008legendre,danielson2011reliable,duczek2014numerical,ArticleDuczek2019b,zhang2021massively,danielson2022curved} \reviewerII{and in industry, see e.g. the commercial codes IMPETUS} \cite{impetus2012impetus} \reviewerII{and LS-Dyna} \cite{teng2016recent}, \reviewerII{which in addition to standard linear elements also promote second- and third-order elements.}

Explicit schemes, however, are only conditionally stable. Therefore, the step size, with which they march forward in time, cannot exceed a maximum critical time step size \cite{Belytschko:06.1}, which is inversely proportional to the maximum eigenfrequency $\omega_{\text{max}}$ of the discretization: $ \Delta \, t_{\text{crit}}^n  \sim \, 1 / {\omega^n_{\text{max}}}$.
Due to this dependence,  
control of the highest frequencies in the spectrum is 
indispensable for guaranteeing acceptable time step sizes and hence efficient explicit dynamics calculations. 
In FEA, the impact of $\omega_{\text{max}}$ can be alleviated, e.g., via selective mass scaling \cite{tkachuk2014local} or time step subcycling \cite{casadei2009binary}.

\subsection{Explicit dynamics and mass lumping in IGA}

IGA is particularly attractive for higher-order accurate structural analysis. In addition to mesh refinement, spline functions naturally enable the increase of the polynomial degree ($p$-refinement), or both at the same time ($k$-refinement). For elastostatic-type problems, it has been consistently shown across the pertinent literature that higher-order IGA is superior to \reviewerII{FEA based on nodal basis functions} in terms of per-degree-of-freedom accuracy \cite{Schillinger:13.2}. A variety of advanced formulations for isogeometric structural analysis have been developed over the past decade, in particular for shells based on Kirchhoff-Love \cite{Kiendl_shell_2009,Echter:10.1}, Reissner-Mindlin \cite{Benson:10.1,Benson:11.1,Hsu:12.2,Dornisch:13.1}, and solid-type models \cite{Hosseini:13.1,Bouclier:15.1}, and hierarchic combinations thereof \cite{Benson:13.1,Echter:13.1}. 
Isogeometric shell formulations have been applied for explicit dynamics calculations, based on the adoption of standard row-sum lumping to diagonalize the mass matrix \cite{chen2014explicit,Leidinger2019}. In addition, selective mass scaling \cite{hartmann2015mass} and different methods to obtain time step estimates \cite{hartmann2015mass,adam2015stable} known from \reviewerII{FEA based on nodal basis functions} have been applied in an IGA context. But studies on explicit dynamics with shell IGA and standard mass lumping 
have reported sub-optimal results. For example, Hartmann and Benson report in \cite{hartmann2015mass} that ``increasing the degree on a fixed mesh size increases the cost without a commensurate increase in accuracy'', which they call a ``disappointing result.''

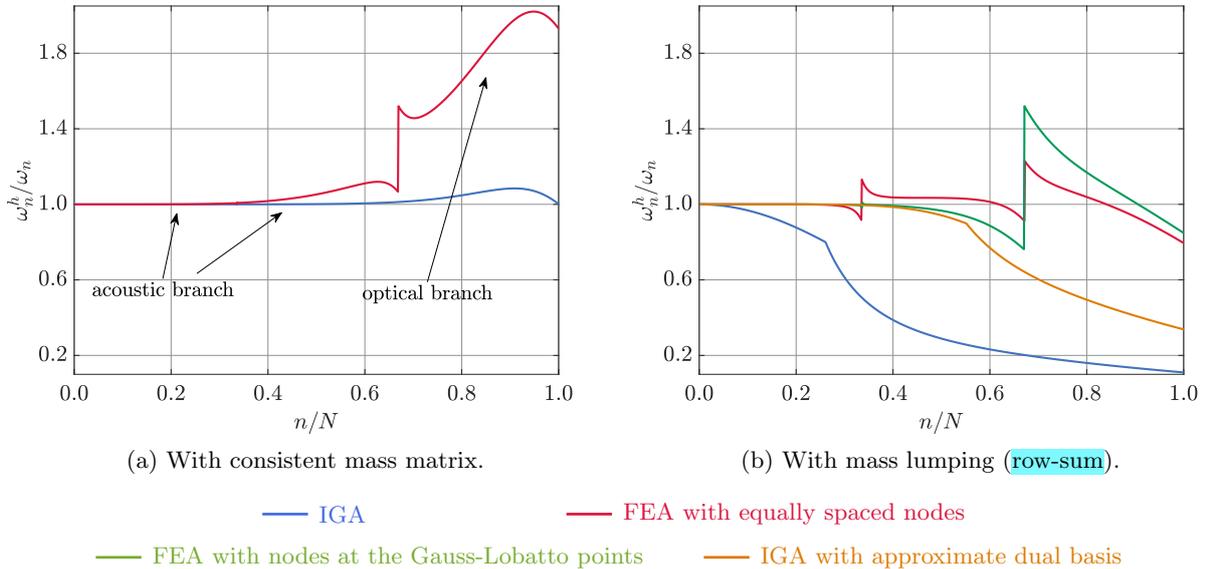
\begin{figure}[t!]
 	\centering
	\subfloat[With consistent mass matrix.]{\includegraphics[width=0.5\textwidth]{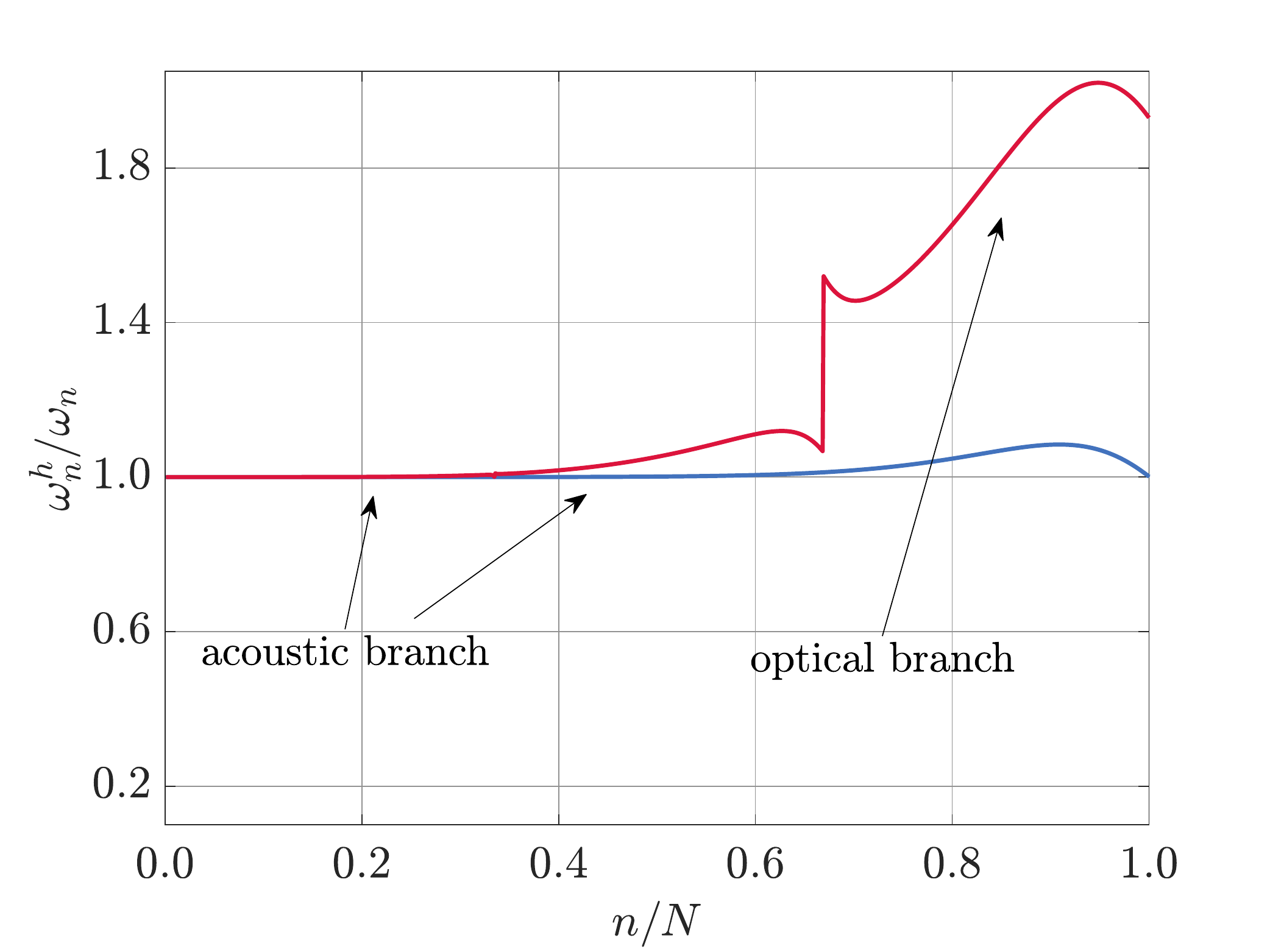}}
	\subfloat[With mass lumping (\changed{row-sum}).]{\includegraphics[width=0.5\textwidth]{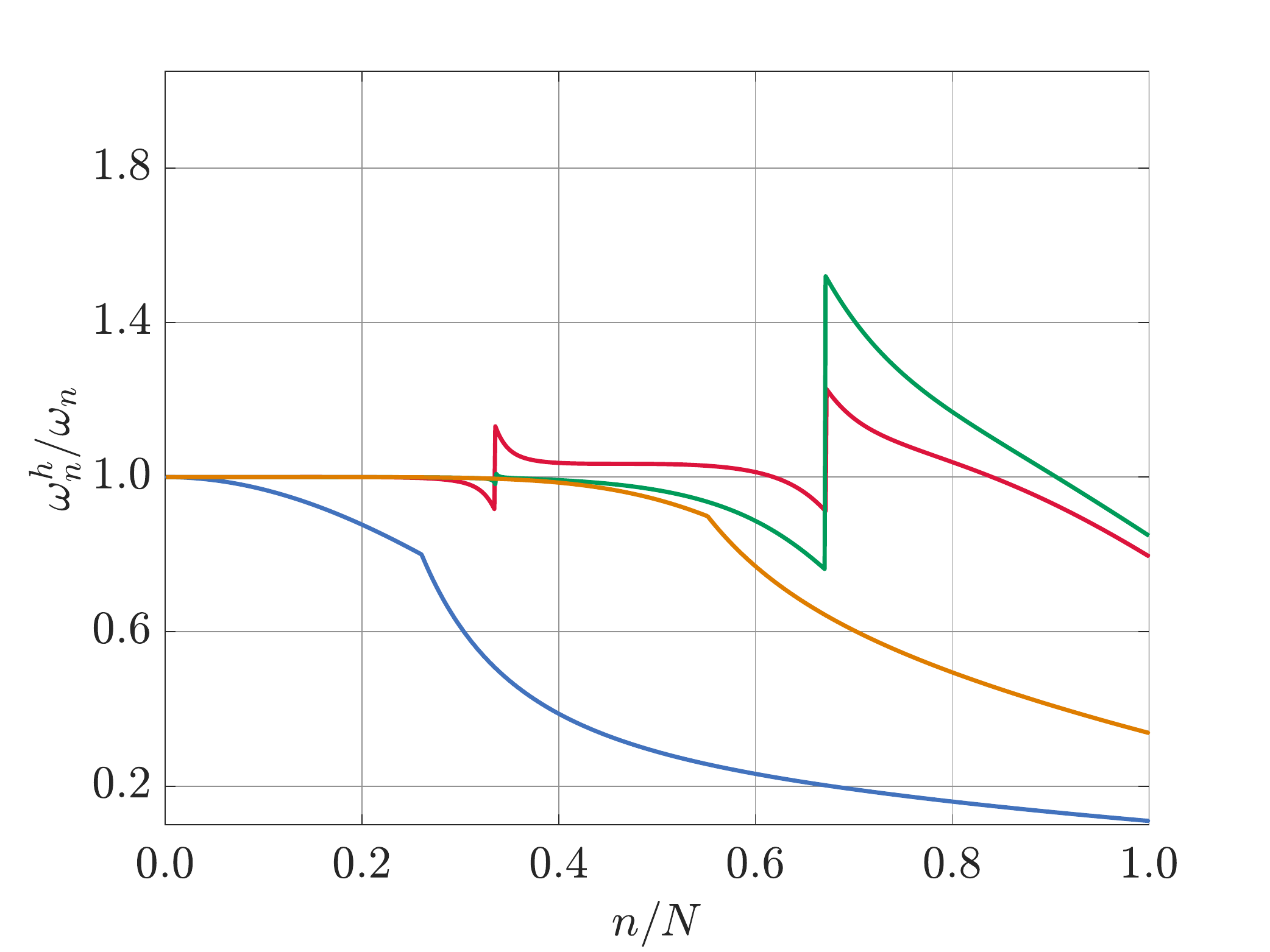}}

	\vspace{0.2cm}
	\begin{tikzpicture}
    \filldraw[blue1,line width=1pt, solid] (0,0) -- (0.6,0);
    \filldraw[blue1,line width=1pt] (0.6,0) node[right]{\footnotesize IGA};
    \filldraw[red1,line width=1pt, solid] (4,0) -- (4.6,0);
    \filldraw[red1,line width=1pt] (4.6,0) node[right]{\footnotesize FEA with equally spaced nodes};
\end{tikzpicture}

\begin{tikzpicture}
    \filldraw[green1,line width=1pt, solid] (0,0) -- (0.6,0);
    \filldraw[green1,line width=1pt] (0.6,0) node[right]{\footnotesize FEA with nodes at the Gauss-Lobatto points};
    \filldraw[burntorange,line width=1pt, solid] (8,0) -- (8.6,0);
    \filldraw[burntorange,line width=1pt] (8.6,0) node[right]{\footnotesize IGA with approximate dual basis};
\end{tikzpicture}
	 \caption{Normalized discrete spectrum of \changed{eigenfrequencies} obtained from a generalized eigenvalue problem for a bar, computed with FEA and IGA, both with 1,000 cubic basis functions. We note that in (a), the orange line and the green line overlap with the blue line and the red line, respectively.}
	 	\label{fig1}
\end{figure}

To put this statement in perspective and illustrate the negative impact of mass lumping on a higher-order Galerkin method, 
we consider the classical result obtained from the generalized eigenvalue problem of a bar, discretized with cubic Lagrange polynomials (FEA) and cubic smooth splines (IGA) \cite{Cottrell:09.1,Cottrell:06.1}. The numerically obtained frequencies, $\omega_n^h$, are normalized with respect to the exact solution, $\omega$, and plotted versus the mode number, $n$, normalized by the total number of degrees of freedom, $N$.
Figure~\ref{fig1}a shows that for FEA, the spectrum (curve plotted in red) with consistent mass matrix contains so-called ``optical'' branches, separated by distinct jumps, which compromise the accuracy of the high frequency range. In contrast, we observe that for IGA (curve plotted in blue) with consistent mass matrix, the accurate ``acoustic branch'' of the spectrum covers the entire frequency range. We note that we removed the outliers at the end of the spectrum \cite{Hiemstra_outlier_2021}. 
\reviewerII{We now apply standard row-sum lumping to the mass matrix in both schemes.
Figure~\ref{fig1}b shows that for FEA with equally spaced nodes, the spectrum exhibits a loss of accuracy in the first optical branch. When we apply FEA with nodes at the Gauss-Lobatto points and Gauss-Lobatto quadrature \cite{ArticleFried1975}, we naturally arrive at a lumped mass matrix, whose spectrum is more accurate in the medium frequency range. With standard row-sum lumping of the mass matrix, the results for IGA exhibit a significant decrease in spectral accuracy across the entire frequency range.}
The results shown in Fig.~\ref{fig1}b thus indicate that one reason for the decrease in accuracy lies in the detrimental impact of row-sum lumping when employed in a standard IGA context, 
outlined already in the first contribution on IGA for structural dynamics 
\cite{Cottrell:06.1}. 

\subsection{Mass lumping based on approximate dual spline functions}

The removal of this gap 
requires the development of novel spline-centered technology that leverages the additional opportunities provided by this class of basis functions. A first attempt was presented in \cite{Anitescu_dual_2019}, based on the idea of using dual spline functions. This set of spline functions is defined ``dual'' to a set of standard B-splines, such that the inner product of the two integrated over a domain yields a diagonal matrix (bi-orthogonality). Hence, if a dual basis is used as the test space in a finite element formulation, we naturally end up with a consistent diagonal mass matrix \cite{Anitescu_dual_2019}. In CAD research, the bi-orthogonality property has been well known for decades, and several approaches for the construction of the underlying dual basis exist, see e.g.\ \cite{schumaker_spline_2007}. In the computational mechanics community, this property has been used for different purposes, e.g.\ for dual mortar methods 
\cite{seitz_mortar_2016,zou_mortar_2018} or for unlocking of Reissner-Mindlin shell formulations \cite{zou_dual_locking_2020}.

Dual spline basis functions still suffer from shortcomings when applied as test functions in the variational formulation of elastodynamics as envisioned in \cite{Anitescu_dual_2019}, to the extent that their use in explicit dynamics seems not straightforward: Firstly, bi-orthogonality in general holds only on the parametric domain, but is in general lost under non-affine geometric mappings. 
Secondly, dual spline functions with the same smoothness as their B-spline counterparts have global support on each patch, thus producing fully populated stiffness forms. Their support can be reduced, but at the price of losing continuity, leading to discontinuous functions in the fully localized case. Both are prohibitive in a finite element context. And thirdly, dual basis functions are not interpolatory at the boundaries. Therefore, the identification of a kinematically admissible set of test functions that allows the variationally consistent strong imposition of Dirichlet boundary conditions is not straightforward.

In this paper, we present an 
isogeometric Petrov-Galerkin method that 
combines the dual basis concept and standard row-sum lumping and achieves higher-order accuracy in explicit dynamics irrespective of the polynomial degree of the spline approximation. 
Its key ingredients are a class of ``approximate'' dual spline functions that was introduced in \cite{chui_wavelet_2004}. It only approximately satisfies the discrete bi-orthogonality property, but preserves all other properties of the original B-spline basis, such as $C^{p-1}$ smoothness, polynomial reproduction and local support. We employ the approximate dual functions to discretize the test function space in the variational formulation of elastodynamics. This leads to a semidiscrete Petrov-Galerkin formulation that can be written in the format \eqref{eq1}, whose consistent mass matrix $\mat{M}$ is not diagonal, but is, in some sense, ``close'' to the identity matrix. Specifically, its row-sum lumped mass preserved all polynomials up to degree $p$ when used in a Galerkin projection. Consequently, the row-sum lumped mass matrix may be used in explicit dynamics without compromising higher-order spatial accuracy. \reviewerII{In Fig.~\ref{fig1}b, we anticipate the result of our isogeometric Petrov-Galerkin formulation with approximate dual basis functions and row-sum lumping for the classical example of the one-dimensional bar. We observe that the spectrum of our method maintains accuracy in the low and medium range of modes, which is a significant improvement over the standard IGA Galerkin formulation with mass lumping. We note that when the mass matrix is evaluated consistently without lumping, the two methods yield exactly the same spectrum shown in Fig.~\ref{fig1}a.}

We note that a similar approach is presented in \cite{ArticleHeld2024}. It consists of a minimally invasive technique, based on approximate dual test functions, which can be applied in the sense of preconditioner by a simple matrix pre-multiplication. They show high-order convergent results and a straightforward application of Dirichlet boundary conditions by manipulating this transformation matrix. An additional row-summing step is required to convert the banded into a diagonal mass matrix. 

The structure of our paper is as follows: 
In Section \ref{sec_dualspace}, we briefly review the definition and construction of approximate dual functions of B-splines. In addition, we present an iterative scheme for improving the approximate dual basis towards exactly satisfying bi-orthogonality. In Section \ref{sec_lumpingscheme}, we derive our isogeometric Petrov-Galerkin formulation with mass lumping for higher-order accurate explicit dynamics. In Section \ref{sec:dynamics_plates}, we present numerical examples of beam and plate problems, demonstrating that our Petrov-Galerkin scheme indeed preserves higher-order accuracy in explicit dynamics calculations.
In Section \ref{sec_conclusion}, we provide a critical discussion of the potential significance of our results, highlight by means of a more involved shell benchmark that extending the demonstrated benefits to practical problems might not be immediate, and motivate potential future research directions.

\section{Approximate dual spline basis}\label{sec_dualspace}

In this section, we briefly review the construction and the relevant properties of the approximate dual functions for univariate B-splines introduced in \cite{chui_wavelet_2004}. We also present an iterative algorithm to improve an approximate dual basis in terms of satisfying the bi-orthogonality constraint. 
We start this section with a brief recap of univariate B-splines and their dual bases.
%---------------------------------------------------
\subsection{B-spline basis functions}

A B-spline function is a piecewise polynomial that is characterized by the polynomial degree, a given series of segments that we call B\'ezier elements, and the regularity at the interfaces of the B\'ezier elements. 
Let $\mathcal{P}^p$ denote the space of piecewise polynomials of degree $p \, \geq \, 0$ and consider a partitioning of an interval $\pardomain \, = \, [a, \,b] \, \subset \, \mathbb{R}$ into an increasing sequence of \textit{breakpoints} that define B\'ezier elements:
\begin{align}
    a \, = \, t_0 \, < \, \ldots \, < \, t_{k-1} \, < \, t_k \, < \, \ldots \, < t_m \, = \, b \, .  \label{eq:breakpts}
\end{align}
Let $\sspace{}{}$ denote the space of $C^{p-1}$ smooth splines that is:
\begin{align}
    \sspace{}{} \, := \, \left\{ \, s \, : \, [a,\, b] \, \mapsto \, \mathbb{R} \, : \, s \rvert_{t_{k-1},\,t_k} \, \in \, \mathcal{P}^p \, , s \text{ is } C^{p-1} \text{ smooth at } \hat{x} \, = \, t_1 \, \ldots \, t_{m-1} \, \right\} \, . \label{eq:sspace}
\end{align}
Consider B-spline functions $\trialf_i$, $i=1,\ldots,\ndofs$, of degree $p$ with $C^{p-1}$-continuity defined on an open \textit{knot vector}, $\knotvect \, := \left\{ \hat{x}_1, \, \ldots, \, \hat{x}_{N+p+1} \right\}$, 
which is the partition \eqref{eq:breakpts} with $p+1$ multiplicity of the first and last breakpoints and non-repeated internal knots. 
Such B-splines have interpolating end-conditions, reproduce polynomials in $\mathcal{P}^p$, and 
form a basis for $\sspace{}{}$. B-splines can for instance be defined recursively by using the Cox-de Boor formula \cite{Piegl_nurbs_1996}. They can be extended to multiple dimensions by constructing tensor products of univariate B-splines.

\begin{remark}
In the current paper, we use B-spline functions with maximum smoothness of $C^{p-1}$. But dual basis functions, approximate dual basis functions and hence the explicit dynamics technology to be presented do not require maximum smoothness, but can also operate with B-splines of any continuity. In particular, the extension to (tensor-product) multi-patch discretizations seems straightforward.
\end{remark}

%---------------------------------------------------
\subsection{Dual basis functions}

We consider the spline space $\sspace{}{}$ in \eqref{eq:sspace} that is spanned by $C^{p-1}$-smooth B-splines of degree $p$, $\trialf_i$, $i=1,\,\ldots,\, \ndofs$, defined on an open knot vector $\knotvect$.  
Let $\pardomain \subset \mathbb{R}$ denote the parametric domain with local coordinates represented in $\knotvect$, i.e. the breakpoints in \eqref{eq:breakpts}.  
Given an inner product $(\cdot, \, \cdot)_{\pardomain} \, : \, \sspace{}{} \times \sspace{}{} \, \mapsto \, \mathbb{R}$ in $\pardomain$. 
The functions 
$\dual_i$, $i=1,\,\ldots,\, \ndofs$, that satisfy the bi-orthogonality constraint:
\begin{align}
    \left(\dual_i, \, \trialf_j\right)_{\pardomain} \, = \, \delta_{ij} \, , \label{eq:duality_constraint}
\end{align}
where $\delta_{ij}$ is the Kronecker delta, 
form the dual basis of $\sspace{}{}$, corresponding to the B-splines $\trialf_i$.

\begin{figure}[h!]
    \centering
    \subfloat[$C^1$-continuous B-spline function.]{{\includegraphics[width=0.45\textwidth]{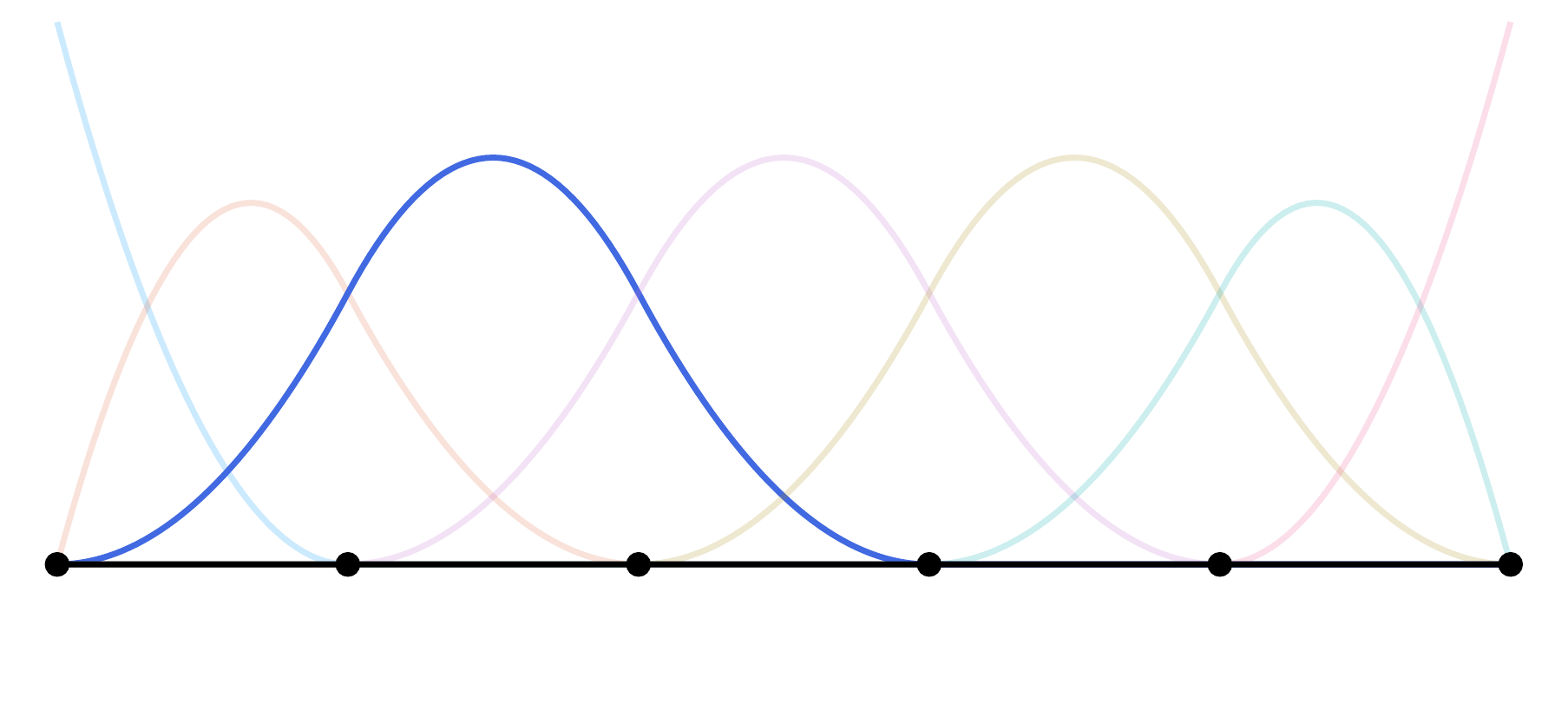} }}
    \subfloat[Associated $C^1$-continuous dual function, but with global support.]{{\includegraphics[width=0.45\textwidth]{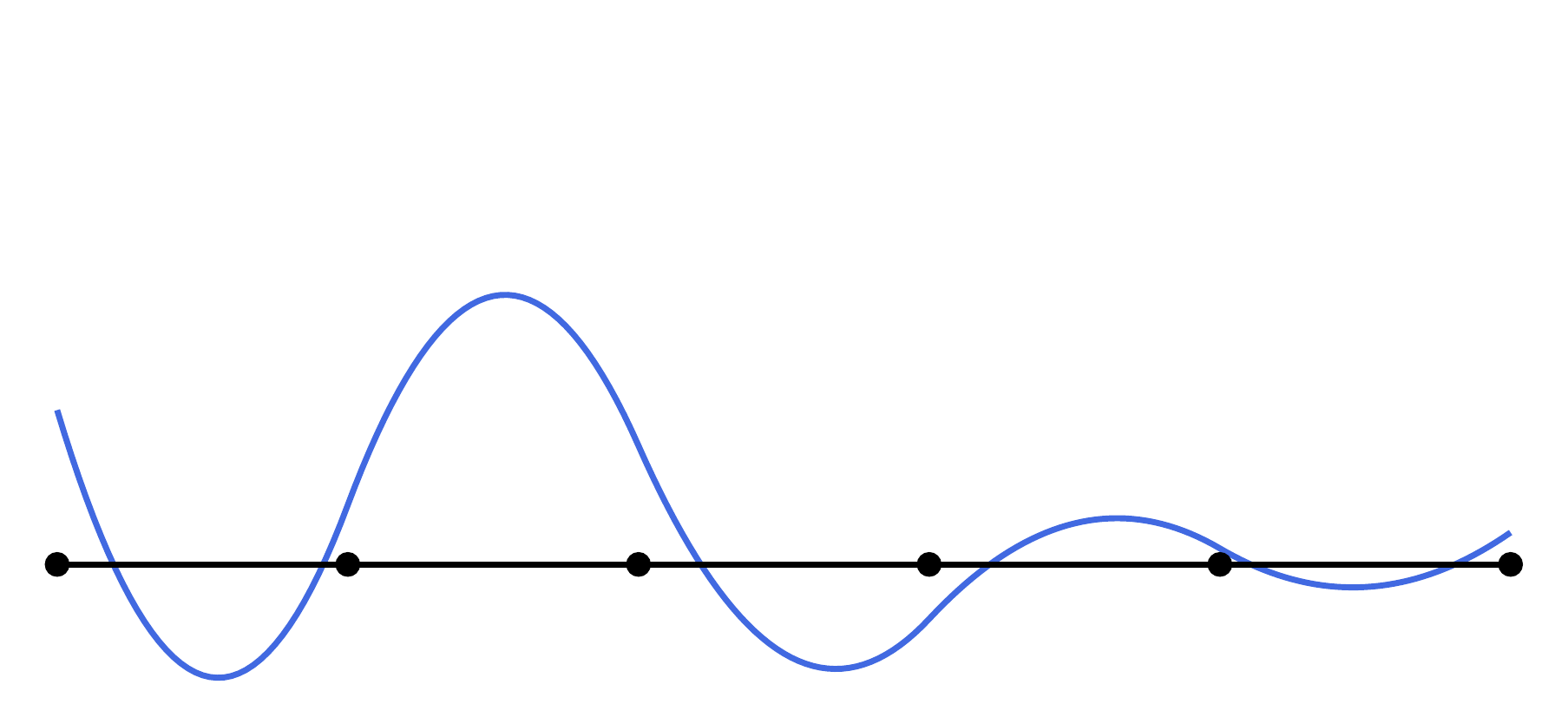} }}

    \subfloat[Associated dual function with minimal support, but discontinuous.]{{\includegraphics[width=0.45\textwidth]{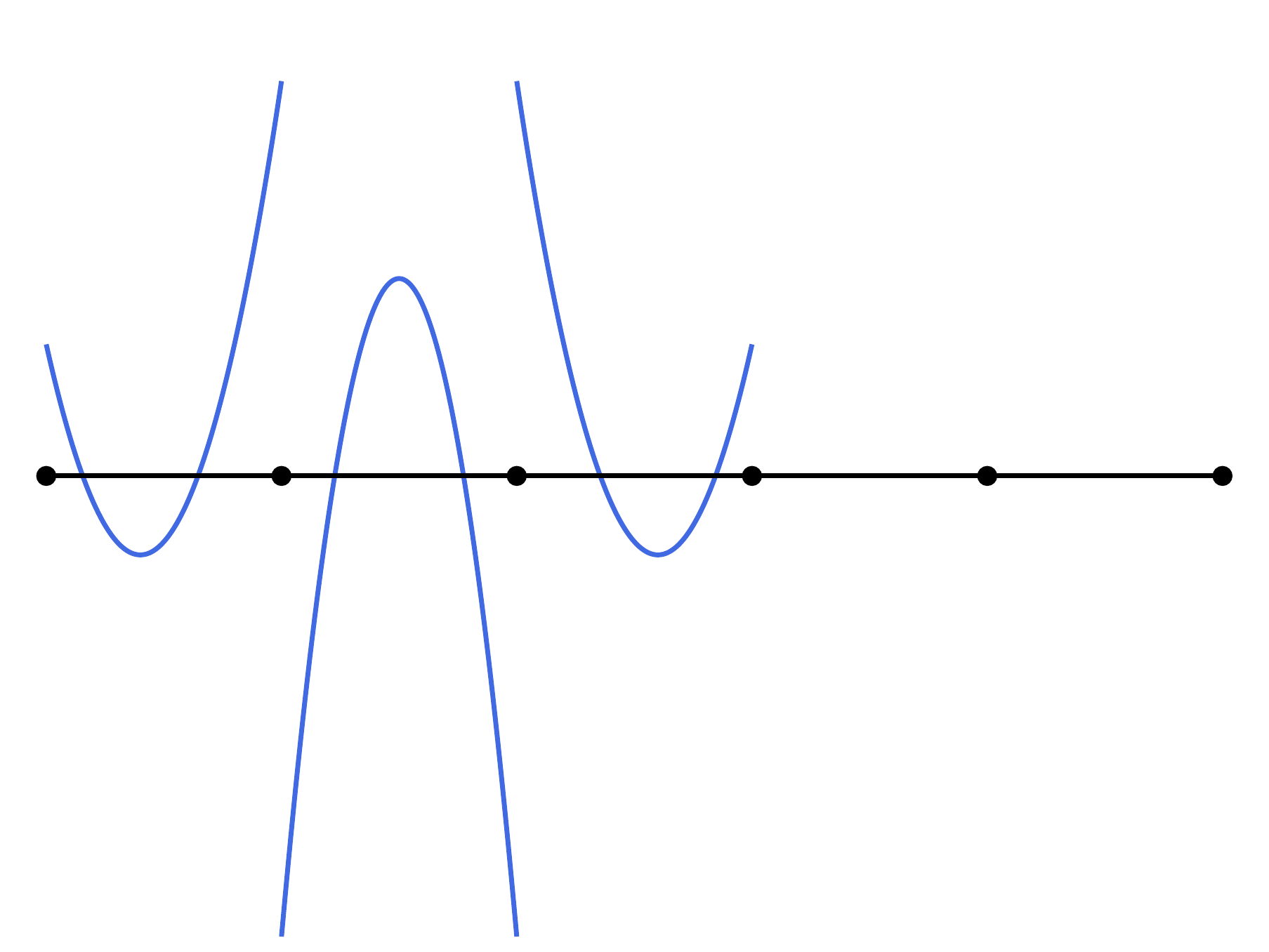} }}
    \caption{Dual functions for a quadratic $C^1$-continuous B-spline function.} 
    \label{fig:bspline_dual}
\end{figure}

Let $\trialfv = \left[ \, \trialf_1 \; \ldots \; \trialf_\ndofs \, \right]^T $ and $\dualv = \left[ \, \dual_1 \; \ldots \; \dual_\ndofs \, \right]^T$ denote the function vector of the B-spline basis $\trialf_i$ and its associated dual functions $\dual_i$, respectively. The bi-orthogonality constraint \eqref{eq:duality_constraint} can then be expressed in matrix form:
\begin{align} 
    \fcolorbox{pink}{pink}{$\left(\dualv \, \trialfv \right)_{\pardomain} \, = \, \mat{I}$}  \, ,
    \label{eq:duality_constraint2}
\end{align}
where $\mat{I}$ is the identity matrix. 
We recall the Gramian matrix $\mat{G}$ of the B-spline basis:
\begin{align}
    \fcolorbox{pink}{pink}{$\mat{G} = \left(\trialfv \, \trialfv \right)_{\pardomain}$}
    \, .
\end{align}
\reviewerI{Equation~\eqref{eq:duality_constraint2} is now reformulated as:}
\begin{align}
    \fcolorbox{pink}{pink}{$\left(\dualv \, \trialfv \right)_{\pardomain} \, = \, \mat{I} = \mat{G}^{-1}\,\mat{G} = \mat{G}^{-1} \, \left(\trialfv \, \trialfv \right)_{\pardomain} $}
\end{align}
The dual functions 
$\dual_i$, $i=1,\,\ldots,\,\ndofs$ can then be formulated via the inverse of the Gramian matrix:
\begin{align}
    \dualv \, = \, \mat{G}^{-1} \, \trialfv 
    \, . \label{eq:dual_func}
\end{align}
These dual functions are linear combinations of the corresponding B-splines, thus span the same spline space $\sspace{}{}$, and also reproduce polynomials in $\mathcal{P}^p$. 
The duals \eqref{eq:dual_func} computed with the global Gramian matrix have global support and generally do not preserve the partition of unity and non-negativity properties of the associated B-splines. 
For illustration purposes, we plot quadratic $C^1$-continuous B-splines and their global dual functions \eqref{eq:dual_func} in Figs.~\ref{fig:bspline_dual}a and b, respectively. 
For other possibilities to compute the global dual basis of spline spaces, we refer to \cite{coox_global_dual_2017,dornisch_global_dual_2015} and the references therein. 

Alternatively, one can construct dual functions with the same minimal support as the corresponding B-splines \cite{Anitescu_dual_2019,miao_dual_2020,wunderlich_dual_2019,zou_mortar_2018}. One approach is to compute the dual function of the Bernstein polynomials, using the inverse of the Gramian matrix for a B\'ezier element, 
and then to apply the B\'ezier extraction operator to obtain the dual of the B-spline basis \cite{miao_dual_2020,zou_mortar_2018}. 
For illustration purposes, we plot a quadratic $C^1$-continuous B-spline function in \changed{Fig.~\ref{fig:bspline_dual}a}. 
We observe that while the global dual \eqref{eq:dual_func} is $C^{p-1}$-continuous, the dual basis with minimal support is discontinuous at the internal knots.

%---------------------------------------------------
\subsection{Approximate dual functions}\label{sec:approx_dual}

In this paper, we consider approximate dual functions of B-splines originally introduced in \cite{chui_wavelet_2004}. Let $\adual_i$, $i=1,\,\ldots,\,\ndofs$, denote the approximate dual functions of $C^{p-1}$-continuous B-splines of degree $p$ in the space $\sspace{}{} \, \subset \, \mathcal{P}^p$, 
and let $\adualv = \left[ \, \adual_1 \; \ldots \; \adual_\ndofs \, \right]^T$ be their function vector.
The approximate dual functions $\adual_i$ can again be constructed as linear combinations of the corresponding B-splines:
\begin{align}
    \adualv = \invmassMf \, \trialfv \, , \label{eq:approx_dual_func}
\end{align}
and thus span the same space $\sspace{}{}$ and reproduce polynomials in $\mathcal{P}^p$.

The approximate dual basis satisfies the bi-orthogonality constraint \eqref{eq:duality_constraint2} ``approximately'', in the sense that 
the matrix $\invmassMf$ is an approximate inverse of the Gramian matrix:
\begin{align}
    \invmassMf \approx \mat{G}^{-1} \, . \label{eq:approx_inv}
\end{align}
We note that the notation $\invmassMf$ is to imply that it is an approximate inverse. The approximate dual basis preserves important properties of the underlying B-spline basis, such as $C^{p-1}$ regularity and local support. \reviewerI{The construction of $\invmassMf$ does not require any matrix inversion, see \cite{chui_wavelet_2004} or the worked-out example in \ref{appendix_approx_inverse}}. 
It is a recursive computation, based on fundamental properties of B-splines such as the nestedness of spaces under knot insertion. 
For a 1D patch of splines, the approximate inverse matrix $\invmassMf$ is symmetric and positive definite with a bandwidth of at most $p+1$ \cite{chui_wavelet_2004}.

\begin{figure}[h!]
    \centering
    \subfloat[$C^1$-continuous B-spline basis.]{{\includegraphics[width=0.45\textwidth]{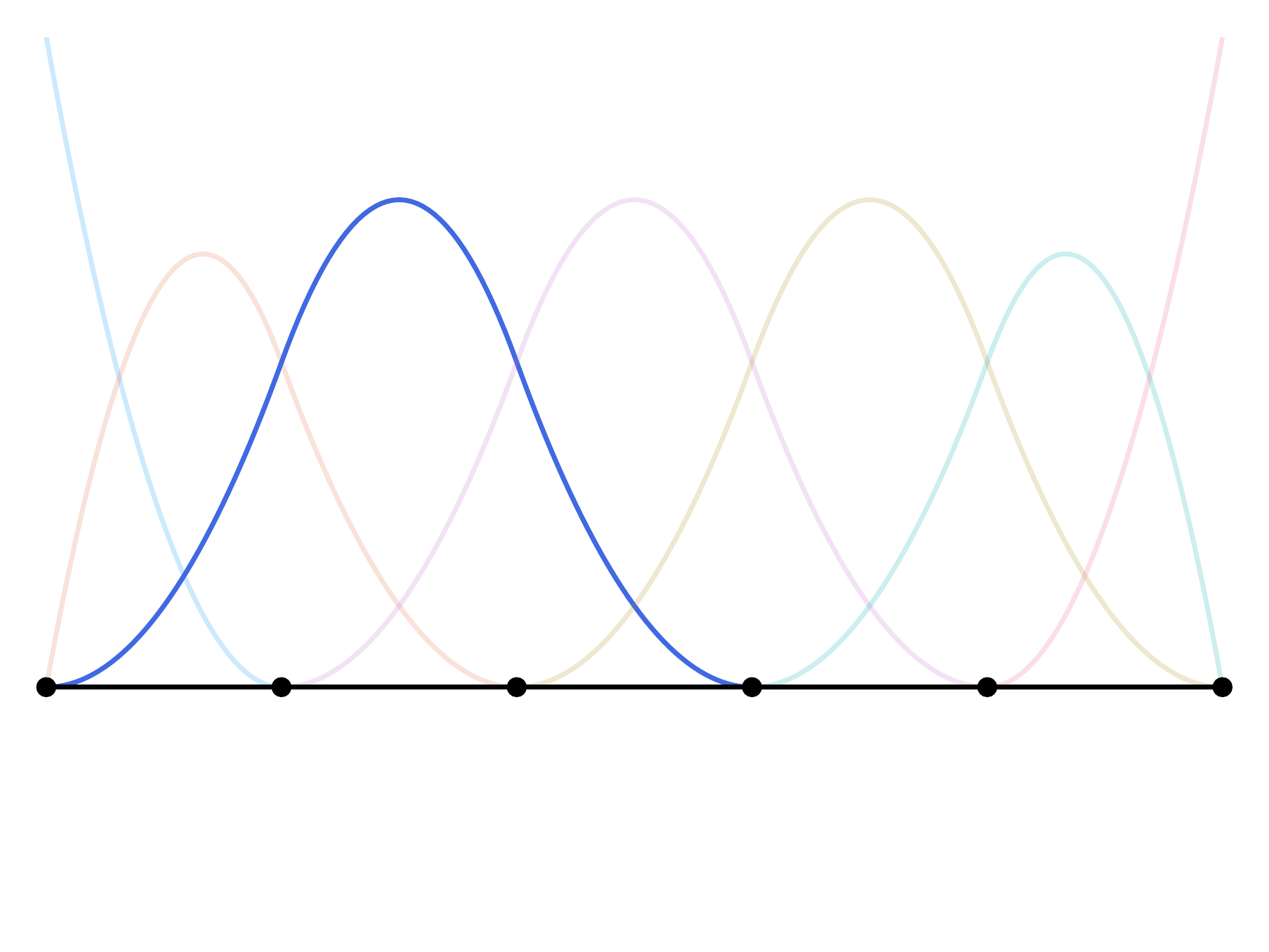} }}
    \subfloat[Approximate dual basis.]{{\includegraphics[width=0.45\textwidth]{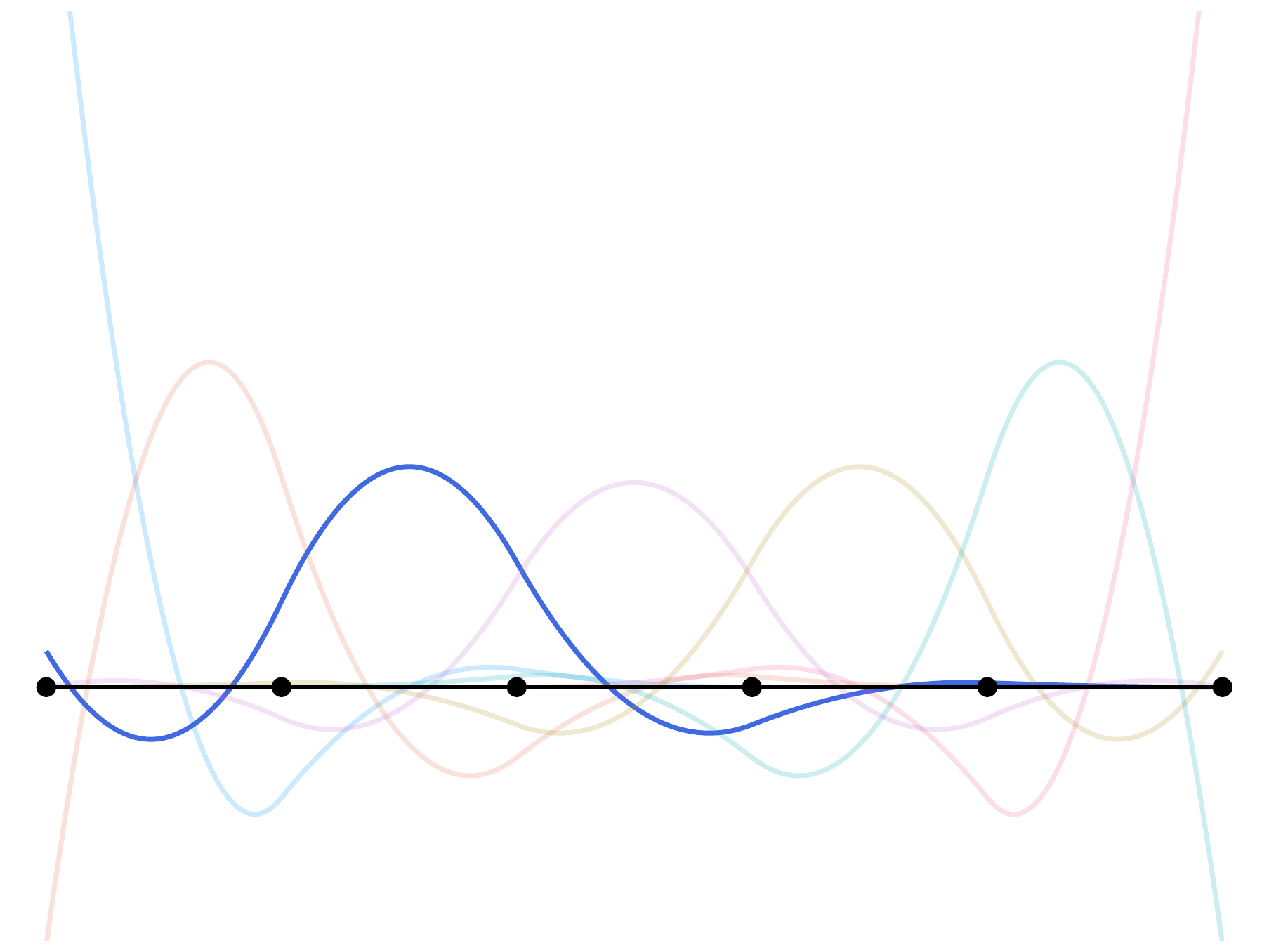} }}
    \caption{Approximate dual functions for a quadratic $C^1$-continuous B-spline basis.}
    \label{fig:approx_dual}
\end{figure}

For illustration purposes, we plot the approximate dual functions \eqref{eq:approx_dual_func}
next to the corresponding quadratic $C^1$-continuous B-spline basis.
We observe that the approximate dual functions preserve the $C^{p-1}$-continuity of the B-spline functions and have local support. 
Their support, however, is larger than that of the corresponding spline functions due to the band structure of $\invmassMf$. 

\begin{remark}
We note that in general, the approximate dual basis does not preserve the partition of unity and non-negativity properties of B-splines. 
    Partition of unity, however, could be restored by scaling the approximate dual with the inverse of its function value sum.  
\end{remark}

In this paper, 
we consider tensor-product extensions to multivariate dual B-spline functions. 
To this end, we use the same spline space $\sspace{}{}$ in every coordinate direction such that the multivariate spaces in the two-dimensional case is $\sspace{}{} \, \otimes \, \sspace{}{}$. 
In the parametric domain, the tensor product structure leads to matrices with Kronecker structure due to affine mapping.  
We note that the approximate dual functions can also be extended to non-uniform rational B-splines (NURBS) as shown in \cite{Dornisch_dual_basis_2017}. 

\subsection{An iterative approach for improving bi-orthogonality}\label{sec_multicorrector}

We outline a method to iteratively improve the accuracy of the approximation \eqref{eq:approx_inv} based on the predictor-multicorrector scheme introduced in \cite{Evans_corrector_scheme_2018}. 
We consider the linear system of equations:
\begin{align}
    \mat{G} \, \vect{u} \, = \, \vect{c} \, , 
\end{align}
where $\vect{u}$ and $\vect{c}$ denote the vector of unknowns and a known right-hand side vector, respectively. 
Replacing the Gramian matrix $\mat{G}$ by an approximation $\massMf$ with a simpler band structure leads to:
\begin{align}
    \massMf \, \vect{u} \, = \, \vect{c} \, , \label{eq:mxc}
\end{align}
which can be iteratively solved to obtain a more accurate solution as follows:
\begin{equation} \label{eq:iterative_scheme}
    \begin{cases}
        & \vect{u}^{(0)} = \boldsymbol{0} \, ,  \\
        & \text{for } i = 0, \, \ldots, \, r \,  \\
        & \quad \, \massMf \, \Delta \, \vect{u}^{(i)} = \vect{c} - \mat{G} \, \vect{u}^{(i)} \,  \\
        & \qquad \, \vect{u}^{(i+1)} = \vect{u}^{(i)} + \Delta \, \vect{u}^{(i)} \, \\
        & \text{end.}
    \end{cases}
\end{equation}
Here, $r$ denotes the number of iterations, which corresponds to the number of corrector passes, $r\,\geq\,0$. 
This iterative scheme is based on the recurrence relation\footnote{\reviewerI{This relation appears in each loop of \eqref{eq:iterative_scheme} when computing: $\massMf \, \vect{u}^{(i+1)}$ $ = \massMf \, \left(\vect{u}^{(i)} + \Delta \, \vect{u}^{(i)} \right)$ $= \massMf \, \vect{u}^{(i)} + \vect{c} - \mat{G} \, \vect{u}^{(i)}$ $= \vect{c} - \left[\mat{G} \, \invmassMf - \mat{I} \right]\massMf \,\vect{u}^{(i)}$ $= \vect{c} - \mat{A} \, \massMf \, \vect{u}^{(i)}$.}}: 
\begin{align}
    \massMf \, \vect{u}^{(i+1)} = \vect{c} - \mat{A} \, \massMf \, \vect{u}^{(i)} \, , \label{eq:relation_corrector_scheme}
\end{align}
where $\mat{A} = \mat{G} \, \invmassMf - \mat{I}$. 
Combining \eqref{eq:iterative_scheme} and \eqref{eq:relation_corrector_scheme} results in: 
\begin{align}
    \massMf \, \vect{u}^{(r)} & = \vect{c} - \mat{A} \, \massMf \, \vect{u}^{(r)} \, \nonumber \\
    & = \vect{c} - \mat{A} \, \left( \, \vect{c} - \mat{A} \, \massMf \, \vect{u}^{(r-1)} \, \right) \, = \, \vect{c} - \mat{A} \vect{c} + \underbrace{\mat{A} \, \mat{A}}_{\mat{A}^2} \, \massMf \, \vect{u}^{(r-1)} \, \nonumber \\
    & 
    = \, \ldots \, 
    = \, \left(\, \sum_{i=0}^{r} \, (-1)^{i} \, \mat{A}^{i} \, \right) \, \vect{c} \, .
\end{align}
Here, $\mat{A}^{i}$ denotes the matrix $\mat{A}$ of power $i$, $i=0,\ldots,r$. 
The matrix $\massMf$ in the initial equation \eqref{eq:mxc} can be then replaced by $\massM$:
\begin{align}
    \massM = \massM(r) := \left(\sum_{i=0}^{r} \, (-1)^{i} \, \mat{A}^{i} \right)^{-1} \, \massMf \, ,
\end{align}
to obtain a more accurate solution for $\vect{u}$. Hence, $\massM$ is an improved approximation of the Gramian matrix $\mat{G}$.

\begin{figure}[h!]
    \centering
    \subfloat[$r=0$, bandwidth = $2$]{{\includegraphics[width=0.45\textwidth]{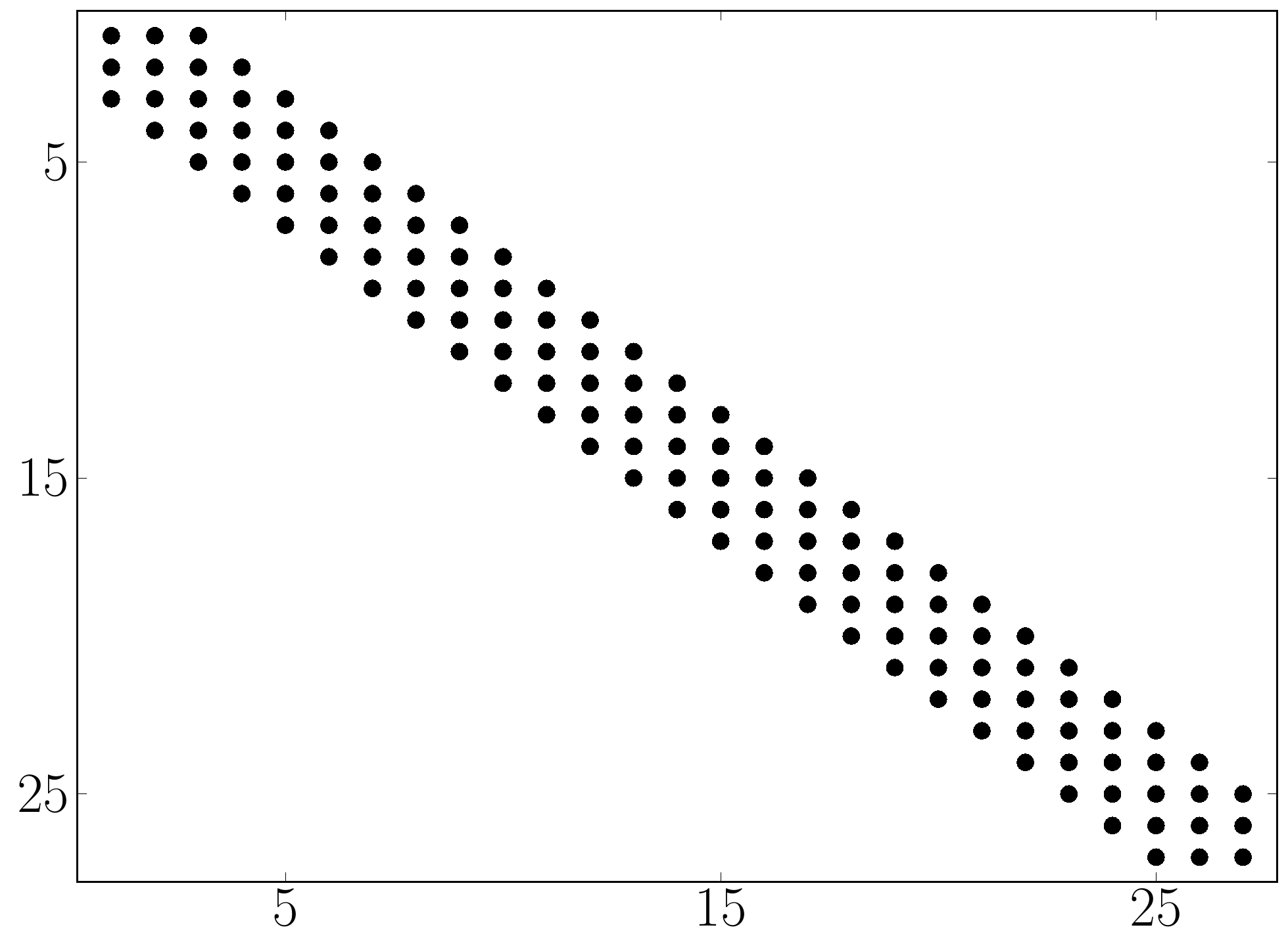} }}
    \subfloat[$r=1$, bandwidth = $6$]{{\includegraphics[width=0.45\textwidth]{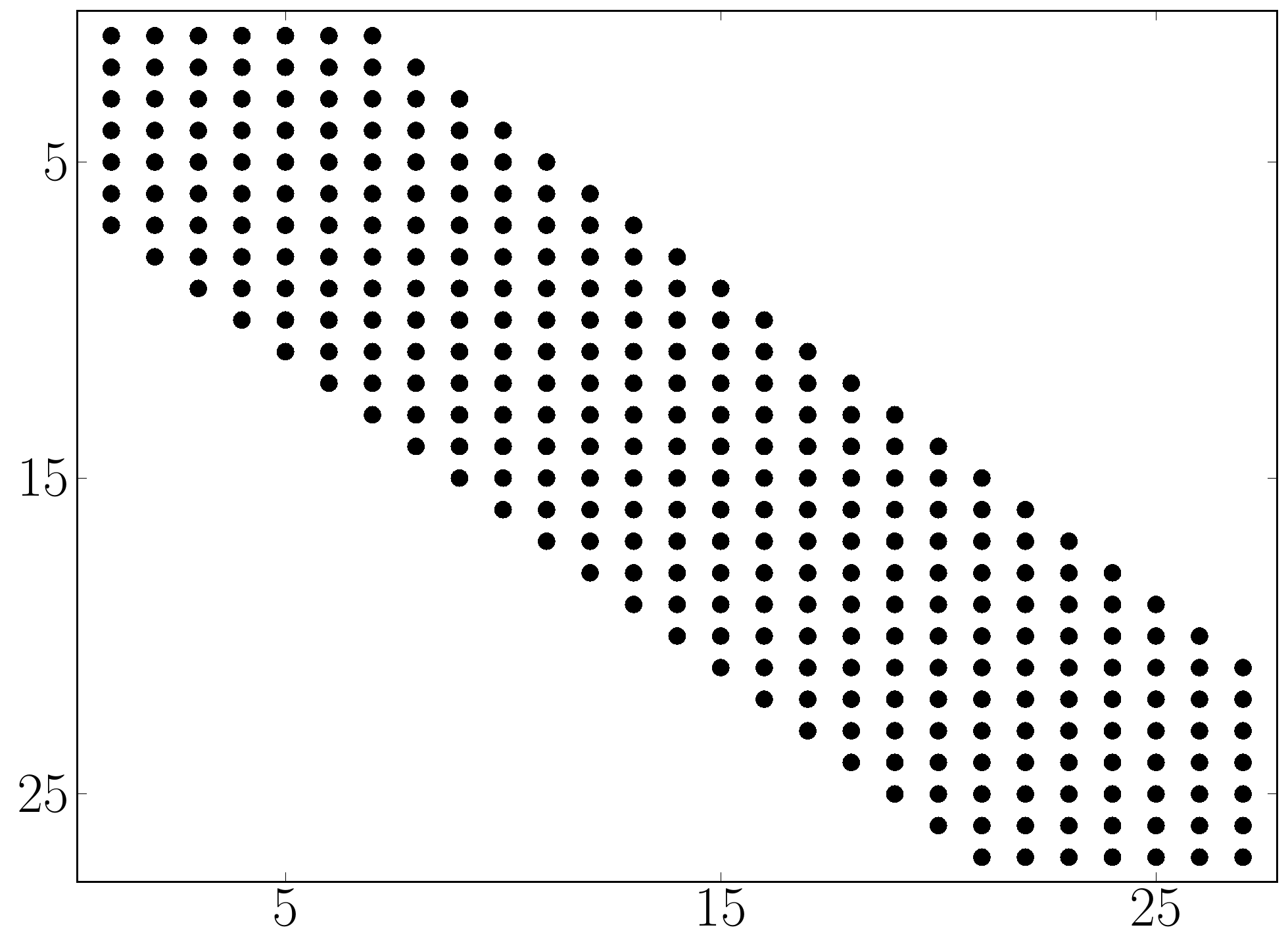} }}

    \subfloat[$r=2$, bandwidth = $10$]{{\includegraphics[width=0.45\textwidth]{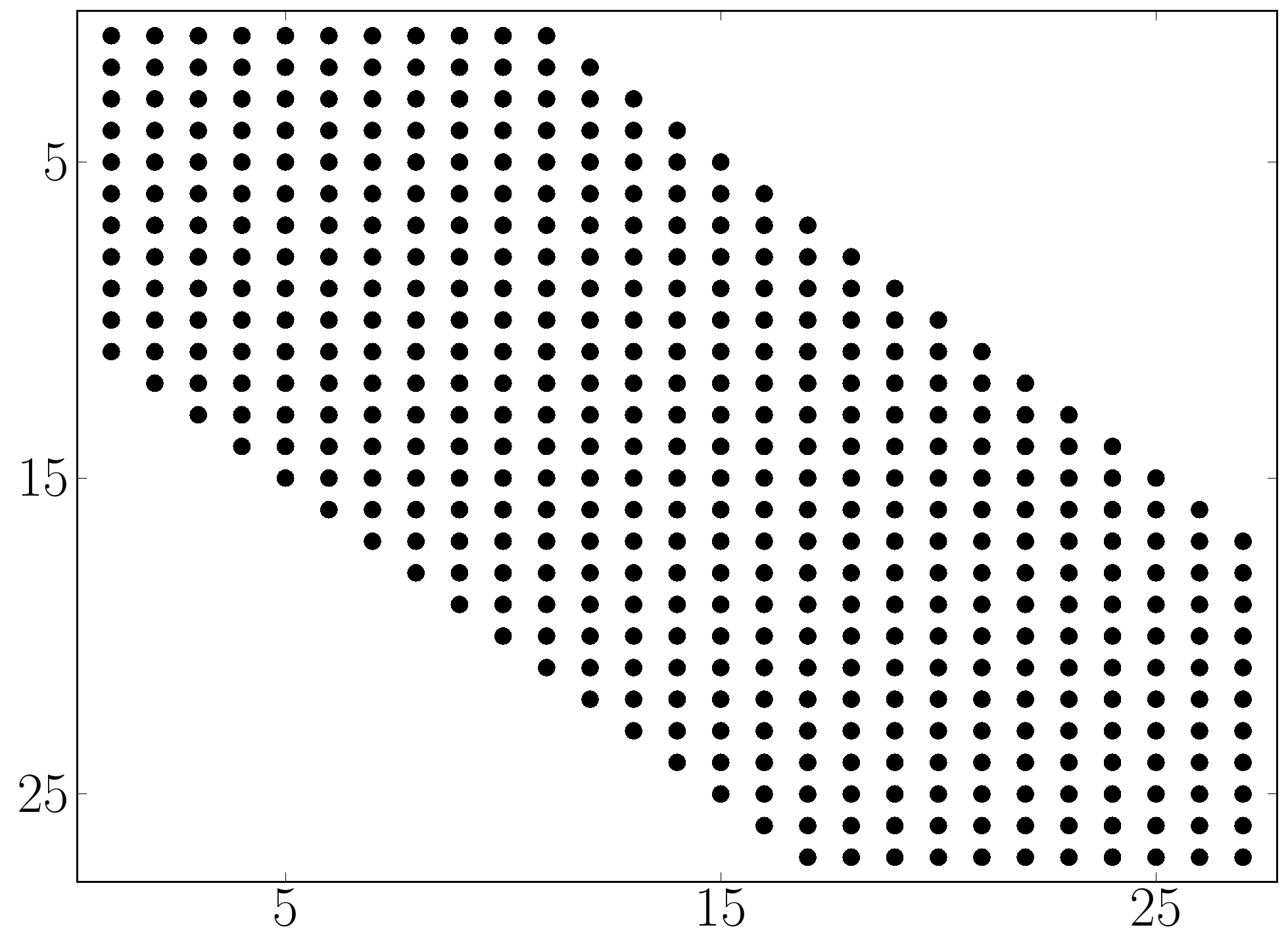} }}
    \subfloat[$r=3$, bandwidth = $14$]{{\includegraphics[width=0.45\textwidth]{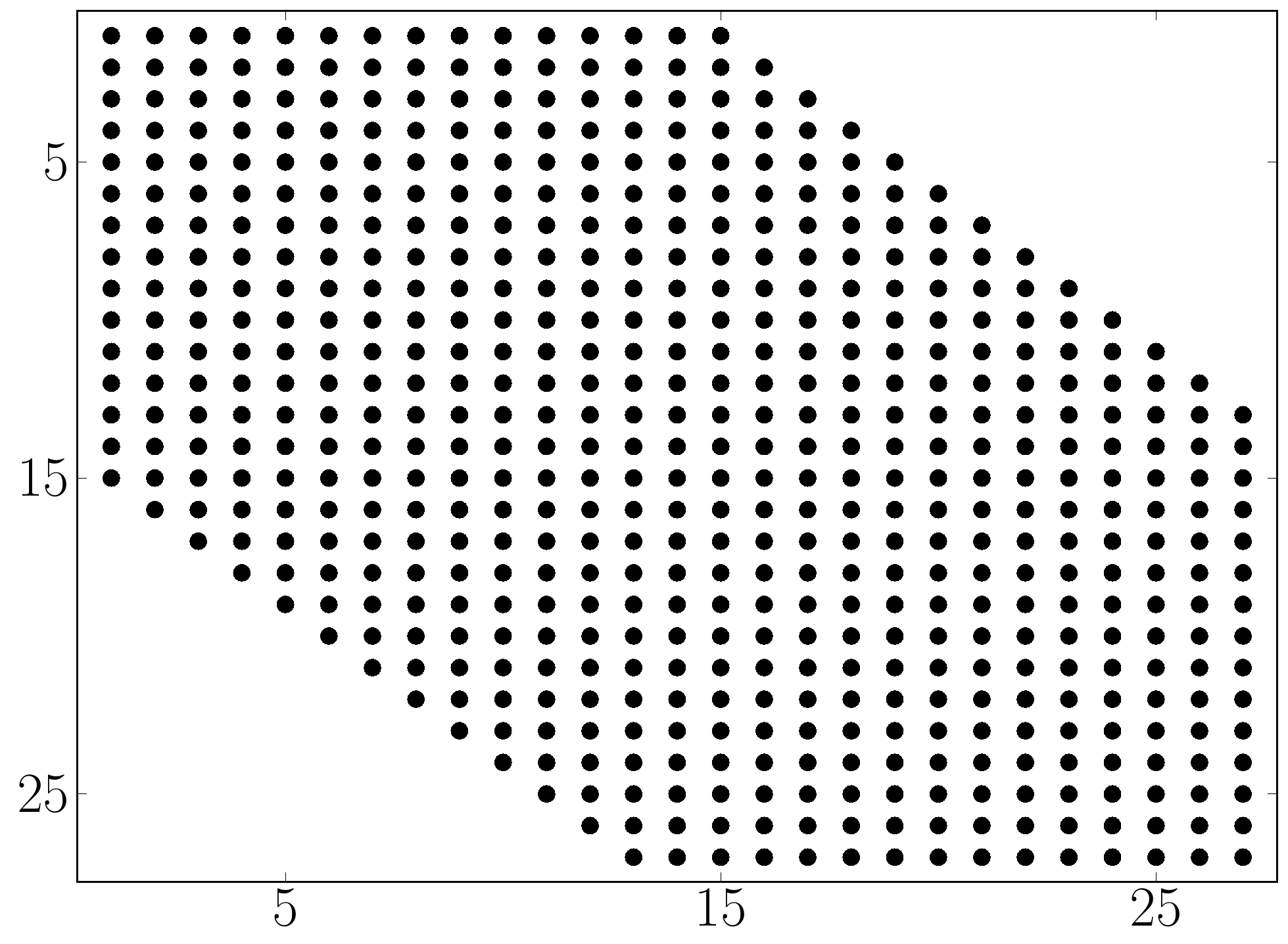} }}
\caption{Structure of the approximate inverse $\invmassM$ for increasing $r$, computed with \eqref{eq:imp_approx_inv}, corresponding to the approximate dual basis of quadratic $C^1$-continuous B-splines in 1D on $25$ B\'ezier elements.}
    \label{fig:bandwidth}
\end{figure}

\begin{table}[h!]
	\centering
	{\small
	\begin{tabularx}{1.0\linewidth}{ >{\centering\arraybackslash\hsize=.1\hsize}X | >{\centering\arraybackslash\hsize=.15\hsize}X >{\centering\arraybackslash\hsize=.15\hsize}X | >{\centering\arraybackslash\hsize=.15\hsize}X >{\centering\arraybackslash\hsize=.15\hsize}X  | >{\centering\arraybackslash\hsize=.15\hsize}X >{\centering\arraybackslash\hsize=.15\hsize}X }
	\toprule
	\multirow{2}{*}{$p$} & \multicolumn{2}{c|}{$r\,=\,0$} & \multicolumn{2}{c|}{$r\,=\,1$} & \multicolumn{2}{c}{$r\,=\,2$} \\
    \cmidrule{2-7}
     & Bandwidth $[-]$ & Support [elements] & Bandwidth $[-]$ & Support [elements] & Bandwidth $[-]$ & Support [elements] \\
    \midrule
    $2$ & $2$ & $7$          & $6$    & $11$         & $10$   & $15$ \\
    $3$ & $3$ & $10$         & $9$    & $16$         & $15$   & $22$\\
    $4$ & $4$ & $13$         & $12$   & $21$         & $20$   & $29$\\
    $5$ & $5$ & $16$         & $15$   & $26$         & $25$   & $36$\\
    \midrule
    $p$ & $p$ & $3\,p\,+\,1$ & $3\,p$ & $5\,p\,+\,1$ & $p\,(2\,r\,+\,1)$ & $2\,p\,r\,+\,3p\,+\,1$ \\
    \bottomrule
\end{tabularx}}
\caption{Bandwidth of $\invmassM$ and the support length of the approximate dual with different degrees $p$ and numbers of iterations $r$.}
\label{tab:band_and_support}
\end{table}

We are interested in the approximate inverse $\invmassM$:
\begin{align}
    \invmassM = \invmassM(r) := \invmassMf \, \left(\sum_{i=0}^{r} \, (-1)^{i} \, \mat{A}^{i} \right) \, . \label{eq:imp_approx_inv}
\end{align}
We note that for $r=0$, $\invmassM = \invmassMf$, i.e.\ no iteration is performed. 
On the one hand, increasing the number of iterations $r$ increases the accuracy of the approximation $\invmassM \, \approx \, \mat{G}^{-1}$. \reviewerI{On the other hand, it increases the bandwidth of the iteratively improved approximate inverse $\invmassM$, as illustrated in Fig.~\ref{fig:bandwidth} for 1D quadratic B-splines, 
and hence also increasing the support of the corresponding approximate dual functions}. 
We report the bandwidth of $\invmassM$ and the maximum support of the improved approximate dual functions in terms of the B\'ezier elements as a function of $p$ and $r$ in Table \ref{tab:band_and_support}. 
We emphasize that the computation of $\invmassM$ does not require any matrix inversion and its notation is to imply that $\invmassM$ is an approximate inverse of the Gramian matrix.

\section{A Petrov-Galerkin formulation with higher-order accurate mass lumping}\label{sec_lumpingscheme}

	In this section, we motivate and describe an isogeometric Petrov-Galerkin scheme that employs standard B-splines as trial functions and the corresponding approximate dual functions as test functions. Due to the approximate bi-orthogonality of the two function spaces, it enables higher-order mass lumping via the standard row-sum technique. We discuss several further relevant aspects, such as maintaining bi-orthogonality on mapped domains, strong enforcement of boundary conditions, and computational cost in explicit dynamics.

%---------------------------------------------------
\subsection{Model problem: Kirchhoff plate}\label{sec:petrov-galerkin}

We consider the vibration of an undamped Kirchhoff plate given by the following fourth-order partial differential equation:
\begin{align}\label{eomb}
    \rho \, d \, \ddot{u}(\vect{x}, t) \,+\, \frac{E\,d^3}{12\left(1-\nu^2\right)}\,\laplace^2 u(\vect{x}, t) \,=\, f(\vect{x}, t) \, , \quad \vect{x} \in \Omega \,, \; t \in \left[0, T \right] \, ,
\end{align} 
with mass density $\rho$, plate thickness $d$, Young's modulus $E$, and Poisson's ratio $\nu$. The spatial and time domains are given by $\Omega$ and $T$. To set up a well-defined initial boundary value problem, equation \eqref{eomb} needs to be complemented by suitable initial and boundary conditions. For ease of notation, we consider the case of a simply supported plate here. 
We thus have the following initial conditions at time $t=0$: 
    \begin{align}
        & u(\vect{x}, 0) \,=\, u_0(\vect{x}) \,, \quad  \dot{u}(\vect{x}, 0) \,=\, v_0(\vect{x}) \,,
    \end{align}
and the following boundary conditions:
    \begin{align}
    \label{bound_condi}
        u(\vect{x}, t) \,=\, 0 \,, \quad \nabla\, u(\vect{x}, t) \,\cdot\,\vect{n} \,=\, 0 \, \quad \text{on } \boundary\,,
    \end{align}
where $\vect{n}$ is the outward unit normal to the boundary $\boundary$.

Multiplying \eqref{eomb} by suitable test functions $w$, integrating over $\domain$, performing integration by parts twice, and substituting the boundary conditions \eqref{bound_condi} results in the following weak form: Find  $u \in \mathcal{S}$ such that $\forall \, w \in \mathcal{\tilde{S}}$:
\begin{align} \label{weom}
    \rho\,d \, \int_\domain \, \ddot{u} \, w \, \mathrm{d} \, \domain \,+\, \frac{E\,d^3}{12\left(1-\nu^2\right)} \, \int_\domain \, \laplace u \, \laplace w \, \mathrm{d} \, \domain \,=\, \int_\domain \, f \, w \, \mathrm{d} \, \domain \,.
\end{align}
We assume the spaces $\mathcal{S}$ and $\mathcal{\tilde{S}}$ are sufficiently regular such that the Laplace operator $\laplace$ can be applied. In addition, we assume that all material and geometric parameters are constants.

\begin{remark}
In the following sections, we will also present computational results for a straight Euler-Bernoulli beam and a Kirchhoff-Love shell. 
Based on the exposition for the Kirchhoff plate, we believe it is straightforward to extend our Petrov-Galerkin formulation to the one-dimensional beam and the three-dimensional shell. In the interest of conciseness of the exposition, we therefore do not repeat the derivation for these two models.
\end{remark}

%---------------------------------------------------
\subsection{Spatial discretization on geometrically mapped domains}

The approximate dual basis is constructed such that the bi-orthogonality constraint is approximately satisfied in the parametric domain, see \eqref{eq:approx_dual_func}. The basis functions, however, need to be mapped from the parametric domain to the physical domain, and this mapping will not be affine in the general case. Discretizing the solution $u$ and the test function $w$ in \eqref{weom} with standard B-splines and the original approximate dual functions \eqref{eq:approx_dual_func}, respectively, does not preserve the approximate bi-orthogonality in the physical domain. 

\reviewerI{To preserve the approximate bi-orthogonality in the physical domain}, we discretize $u$ with standard B-splines of degree $p$, $\trialf_i(x)$, $i=1,\,\ldots,\,\ndofs$, 
and \reviewerI{$w$ with modified approximate dual functions that are divided by the determinant of the Jacobian matrix of the mapping}:
\begin{align}
    \testf_i(\hat{\vect{x}}) \, := \, \frac{\adual_i(\hat{\vect{x}})}{\detJ(\hat{\vect{x}})} \, , \qquad \, i=1,\,\ldots,\,\ndofs \, . \label{eq:test_func}
\end{align}
Here, the function $\detJ(\hat{\vect{x}})$ denotes the determinant of the Jacobian matrix. It corresponds to a geometry map $\map: \pardomain \rightarrow \domain$ that maps a point $\hat{\vect{x}}$ from the parametric domain $\pardomain$ to a point $\vect{x}$ in the physical domain $\domain$. 
We assume that $\map$ is sufficiently smooth and invertible such that the Jacobian matrix and its inverse are well-defined.

The modified functions defined in \eqref{eq:test_func} constitute the approximate dual basis of the standard B-spline basis in the physical domain $\domain$, where it approximately satisfies the bi-orthogonality constraint \eqref{eq:duality_constraint2} in the same sense as the original approximate dual basis \eqref{eq:approx_dual_func} in the parametric domain $\pardomain$:
\begin{align}
    \left(\testf_i, \, \trialf_j\right)_\domain & = \int_\domain \, \testf_i(\hat{\vect{x}}) \, \trialf_j(\hat{\vect{x}}) \, \mathrm{d} \, \domain = \int_{\pardomain} \frac{\adual_i(\hat{\vect{x}})}{\detJ(\hat{\vect{x}})} \, \trialf_j(\hat{\vect{x}}) \, \detJ(\hat{\vect{x}}) \, \mathrm{d} \, \pardomain \, \nonumber \\
    & = \int_{\pardomain} \, \adual_i(\hat{\vect{x}}) \, \trialf_j(\hat{\vect{x}}) \, \mathrm{d} \, \pardomain \, 
    = \left(\adual_i,\, \trialf_j\right)_{\pardomain} \, .
\end{align}
The modified functions $\testf_i$ are linearly independent due to the linear independence of the approximate dual functions $\adual_i$ and preserve their local support. 
Their regularity, however, depends on the smoothness of the Jacobian $\detJ(\hat{\vect{x}})$. 

We are now in a position to discretize the weak form \eqref{eomb}. To this end, we first write the discretized displacement solution $u^h$ and the discretized test function $w^h$ as:
\begin{subequations}
    \begin{align}
        & u^h(\vect{x},t) = \begin{bmatrix}
            \trialf_1 (\vect{x}) \; \ldots \; \trialf_\ndofs (\vect{x}) 
        \end{bmatrix} \, \mat{\hat{u}}^h(t) \, ,  \nonumber \\ 
        & w^h(\vect{x}) = \begin{bmatrix}
            \testf_1 (\vect{x}) \; \ldots \; \testf_\ndofs (\vect{x}) 
        \end{bmatrix} \, \mat{\hat{w}}^h \, ,  \nonumber
    \end{align}
\end{subequations}
where $\mat{\hat{u}}^h(t)$ and $\mat{\hat{w}}^h$ are the unknown time-dependent displacement coefficients and the coefficients of the discrete test function, respectively. 
The discrete trial space $\mathcal{S}^h$ and the discrete test space $\mathcal{\tilde{S}}^h$ are then:
\begin{align}
    \mathcal{S}^h \,:=\, \Span{\trialf_i(\hat{\vect{x}})}_{i\in [1,\,N]} \,, \quad \mathcal{\tilde{S}}^h \,:=\, \Span{\testf_i(\hat{\vect{x}})}_{i\in [1,\,N]} \,=\, \Span{\frac{\adual_i(\hat{\vect{x}})}{\detJ(\hat{\vect{x}})}}_{i\in [1,\,N]} \,. 
 \end{align}
The resulting semidiscrete formulation of \eqref{weom} is: Find $u^h \in \mathcal{S}^h \, \subset \, \mathcal{S}$ such that $\forall \, w^h \in \mathcal{\tilde{S}}^h \, \subset \, \mathcal{\tilde{S}}$:
\begin{align}\label{dweom}
    \rho\,d \, \int_\domain \, \ddot{u}^h \, w^h\, \mathrm{d} \, \domain \,+\, \frac{E\,d^3}{12\left(1-\nu^2\right)} \, \int_\domain \, \laplace u^h \, \laplace w^h \, \mathrm{d} \, \domain \,=\, \int_\domain \, f \, w^h \, \mathrm{d} \, \domain \,.
\end{align}

\begin{figure}[t!]
    \centering
    \subfloat[$\mathcal{S}^h \,:=\, \Span{\trialf_i(\hat{x})}_{i\in [1,\,7]}$]{{\includegraphics[width=0.45\textwidth]{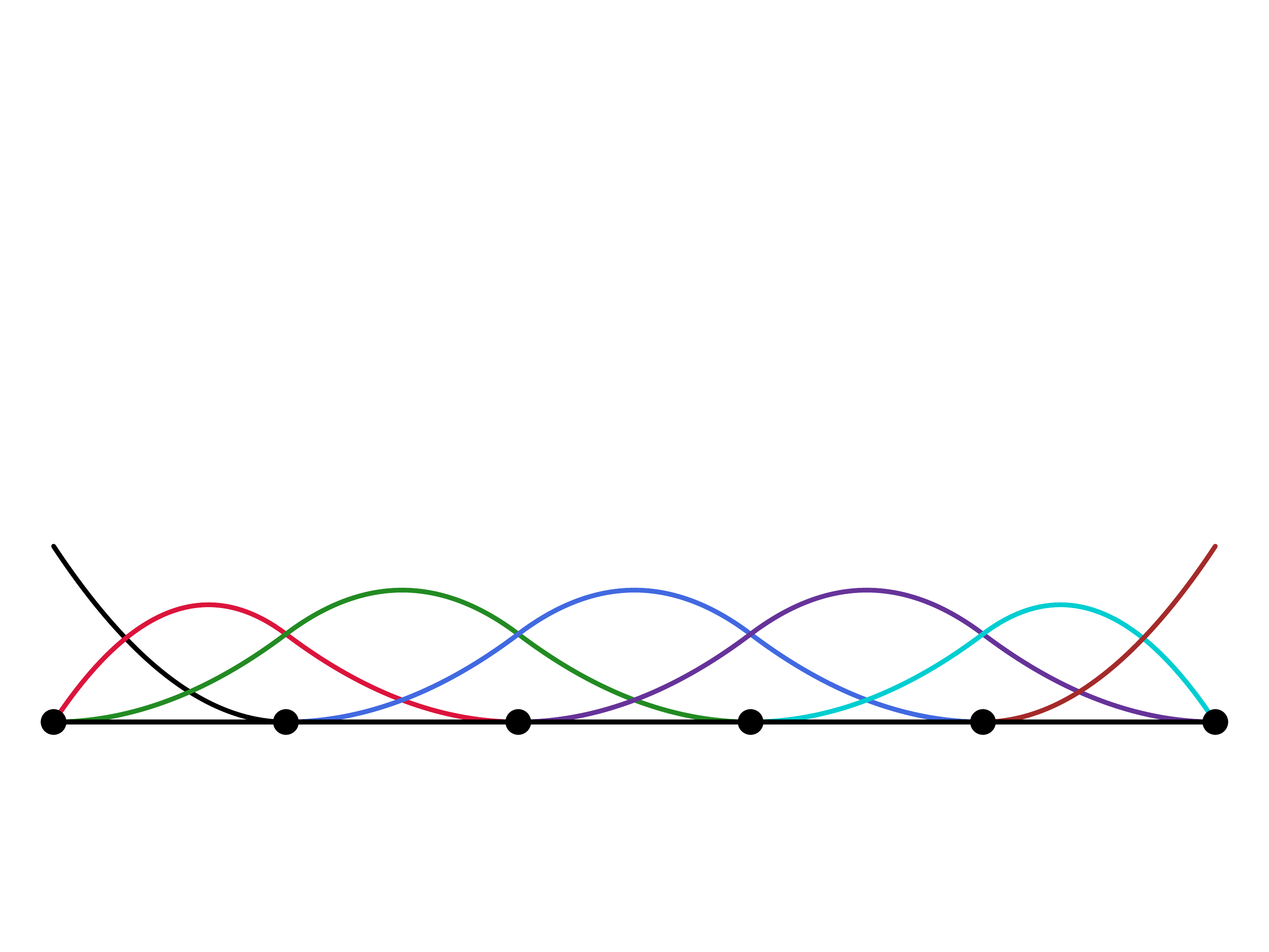} }}
    \subfloat[$\mathcal{\tilde{S}}^h \,:=\, \Span{\frac{\adual_i(\hat{x})}{\detJ(\hat{x})}}_{i\in [1,\,7]}$]{{\includegraphics[width=0.45\textwidth]{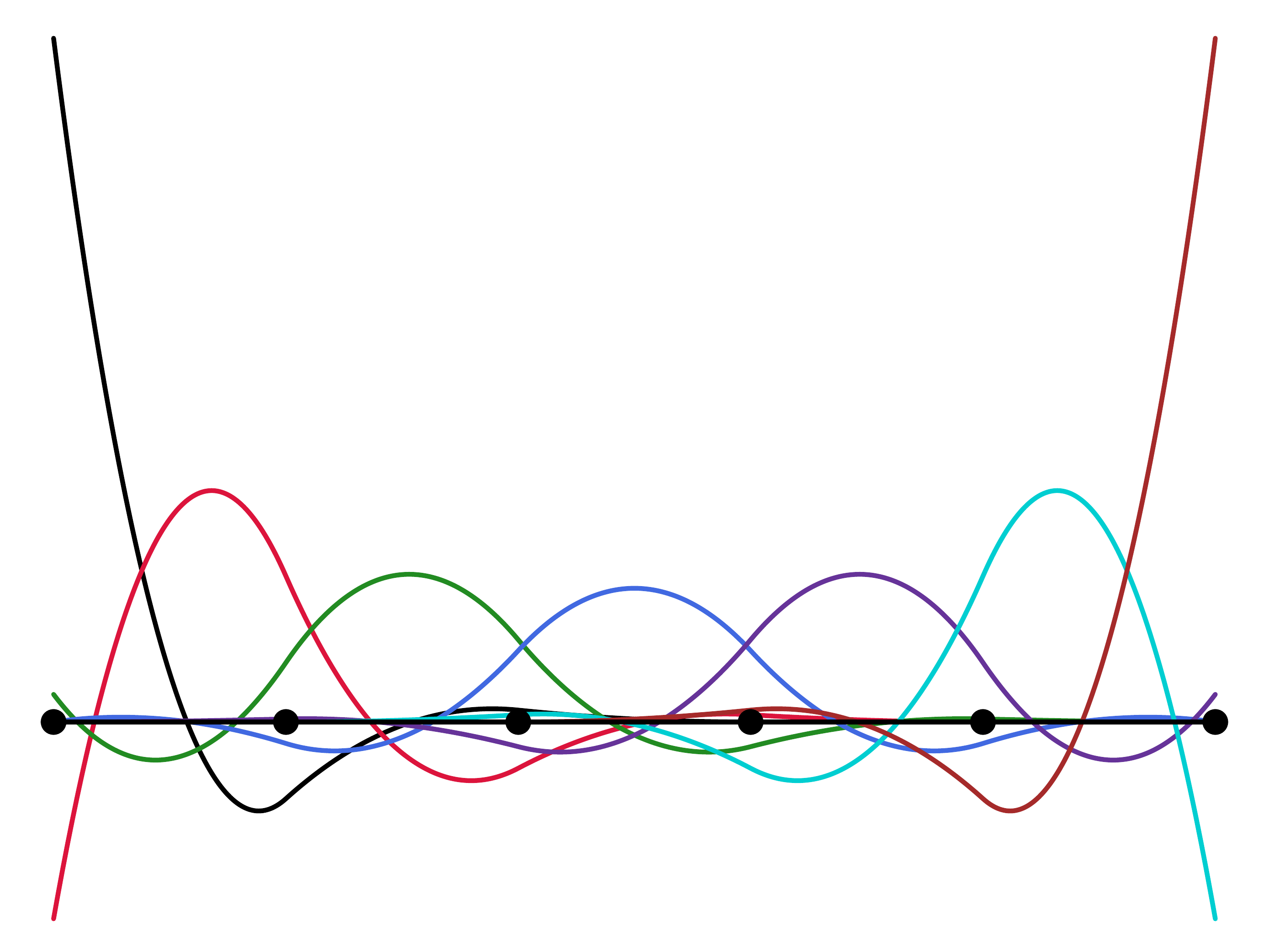} }}

    \subfloat[Jacobian determinant function for a quarter circle.]{{\includegraphics[width=0.45\textwidth]{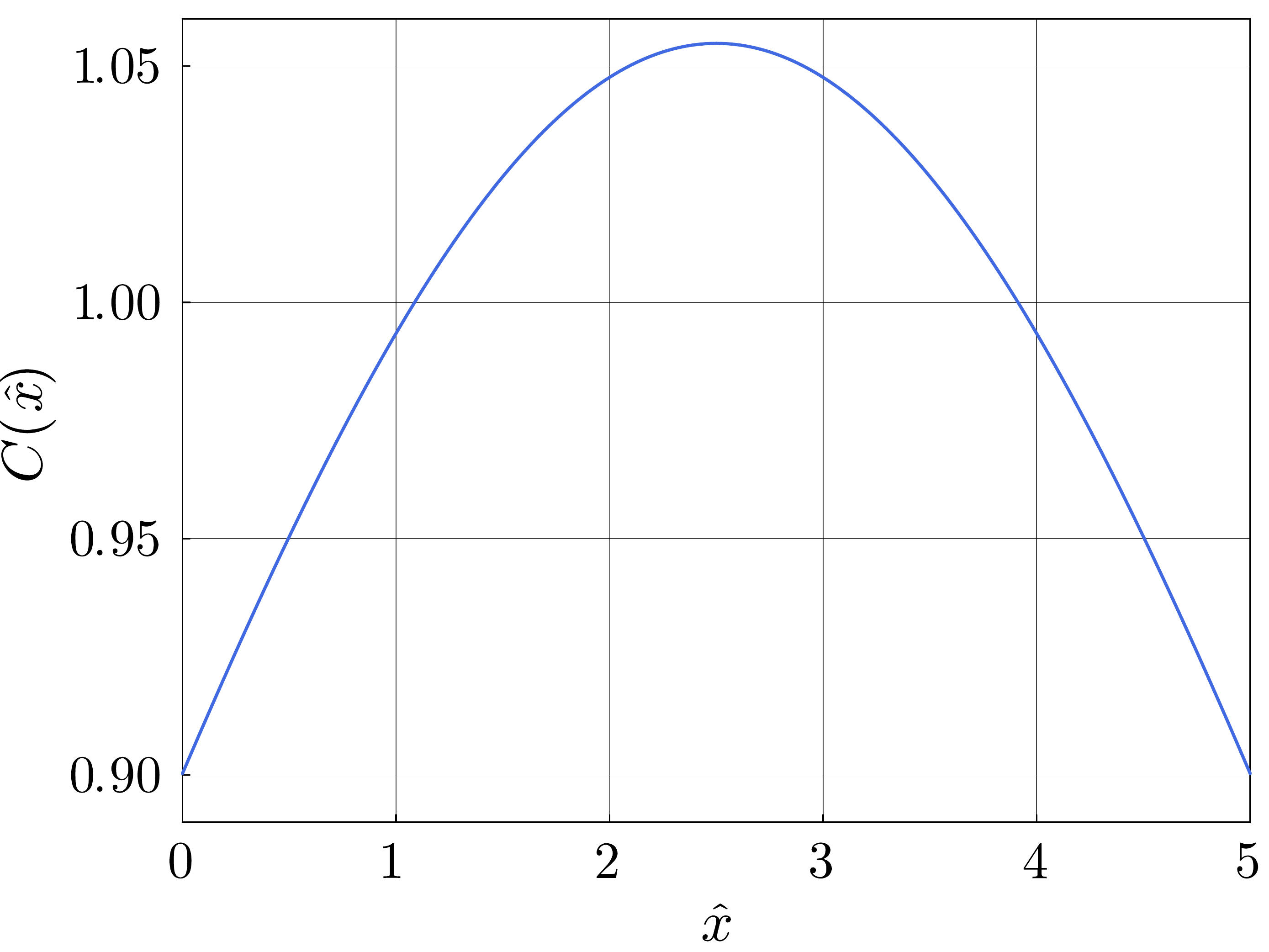} }}

    \vspace{0.3cm}
    \caption{Petrov-Galerkin discretization in space: discrete trial space $\mathcal{S}^h$ of standard B-splines and discrete test space $\tilde{\mathcal{S}}^h$ of modified approximate dual functions for a 1D patch of quadratic B-splines and a non-constant Jacobian determinant.}
    \label{fig:trial_test_funcs}
\end{figure}

Figure~\ref{fig:trial_test_funcs} shows a one-dimensional example: $\mathcal{S}^h$ contains a standard quadratic B-spline patch, and $\mathcal{\tilde{S}}^h$ contains the corresponding modified approximate dual functions. For illustration purposes, we assume a geometric map based on a quarter circle with unit radius represented by NURBS. We also plot the corresponding non-constant function of the Jacobian determinant.

{\remark
In general, due to the Jacobian determinant $\detJ(\hat{\vect{x}})$ in the denominator, the approximate dual functions \eqref{eq:test_func} span a different space as the corresponding B-splines. 
Therefore, $\mathcal{S}^h$ and $\mathcal{\tilde{S}}^h$ represent two different spaces. Hence, the semidiscrete formulation \eqref{dweom} is in general a Petrov-Galerkin formulation. 
Only in the special case of a constant Jacobian determinant, the modified approximate dual function functions \eqref{eq:test_func} span the same space as the original approximate dual functions \eqref{eq:approx_dual_func}, since they are only scaled by a constant factor, and thus \eqref{dweom} falls back to a Galerkin formulation.}

{\remark
The differential operators in \eqref{dweom} require the following derivatives of the modified approximate dual functions \eqref{eq:test_func}: 
\begin{subequations}\label{deriv_testfunc1}
    \begin{align}
        & \testf_{i,\alpha} = \left(\frac{\adual_i}{\detJ} \right)_{,\alpha} = \frac{1}{\detJ} \, \adual_{i,\alpha} - \frac{1}{\detJ^2} \, \adual_i \, \detJ_{,\alpha} \, , \\
        & \testf_{i,\alpha \beta} = \left(\frac{\adual_i}{\detJ} \right)_{,\alpha \beta} = \frac{1}{\detJ} \, \adual_{i,\alpha \beta} - \frac{1}{\detJ^2} \, \adual_{i,\alpha} \, \detJ_{,\beta} - \frac{1}{\detJ^2} \, \adual_{i,\beta} \, \detJ_{,\alpha} - \frac{1}{\detJ^3} \, \adual_i \, \left(\detJ_{,\alpha \beta} \, \detJ - 2 \detJ_{,\alpha} \, \detJ_{,\beta} \right) \, .
    \end{align}    
\end{subequations}
For the plate model considered here, the indices $\alpha$ and $\beta$ take values ${1,2}$ and the notation $(\cdot)_{,\alpha}$ denotes the derivative with respect to the $\alpha^{\text{th}}$ Cartesian coordinate.}

%---------------------------------------------------    
\subsection{Higher-order accurate mass lumping scheme}\label{sec:lumping}

In the following, we again use of the following vector notation: $\testfv = \left[ \, \testf_1 \; \ldots \; \testf_\ndofs \, \right]^T$ is the vector of modified approximate dual functions, and $\adualv$ and $\trialfv$ the corresponding vectors of approximate dual functions and B-splines, respectively. Moving towards explicit dynamics, we now rethink the semidiscrete variational formulation \eqref{fig:trial_test_funcs} in the format \eqref{eq1}, bringing its second term on the right-hand side. Focusing on the left-hand side, we find the following (consistent) mass matrix $\mat{M}$:
\begin{align}
    \mat{M} = \rho\,d \, \int_\domain \, \testfv \, \trialfv^T \, \mathrm{d} \, \domain = \rho\,d \, \int_{\pardomain} \, \frac{\adualv}{\detJ(\hat{\vect{x}})} \, \trialfv^T \, \detJ(\hat{\vect{x}}) \, \mathrm{d} \, \pardomain = \rho\,d \, \int_{\pardomain} \, \adualv \, \trialfv^T \, \mathrm{d} \, \pardomain \, .  \label{eq:mass1}
\end{align}
integrated over the physical domain $\Omega$ and the parametric domain $\pardomain$. 
Employing \eqref{eq:approx_dual_func} leads to:
\begin{align}\label{eq:mass2}
    \mat{M} = \rho\,d \, \invmassMf \, \int_{\pardomain} \, \trialfv \, \trialfv^T \, \mathrm{d} \, \pardomain = \rho\,d \, \invmassMf \, \mat{G} \; = \; \rho\,d \, \mat{\hat{I}} \, ,
\end{align}
and hence to the following semidiscrete form of our model problem:
\begin{align}
\rho\,d \, \mat{\hat{I}} \; \vect{\ddot{u}}^n \; = \;  \mat{F}^n_{\text{ext}} \; - \; \int_{\Omega} \boldsymbol{B}(\vect{x}^n)^T \; \boldsymbol{\sigma}^n \; \mathrm{d} \Omega \; ,
\label{eq29}
\end{align}
where the mass matrix, due to \eqref{eq:duality_constraint2}, corresponds to an approximation of the identity matrix, $\mat{\hat{I}} \approx \mat{I}$, scaled by a scalar that corresponds for our model problem to the density $\rho$ times the thickness $d$ of the plate. The approximation $\mat{\hat{I}}$, however, is not diagonal, and hence precludes efficient explicit dynamics, when e.g.\ a central difference scheme in time is applied to \eqref{eq29}.

Our central idea to enable efficient higher-order accurate explicit dynamics based on \eqref{eq29} is simple: we apply standard row-sum mass lumping to diagonalize the approximate identity matrix $\mat{\hat{I}}$. It is straightforward to show that this leads in fact to the identity matrix $\mat{{I}}$.

\vspace{0.36cm}

\noindent\textbf{Theorem}: \reviewerI{The application of row-sum lumping to the matrix} $\left(\adual_i, \, \trialf_j \right)_{\pardomain}$ \reviewerI{yields the identity matrix}.

\vspace{0.36cm}

\noindent\textbf{Proof}: 
    Let $P \, \in \, \mathcal{P}^p$ be a polynomial of degree $p$ in the parametric domain $\pardomain$,  
    and consider the spline basis $\trialf_i$, $i=1,\, \ldots, \, \ndofs$ and its approximate dual basis $\adual_i$. 
    There exists a set of coefficients $u_i$ such that:
    \begin{align}
        \left(\adual_i, \, \trialf_j \right)_{\pardomain} \, u_j = \left(\adual_i, \, P \right)_{\pardomain} \, .
    \end{align}
    The spline basis $B_i$ reproduces polynomials $P \, \in \, \mathcal{P}^p$, i.e. $\sum_{j=1}^\ndofs \, \trialf_j \, u_j = P$. 
    If $f=1$, $u_j = 1$ due to the partition of unity $\sum_{j=1}^\ndofs \, \trialf_j(\hat{x}) = 1$. We thus obtain:
    \begin{align}
        \left(\adual_i, \, \trialf_j \right)_{\pardomain} \, 1 = \left(\adual_i, \, 1 \right)_{\pardomain} = 1 \, .
    \end{align}
 
\vspace{0.36cm}

Our hypothesis is that the effect of row-sum lumping does not preclude higher-order accuracy, as  
\changed{$\hat{\mat{I}}$} 
is already close to the true identity matrix, which is of course diagonal. We will show by way of numerical examples that this is indeed the case.

%---------------------------------------------------
\subsection{Computational cost}

The computational cost in explicit dynamics primarily depends on the critical time step size, the number of quadrature points, and the evaluation of the internal force vector per quadrature point. We expect that the first two aspects do not differ significantly between the isogeometric Galerkin method and our isogeometric Petrov-Galerkin method. In this context, we refer to outlier removal techniques to prevent prohibitively small critical time steps \cite{Hiemstra_outlier_2021,deng2021boundary,Nguyen_outlier_2022} and to advanced quadrature schemes for spline discretizations that significantly reduce the number of quadrature points with respect to standard Gauss quadrature per B\'ezier element \cite{schillinger2014reduced,hiemstra2017optimal,hiemstra2019fast}.

\begin{figure}[b!]
    \centering
    \subfloat[Patch of approximate dual functions.]{{\includegraphics[width=0.45\textwidth]{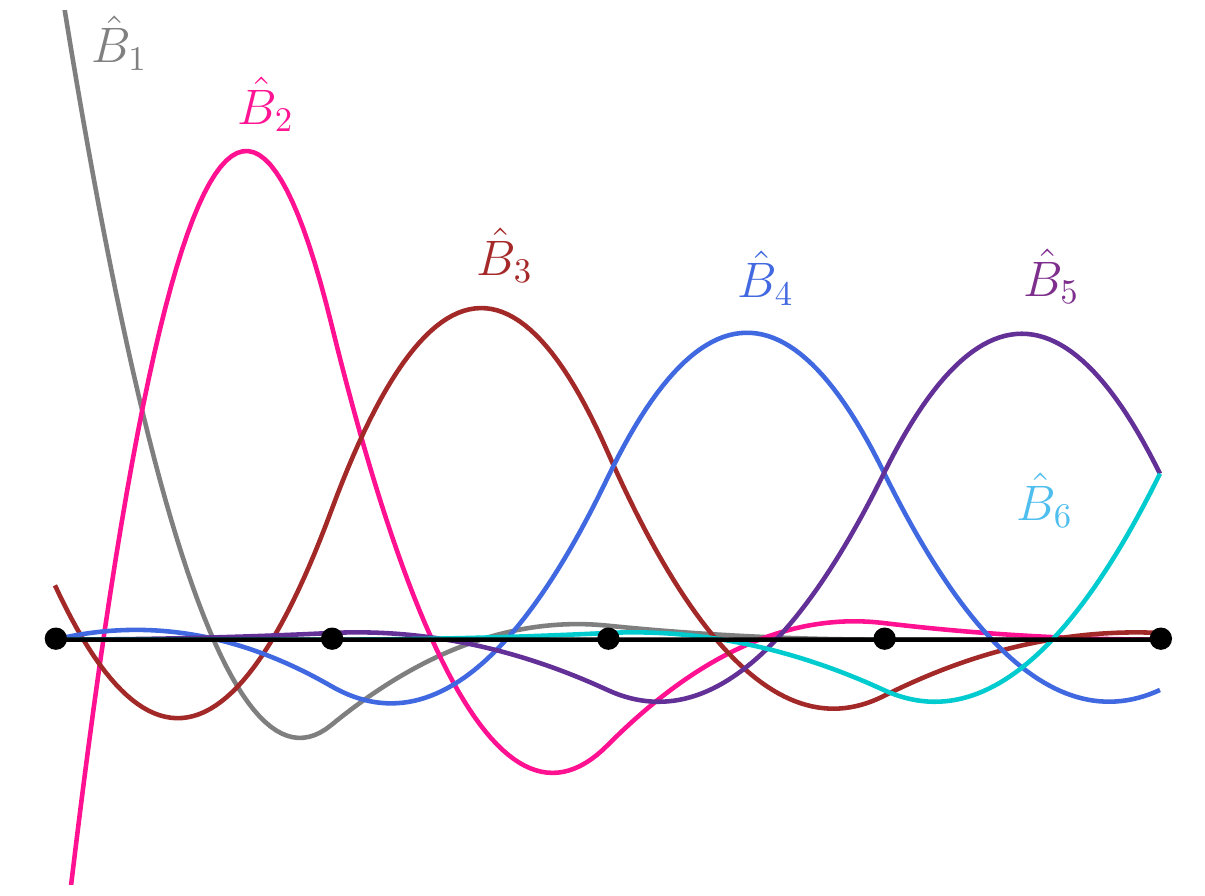} }} \hspace{0.2cm}
    \subfloat[Boundary functions replaced by B-splines.]{{\includegraphics[width=0.45\textwidth]{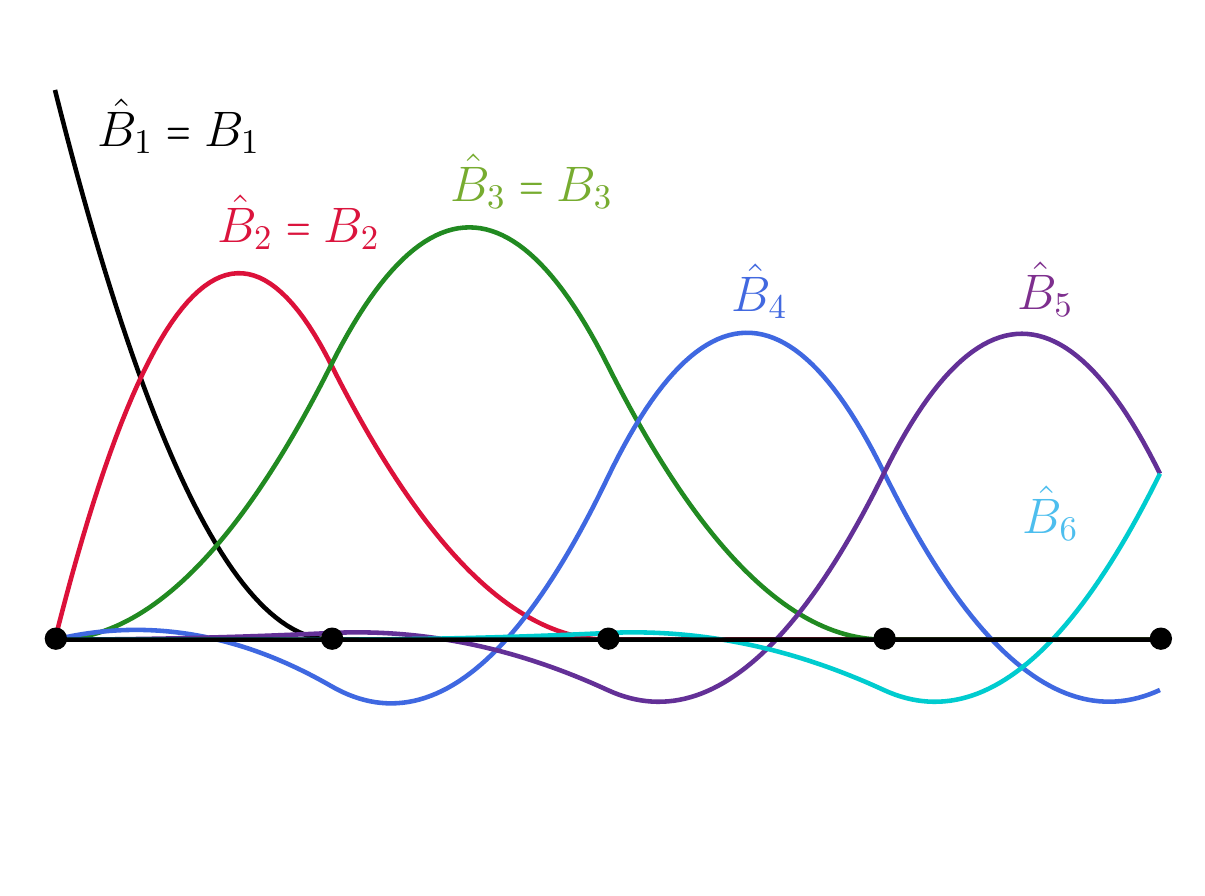} }}
    
    \caption{Replacing approximate dual functions with support at a boundary with interpolatory B-splines, illustrated here for one end of a quadratic B-spline patch, recovers the ability to strongly impose Dirichlet constraints.}
    \label{fig:combined_basis}
\end{figure}

\begin{figure}[t!]
    \centering
    \subfloat[Consistent mass matrix.]{{\includegraphics[width=0.45\textwidth]{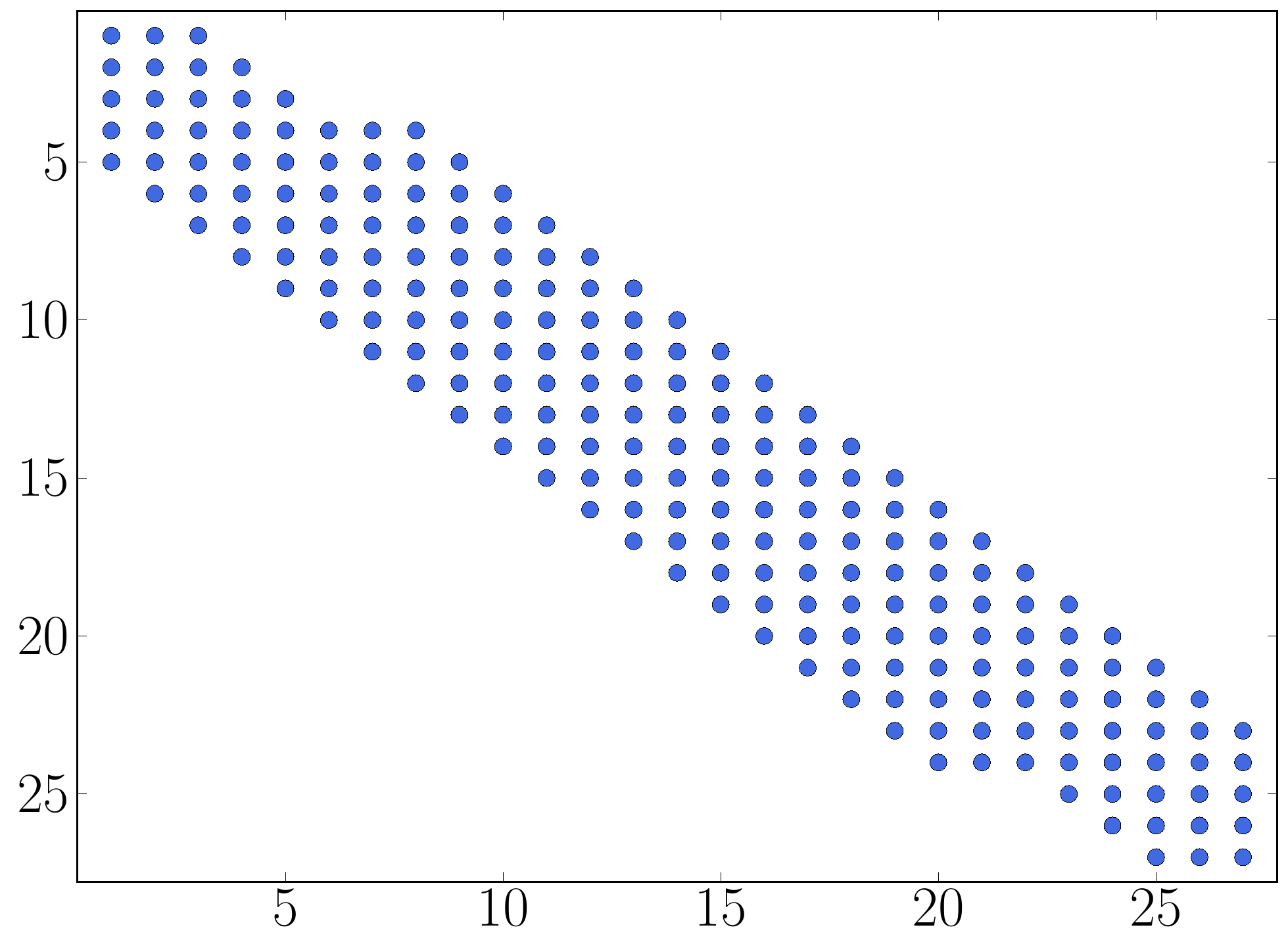} }} \hspace{0.2cm}
    \subfloat[(Partially) Lumped mass matrix.]{{\includegraphics[width=0.45\textwidth]{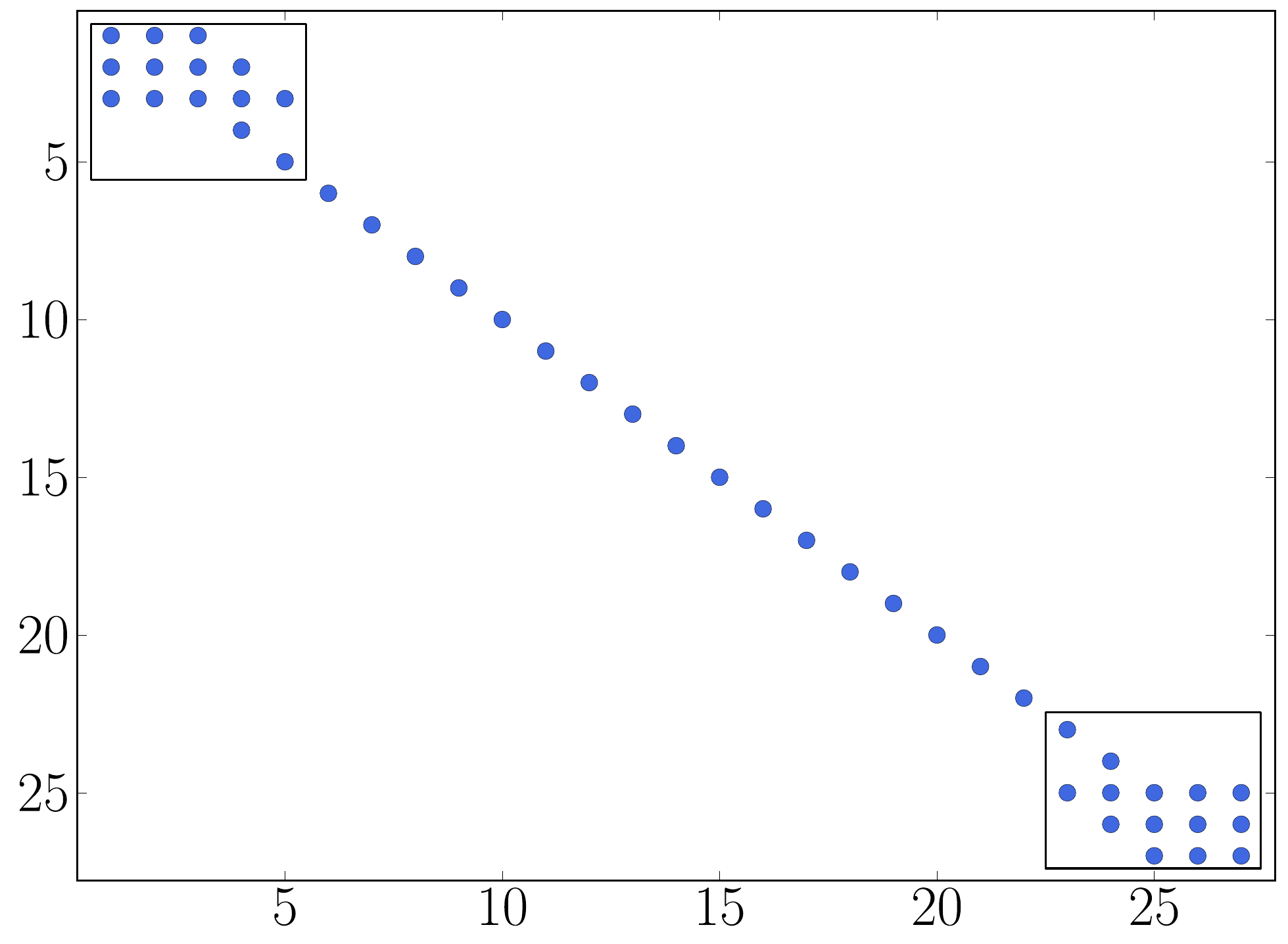} }}

    \subfloat[Inverse of the (partially) lumped mass matrix.]{{\includegraphics[width=0.45\textwidth]{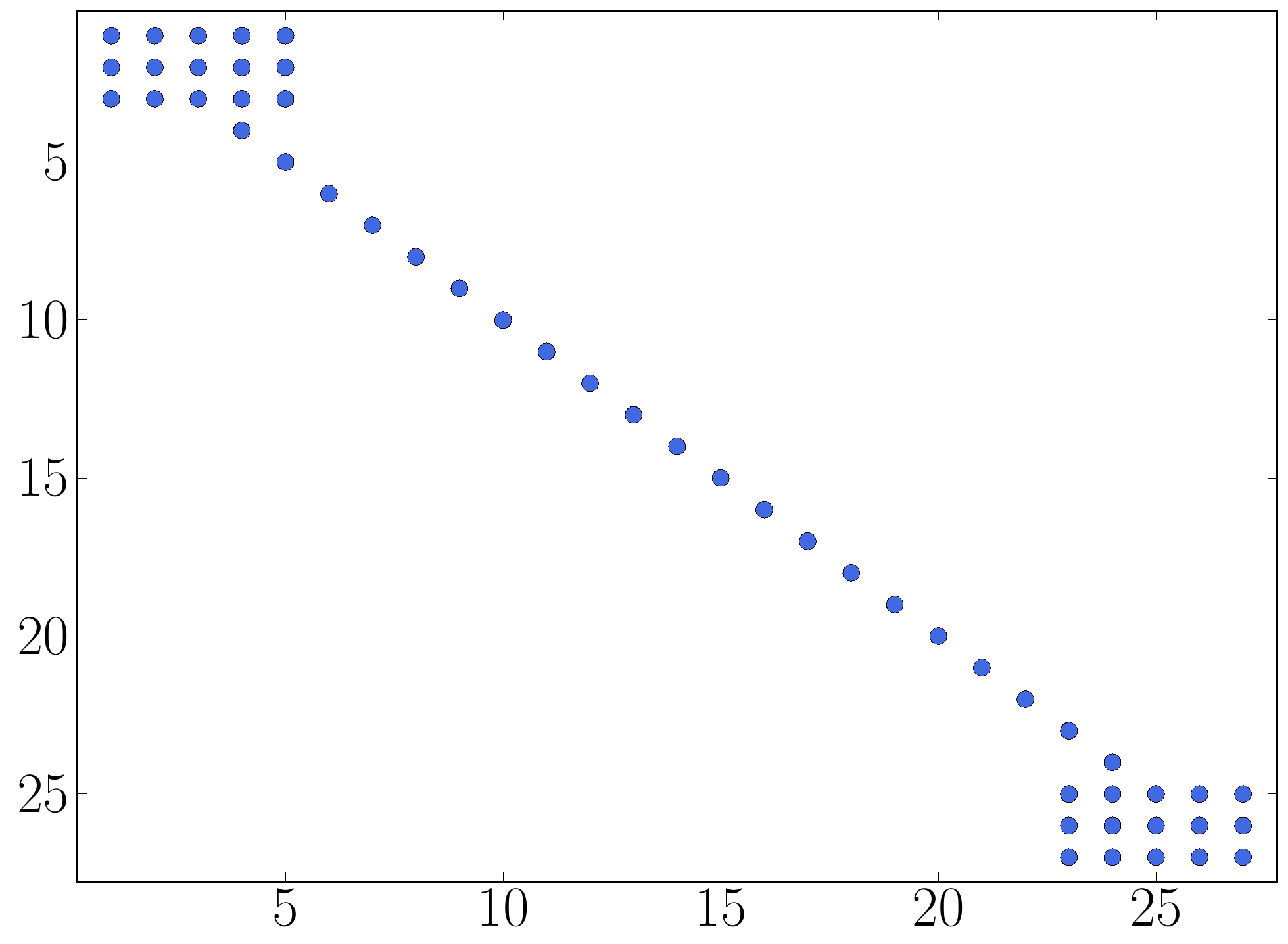} }}
    \caption{Population of the consistent mass matrix, (partially) lumped mass matrix and its inverse for our Petrov-Galerkin scheme for an Euler-Bernoulli beam problem, discretized by quadratic splines on 25 B\'ezier elements.}
    \label{fig:mass_struct}
\end{figure}

The cost of the evaluation of the internal force vector requires a more detailed analysis. Just as standard B-splines (or NURBS), the modified approximate dual basis functions can be cast into a B\'ezier or Lagrange extraction format per element \cite{borden2011isogeometric,schillinger2016lagrange}, whose derivation from relation \eqref{eq:approx_inv} and the appoximate inverse of the Gramian matrix is straightforward. The extraction operators can be computed in an offline step, such that they do not contribute to the online cost of explicit dynamics calculations. In a Petrov-Galerkin sense, bilinear stiffness forms are not symmetric, as we are using different test and trial functions on the right-hand side. In explicit dynamics, this is not an issue per se, as the stiffness matrix is never assembled, stored, or inverted. The central difference to the standard Galerkin scheme, however, is that the support of the modified approximate dual functions is larger than the one of the standard B-splines (or NURBS). \reviewerI{According to Table~\ref{tab:band_and_support}, the support of an approximate dual function is up to $3p+1$ B\'ezier elements in each parametric direction, as compared to $p+1$ for the corresponding B-spline. For two-dimensional elements, such as plates or shells, due to the increased support of the basis functions of the discrete test space, the internal force vector has eight to nine times as many entries, and we have to expect approximately eight to nine times as many basis function related operations to compute it}. In practical scenarios, computationally costly routines to take into account nonlinear material behavior, such as radial return algorithms in plasticity, do not depend on the cost or number of the basis functions of the discrete test space, and hence the net increase in computational cost per quadrature point will be much lower. It remains to be seen how our Petrov-Galerkin scheme can be implemented in a competitive fashion and how much this disadvantage in computational cost then really matters in comparison with its advantage in accuracy.

%---------------------------------------------------
\subsection{Imposition of Dirichlet boundary conditions}\label{sec:boundary_cond}

In general, the approximate dual basis \eqref{eq:test_func} does not preserve the interpolatory ends of the underlying B-spline basis. 
To enable the strong imposition of Dirichlet boundary conditions, we recover the interpolatory ends by simply replacing the original approximate dual functions with support at the Dirichlet boundary by the corresponding interpolatory B-splines. 
Figure~\ref{fig:combined_basis} illustrates this simple idea for one end of a quadratic B-spline patch.

Figure~\ref{fig:mass_struct} illustrates the structure of the resulting mass matrix $\mat{M}$ for an Euler-Bernoulli beam model computed with our Petrov-Galerkin scheme, where we use a 1D patch of quadratic basis functions and their corresponding approximate dual functions. Figure~\ref{fig:mass_struct}a shows the consistent mass matrix that approximates a diagonal matrix, but still exhibits the full bandwidth of non-zero entries according to Table~1. Figure~\ref{fig:mass_struct}b shows the non-zero entries of $\mat{M}$ after row-sum lumping. We note that for now, we employ the row-sum technique only in those rows that associate to approximate dual functions $\adual_j$ in order not to jeopardize full accuracy of the Petrov-Galerkin scheme. In turn, a few rows that associate to interpolatory B-spline functions with support on the Dirichlet boundary remain ``unlumped'', as illustrated in Fig.~\ref{fig:mass_struct}b for our 1D example. 
Hence, we only require the inversion of small blocks, which limits requirements on computational and memory resources. 
The block structure of the inverse mass matrix is illustrated in Fig.~\ref{fig:mass_struct}c. 

\begin{remark}

If one wants to keep the modified approximate dual functions at the boundary to maintain a fully diagonal mass matrix without the need for any inversion operations or associated memory storage, one can resort to imposing Dirichlet boundary conditions weakly, for instance via the penalty method \cite{Leidinger2019,Breitenberger:15.1} or Nitsche's method \cite{embar_nitsche_2010,Schillinger_nitsche_2016,guo2018variationally}. 
\end{remark}

\section{Numerical examples}\label{sec:dynamics_plates}

In this section, we demonstrate the favorable numerical behavior of our Petrov-Galerkin scheme. 
We will consider both explicit dynamics calculations and spectral analysis results, obtained for B-spline discretizations of degrees $p=2$ through $p=5$. In particular, we will compare the performance of our Petrov-Galerkin scheme with the standard isogeometric Galerkin method based on a consistent or row-sum lumped mass matrix. 

%%% simply supported steel beam

\begin{figure}[h!]%[!htb]
	\centering
    \subfloat[Beam-like plate model]{{\includegraphics[width=0.45\textwidth]{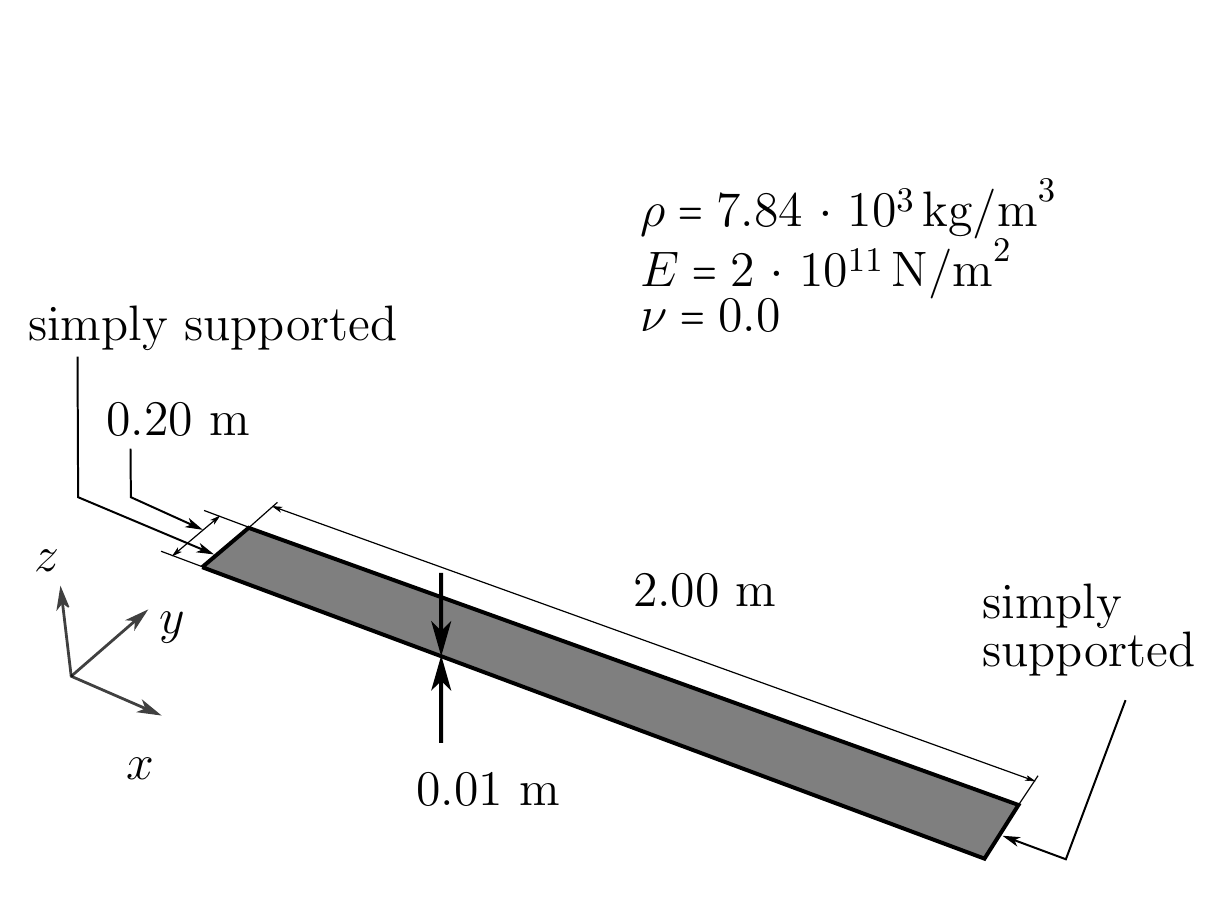} }} \hspace{0.2cm}
    \subfloat[Displacement snapshots]{{\includegraphics[width=0.45\textwidth]{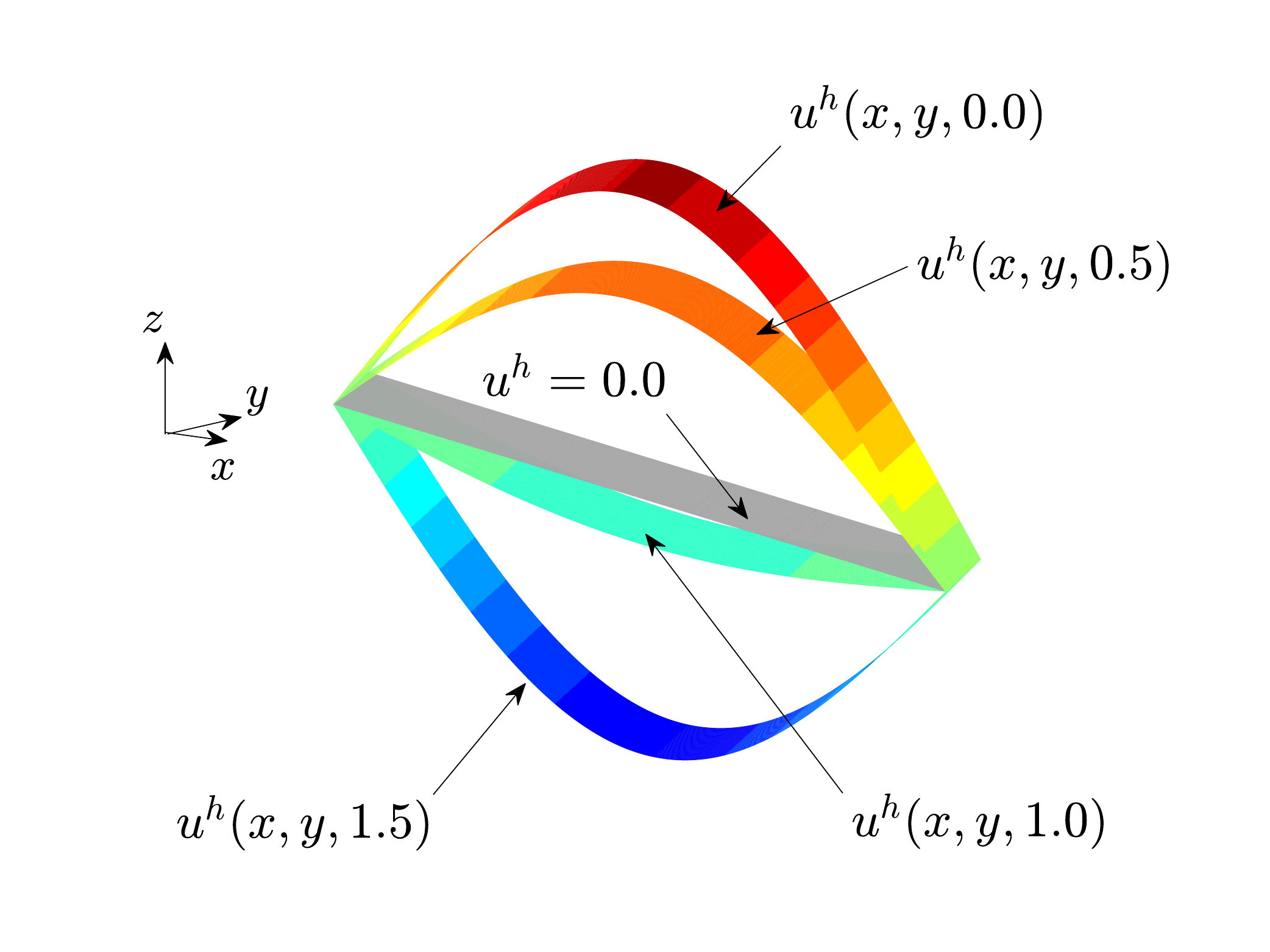} }}

	\caption{Set-up of the simply supported beam benchmark.} \label{fig:beam_geometry}
\end{figure}

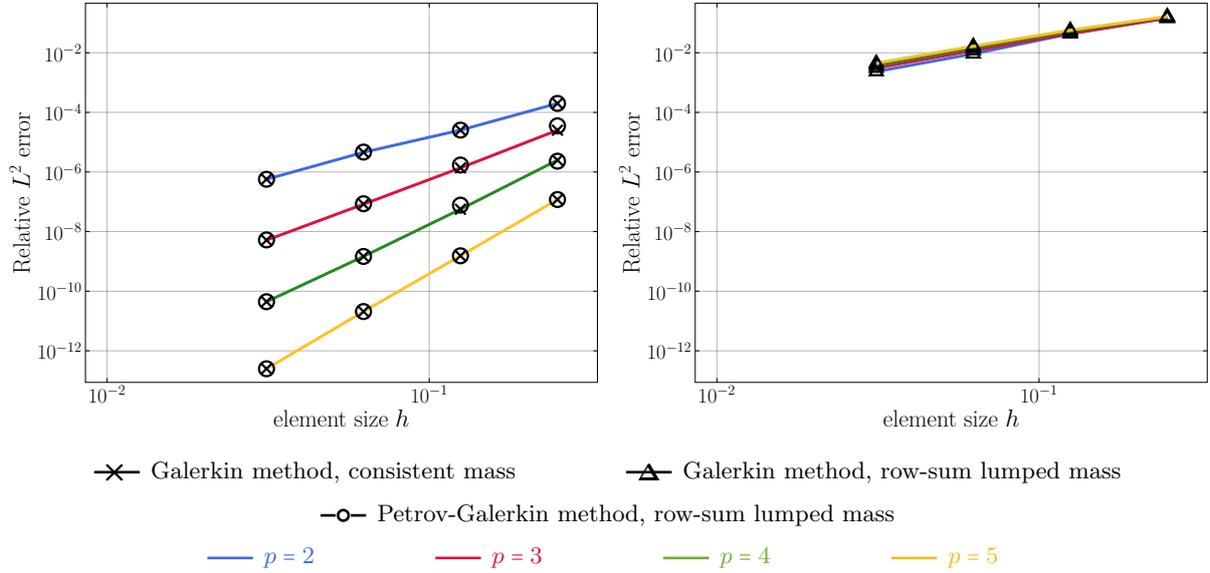
\begin{figure}[h!]
    \centering
    \includegraphics[width=0.48\textwidth]{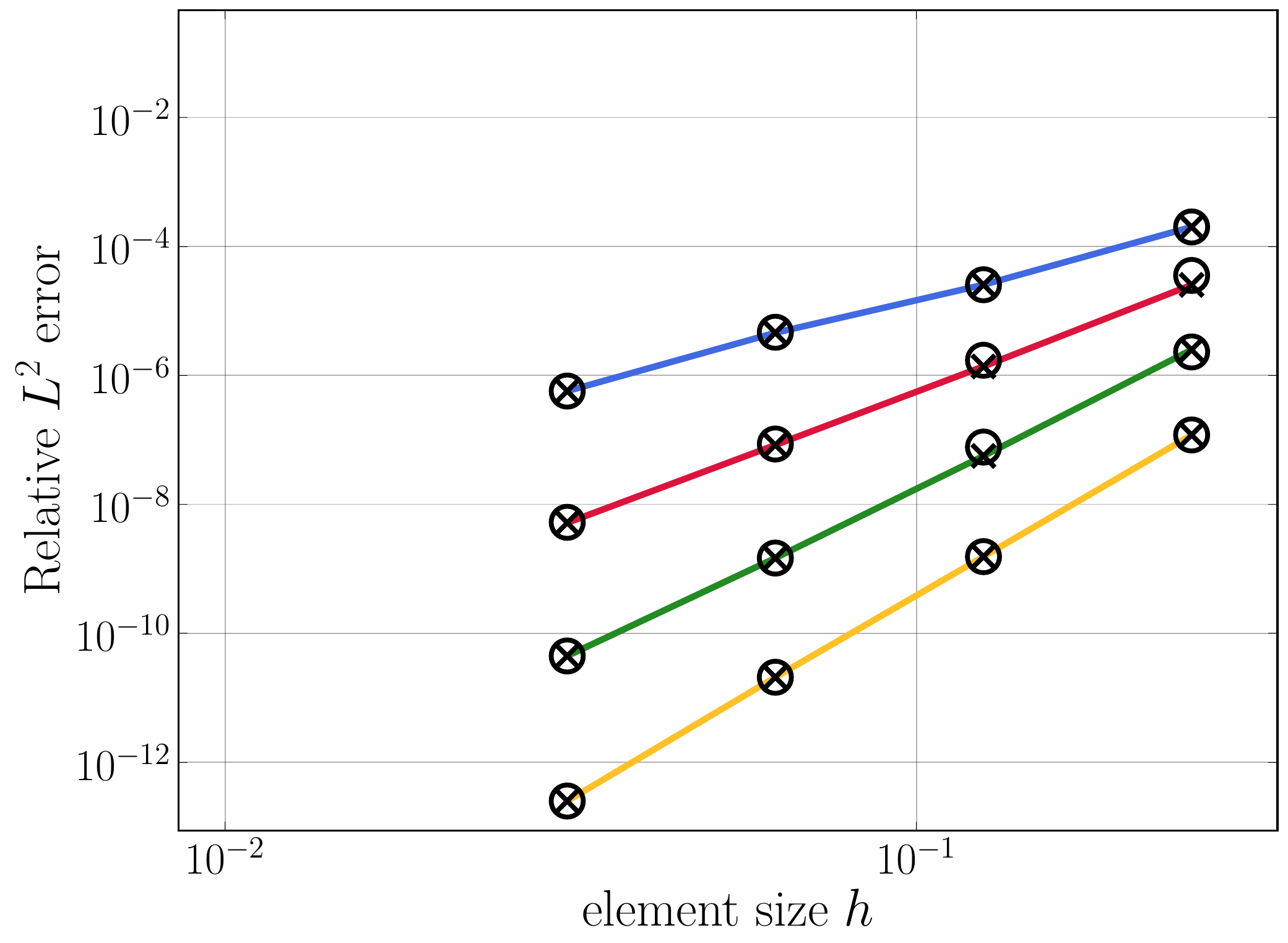}
   \includegraphics[width=0.48\textwidth]{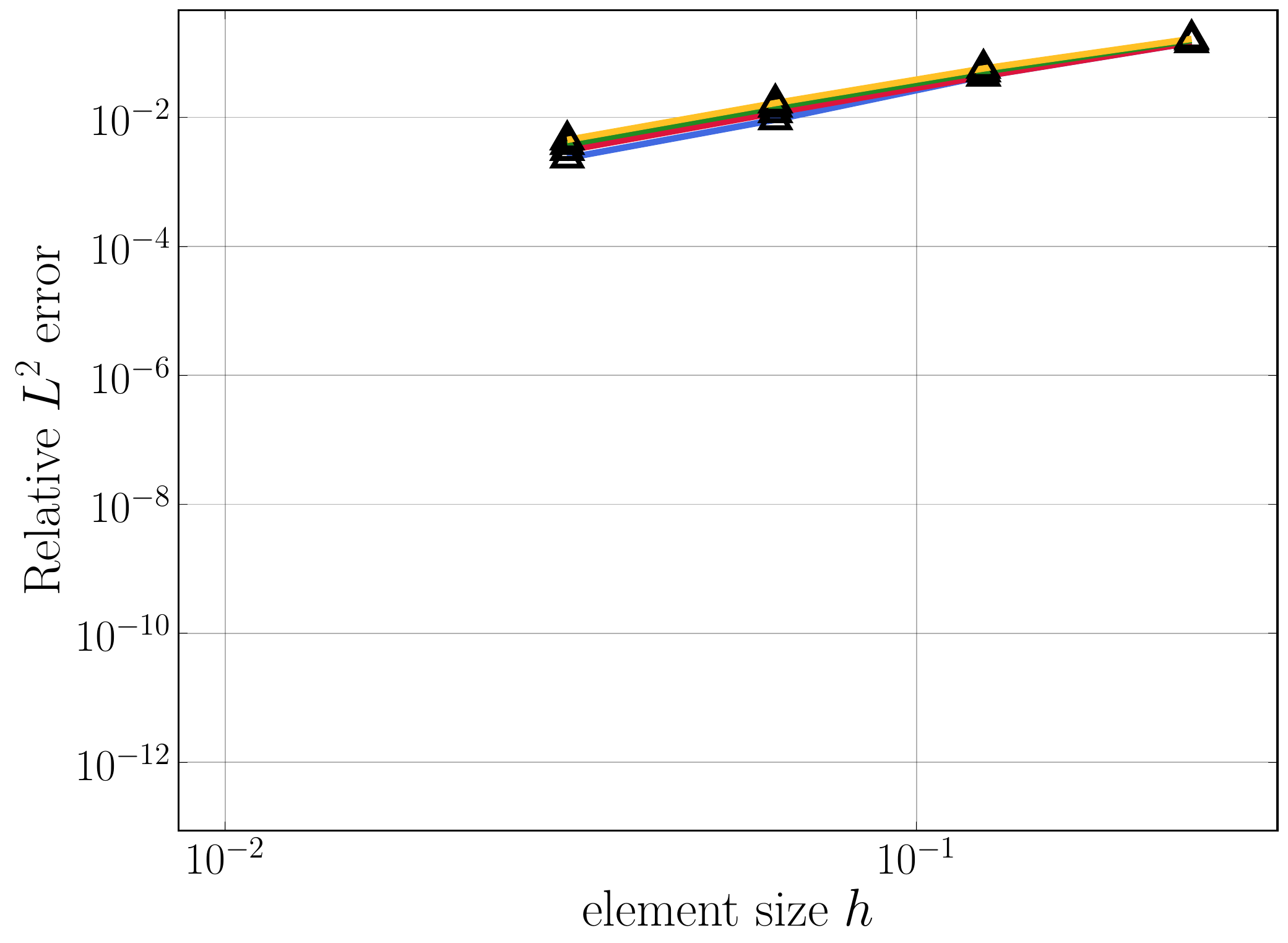}
    \vspace{0.2cm}
    
    \begin{tikzpicture}
    \filldraw[black,line width=1pt, solid] (0.0,0) -- (0.6,0);
    \filldraw[black,line width=1pt] (0.0,0) node[right]{\footnotesize $\boldsymbol{\bigtimes}$};
    \filldraw[black,line width=1pt] (0.6,0) node[right]{\footnotesize Galerkin method, consistent mass};
    \filldraw[black,line width=1pt, solid] (7,0) -- (7.6,0);
    \filldraw[black,line width=1pt] (7.0,0) node[right]{\footnotesize $\boldsymbol{\Delta}$};
    \filldraw[black,line width=1pt] (7.6,0) node[right]{\footnotesize Galerkin method, row-sum lumped mass};
\end{tikzpicture}

\begin{tikzpicture}
    \filldraw[black,line width=1pt, solid] (3.0,0) -- (3.2,0);
    \filldraw[black,line width=1pt] (3.3,0) [fill=none] circle (2pt);
    \filldraw[black,line width=1pt, solid] (3.4,0) -- (3.6,0);
    \filldraw[black,line width=1pt] (3.6,0) node[right]{\footnotesize Petrov-Galerkin method, row-sum lumped mass};
\end{tikzpicture}
    \begin{tikzpicture}
    \filldraw[blue1,line width=1pt, solid] (0,0) -- (0.6,0);
    \filldraw[blue1,line width=1pt] (0.6,0) node[right]{\footnotesize $p=2$};	
    \filldraw[red1,line width=1pt, solid] (3,0) -- (3.6,0);
    \filldraw[red1,line width=1pt] (3.6,0) node[right]{\footnotesize $p=3$};
    \filldraw[green1,line width=1pt, solid] (6,0) -- (6.6,0);
    \filldraw[green1,line width=1pt] (6.6,0) node[right]{\footnotesize $p=4$};
    \filldraw[yellow1,line width=1pt, solid] (9,0) -- (9.6,0);
    \filldraw[yellow1,line width=1pt] (9.6,0) node[right]{\footnotesize $p=5$};
\end{tikzpicture}
    \caption{Beam-like plate: convergence of the relative $L^2$ error in the displacement solution at $t$ = 1.5 s.} \label{fig:beam_dyn_convergence}
\end{figure}

\subsection{Simply supported beam}

Our first example is the explicit dynamics calculation of a freely vibrating simply supported steel beam, which is modeled as a stretched Kirchhoff plate as illustrated in Fig.~\ref{fig:beam_geometry}a. 
We assume the following analytical solution:
\begin{align}\label{eq:analytic_beam_dyn}
    u(x,y,t) \,=\, \sin(\pi \, \frac{x}{L_x}) \, \cos(\left(\frac{\pi}{L_x}\right)^2 \, \sqrt{\frac{EI}{\rho\,A}} \, t)\,,    
\end{align}
where $L_x$, $I$ and $A$ are the longitudinal length, the moment of inertia, and the cross section area, respectively. From \eqref{eq:analytic_beam_dyn}, we can also derive the initial and boundary conditions.  
We perform uniform mesh refinement in $x$-direction with $8,\,16,\,32,\,64$ B\'ezier elements, while we use one B\'ezier element in $y$-direction. 
We apply the central difference method  
\cite{hughes_finite_2003} for explicit time integration, simulating the first 1.5 s of the vibration history.
To guarantee that the time integration error is very small, we choose the following order-dependent time step size: $\Delta \, t \,=\, \left(p\,/\,\left(2\,n_{ele}\right)\right)^p$ \cite{Hiemstra_outlier_2021,Nguyen_outlier_2022}. Figure~\ref{fig:beam_geometry}b illustrates the solution for quadratic splines and 16 B\'ezier elements at three different snapshots in time.

Figure~\ref{fig:beam_dyn_convergence} illustrates the convergence behavior of the relative $L^2$ error for the standard Galerkin method with consistent mass matrix (crosses), the standard Galerkin method with row-sum lumped mass matrix (triangles), and our Petrov-Galerkin method with lumped mass matrix (circles). 
On the one hand, we observe that our approach achieves the same optimal accuracy under mesh refinement as the Galerkin method with consistent mass matrix. Thus, the accuracy of our Petrov-Galerkin method based on test functions discretized by approximate dual functions is not affected by row-sum lumping. On the other hand, we observe that the accuracy of the Galerkin method is significantly affected by row-sum lumping, limiting convergence to a maximum of second order irrespective of the polynomial degree of the spline basis.

%%% annular plate

\begin{figure}[h!]%[!htb]
	\centering
    \subfloat[Coarsest \Bezier mesh with boundary conditions.]{{\includegraphics[width=0.45\textwidth]{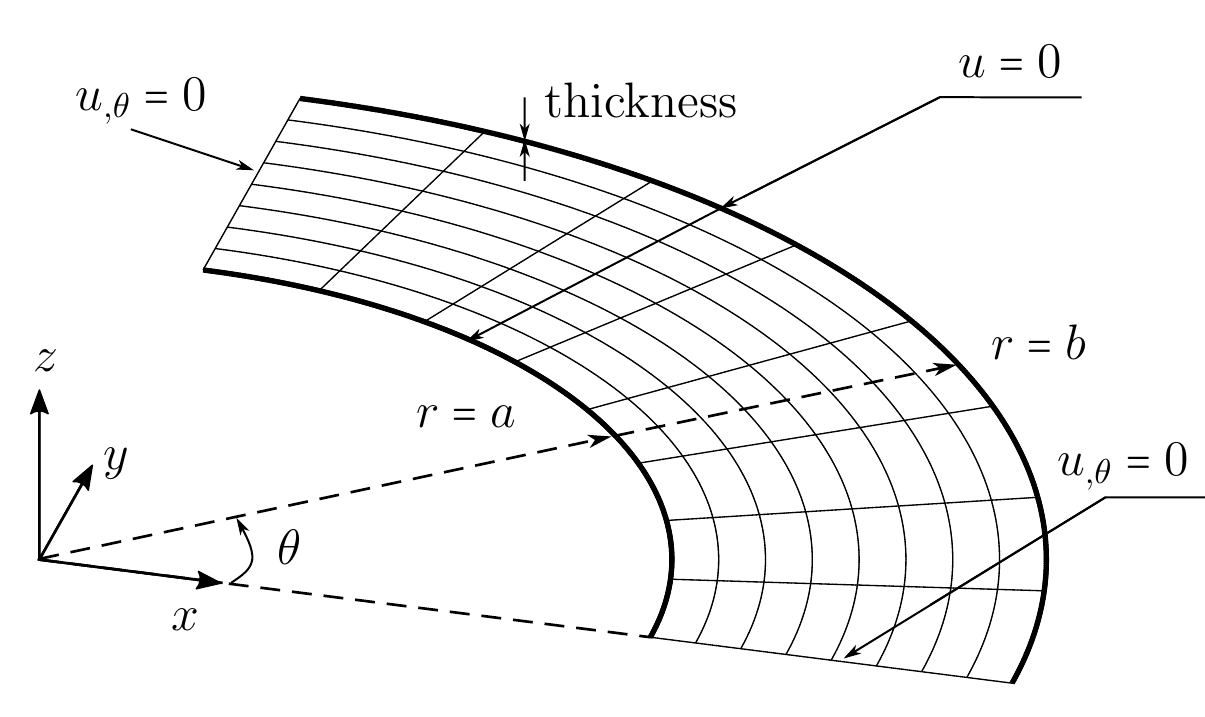} }} \hspace{0.2cm}
    \subfloat[Initial displacement field $u(r,\theta,0)$.]{{\includegraphics[width=0.45\textwidth]{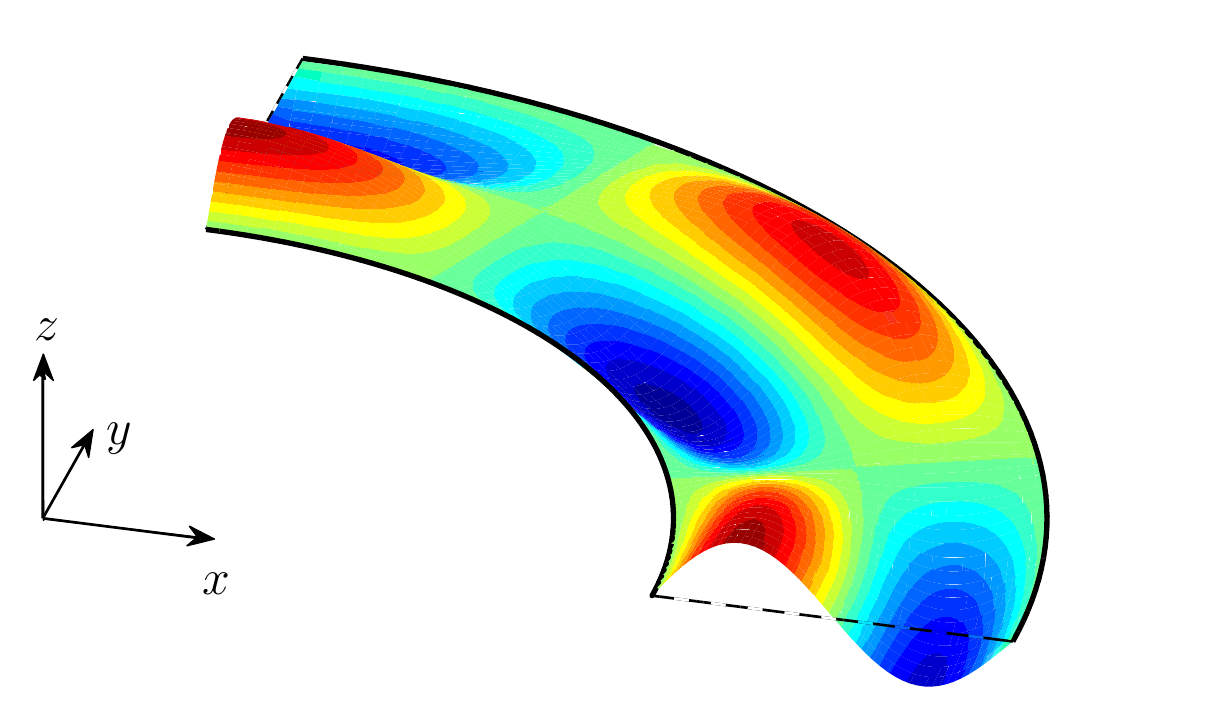} }}

	\caption{Set-up of the annular plate benchmark.} \label{fig:annulus_geometry}
\end{figure}

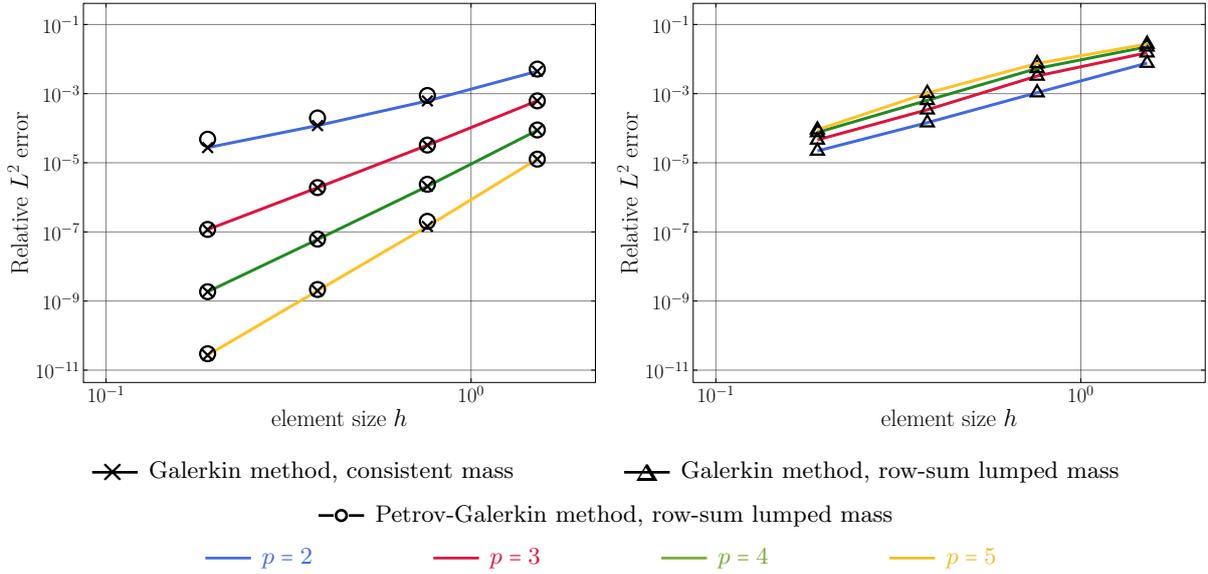
\begin{figure}[h!]
    \centering
   \includegraphics[width=0.48\textwidth]{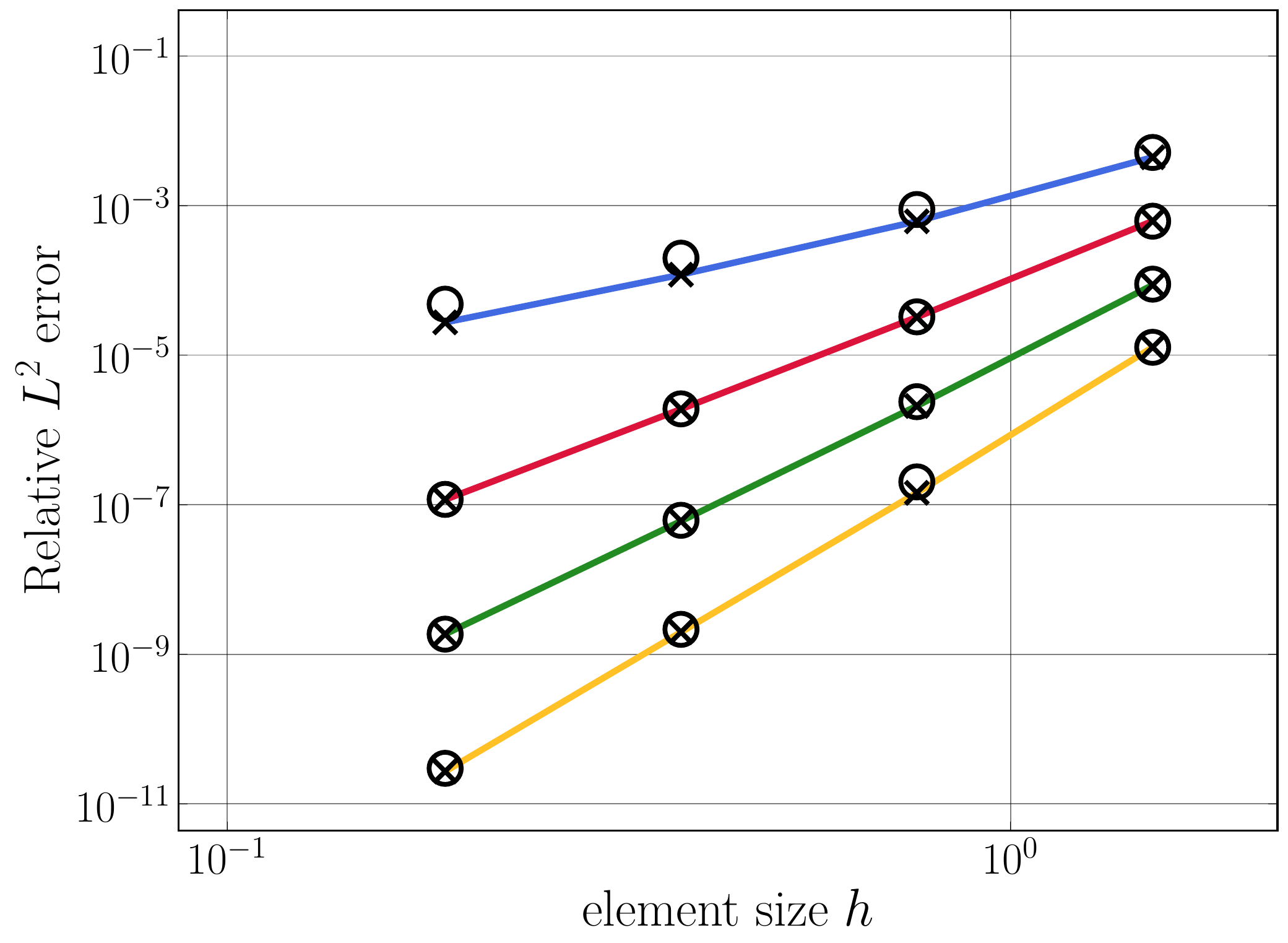} 
   \includegraphics[width=0.48\textwidth]{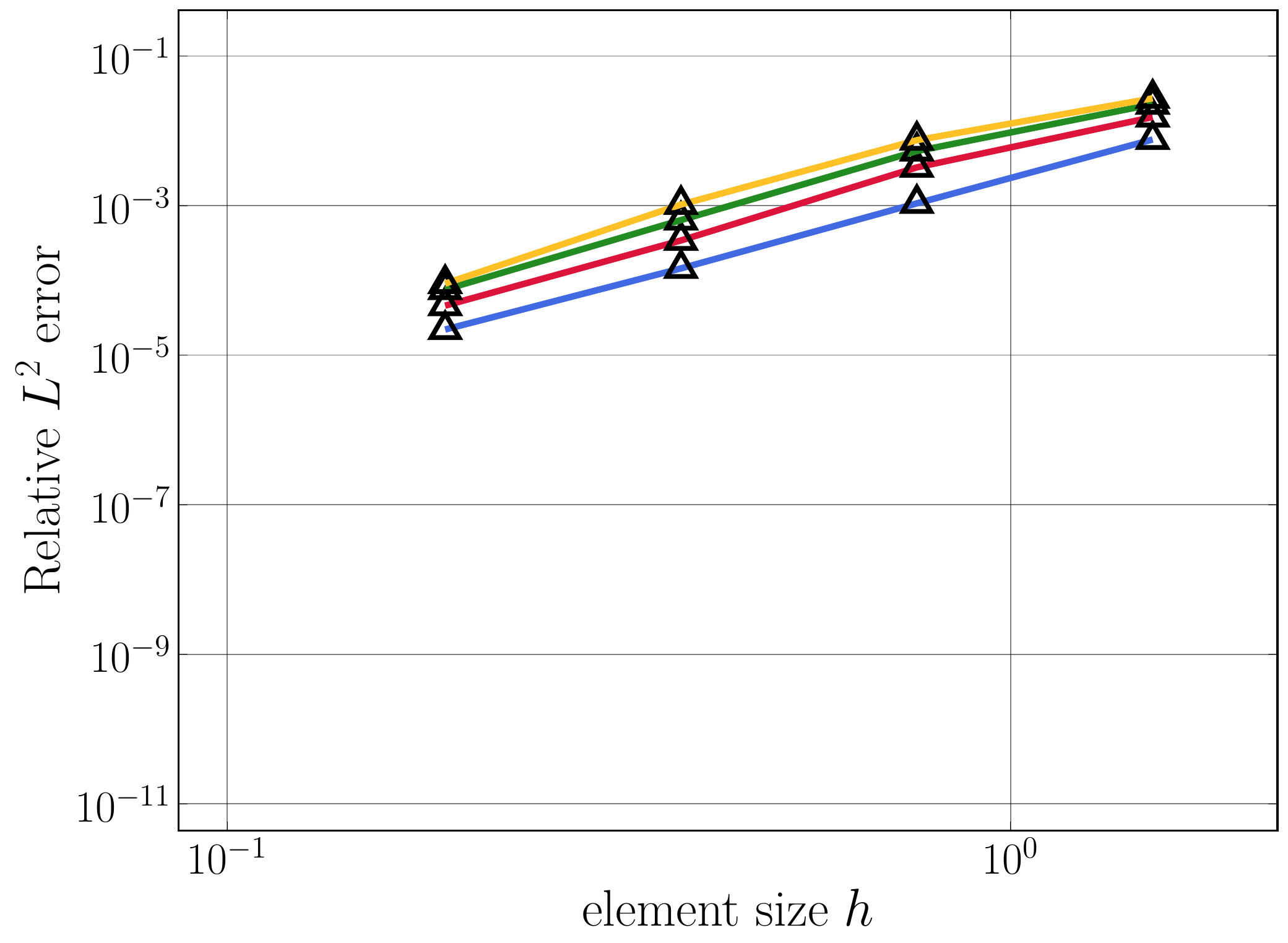}
    \vspace{0.2cm}
    
    \begin{tikzpicture}
    \filldraw[black,line width=1pt, solid] (0.0,0) -- (0.6,0);
    \filldraw[black,line width=1pt] (0.0,0) node[right]{\footnotesize $\boldsymbol{\bigtimes}$};
    \filldraw[black,line width=1pt] (0.6,0) node[right]{\footnotesize Galerkin method, consistent mass};
    \filldraw[black,line width=1pt, solid] (7,0) -- (7.6,0);
    \filldraw[black,line width=1pt] (7.0,0) node[right]{\footnotesize $\boldsymbol{\Delta}$};
    \filldraw[black,line width=1pt] (7.6,0) node[right]{\footnotesize Galerkin method, row-sum lumped mass};
\end{tikzpicture}

\begin{tikzpicture}
    \filldraw[black,line width=1pt, solid] (3.0,0) -- (3.2,0);
    \filldraw[black,line width=1pt] (3.3,0) [fill=none] circle (2pt);
    \filldraw[black,line width=1pt, solid] (3.4,0) -- (3.6,0);
    \filldraw[black,line width=1pt] (3.6,0) node[right]{\footnotesize Petrov-Galerkin method, row-sum lumped mass};
\end{tikzpicture}
    \begin{tikzpicture}
    \filldraw[blue1,line width=1pt, solid] (0,0) -- (0.6,0);
    \filldraw[blue1,line width=1pt] (0.6,0) node[right]{\footnotesize $p=2$};	
    \filldraw[red1,line width=1pt, solid] (3,0) -- (3.6,0);
    \filldraw[red1,line width=1pt] (3.6,0) node[right]{\footnotesize $p=3$};
    \filldraw[green1,line width=1pt, solid] (6,0) -- (6.6,0);
    \filldraw[green1,line width=1pt] (6.6,0) node[right]{\footnotesize $p=4$};
    \filldraw[yellow1,line width=1pt, solid] (9,0) -- (9.6,0);
    \filldraw[yellow1,line width=1pt] (9.6,0) node[right]{\footnotesize $p=5$};
\end{tikzpicture}
    \caption{Annular plate: convergence of the relative $L^2$ error in the displacement solution at $t$ = 0.6 s.} \label{fig:annulus_dyn_convergence}
\end{figure}

\subsection{Annular plate}

We now consider a quarter of a freely vibrating annular steel plate. Its set-up is illustrated in Fig.~\ref{fig:annulus_geometry}a, with thickness $d=0.01$ m, Young's modulus $E = 2\,\cdot\,10^{11} \, \text{N/m}^2$, Poisson's ratio $\nu=0.0$, and mass density $\rho=7.84\,\cdot\,10^{3} \, \text{kg/m}^3$. 
It is simply supported along both the inner radius $r=a$ and the outer radius $r=b$.  
We construct the following analytical solution in polar coordinates ($r$, $\theta$): 
\begin{align}\label{eq:analytic_annulus_dyn}
    u(r,\theta,t) \,=\, J_4(r)\, \cos(4\,\theta) \, \cos\left(\sqrt{\frac{E\,d^3}{12 \left(1-\nu^2\right)\,\rho \, d}} \,t\right)\,,    
\end{align} 
where $J_4(r)$ denotes the $4^{\text{th}}$ Bessel function of the first kind. 
We choose the second and fourth positive zeros of $J_4(r)$ 
as the inner and outer radii of the annulus, i.e., $a\,=\,\lambda_2 \,\approx \, 11.065$ and $b\,=\,\lambda_4 \,\approx\, 17.616$. 
The analytical solution \eqref{eq:analytic_annulus_dyn} at time $t=0$, plotted in Fig.~\ref{fig:annulus_geometry}b, represents the initial displacement field. 
We perform uniform mesh refinement in both radial and angular directions, using 8, 16, 32, and 64 B\'ezier elements. We again apply the central difference method for time integration with the same time step estimate as above.

We focus on the accuracy in space at time $t$ = 0.6 s.
Figure~\ref{fig:annulus_dyn_convergence} shows the convergence behavior of the relative $L^2$ error for the standard Galerkin method with consistent mass matrix (crosses), the standard Galerkin method with row-sum lumped mass matrix (triangles), and our Petrov-Galerkin method with lumped mass matrix (circles). 
The results again confirm that our approach achieves the same optimal accuracy under mesh refinement as the Galerkin method with consistent mass matrix, whereas the accuracy of the Galerkin method is significantly affected by row-sum lumping, limiting convergence to second order irrespective of the polynomial degree of the spline basis.

\begin{figure}[h!]
    \centering
    \subfloat[$p=2$]{{\includegraphics[width=0.48\textwidth]{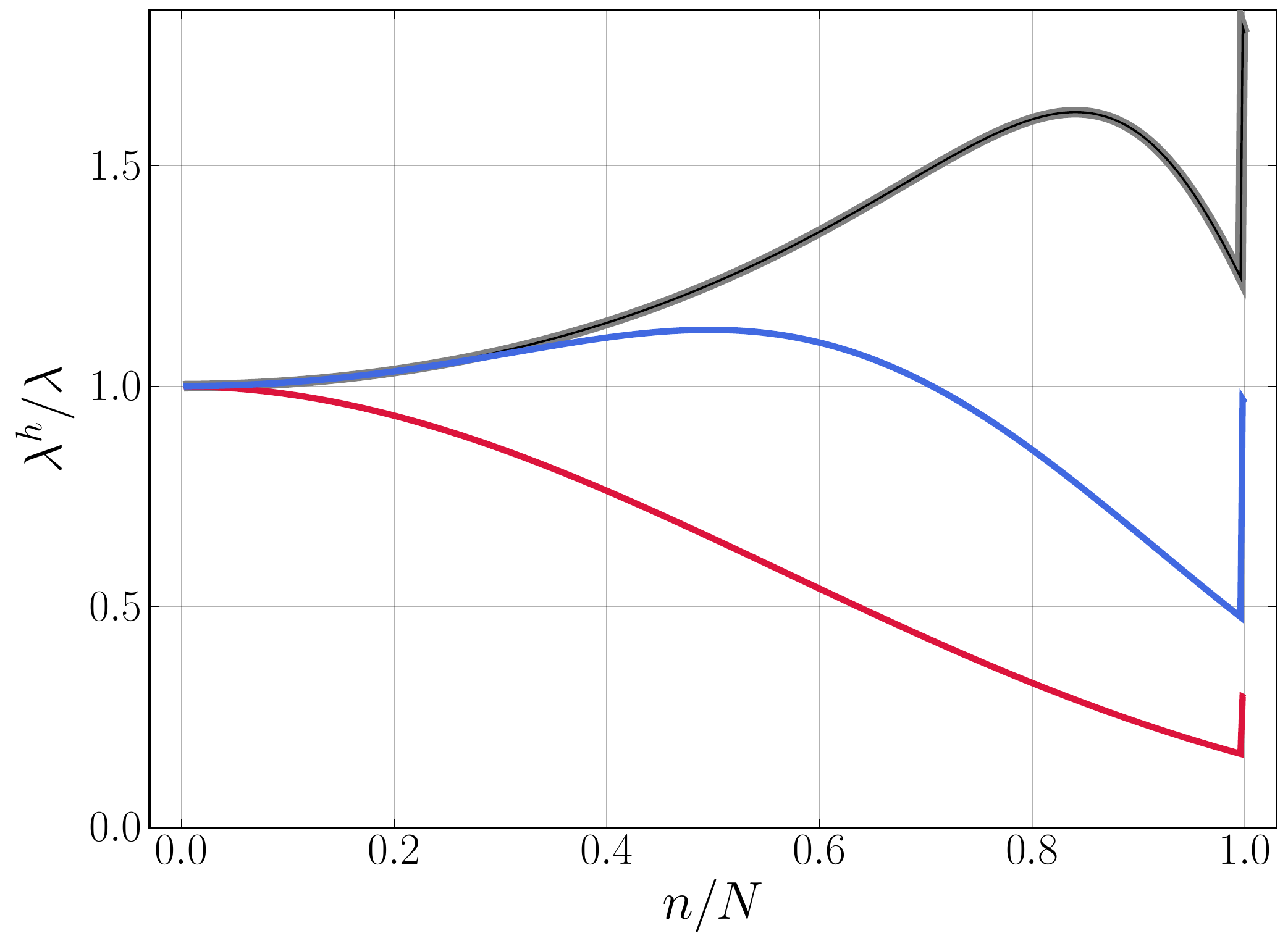} }}
    \subfloat[$p=3$]{{\includegraphics[width=0.48\textwidth]{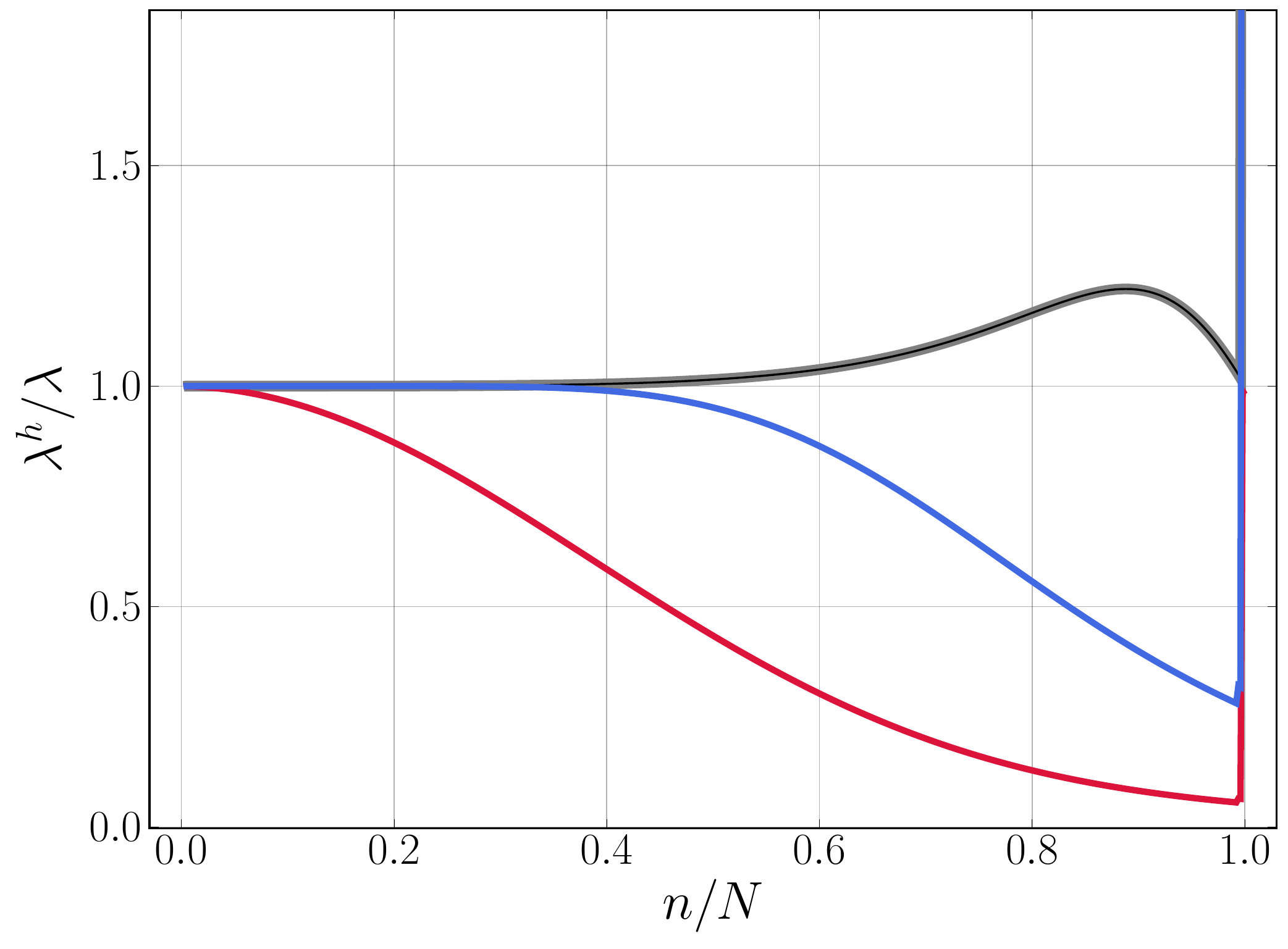} }}

    \subfloat[$p=4$]{{\includegraphics[width=0.48\textwidth]{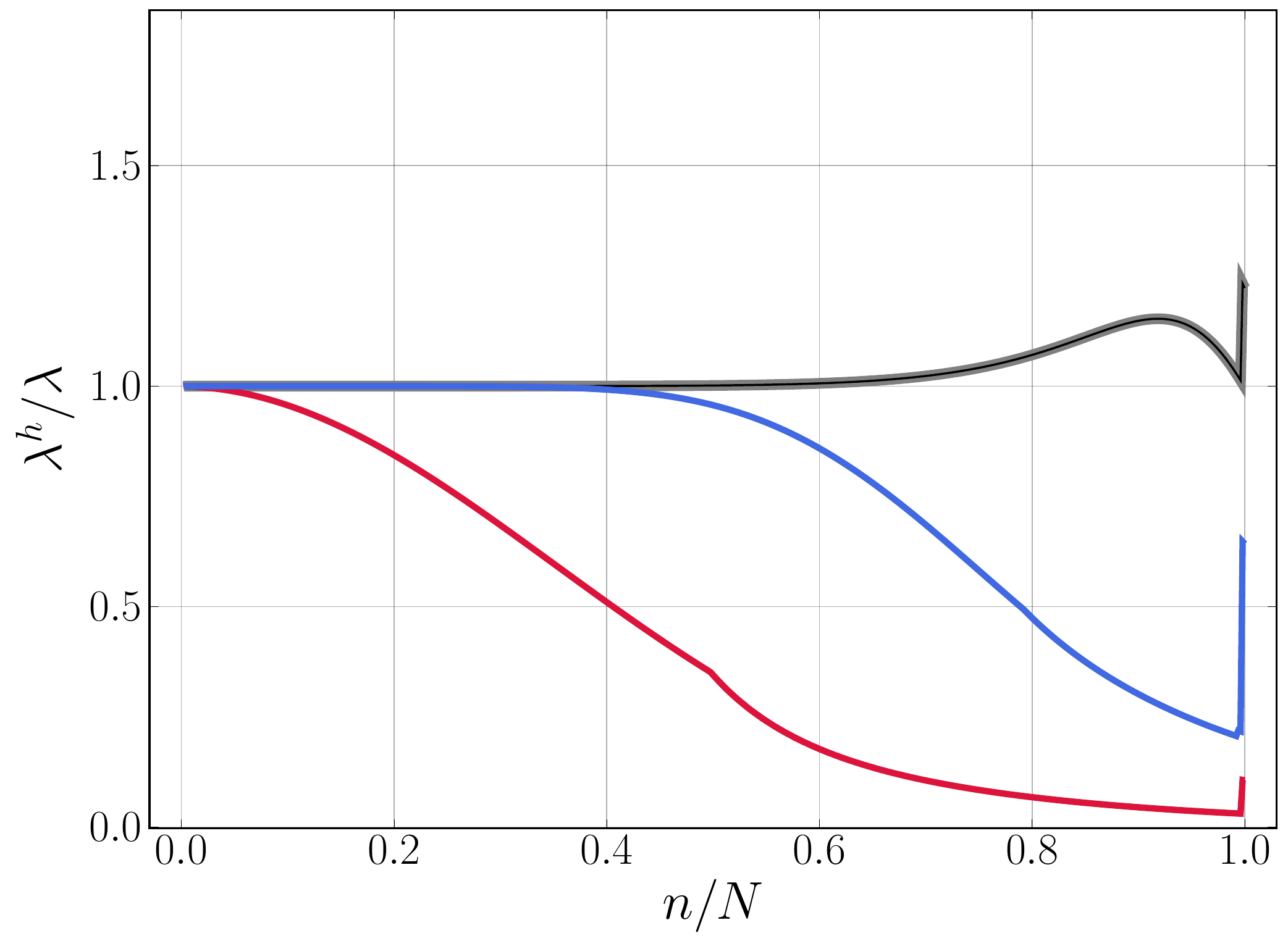} }}
    \subfloat[$p=5$]{{\includegraphics[width=0.48\textwidth]{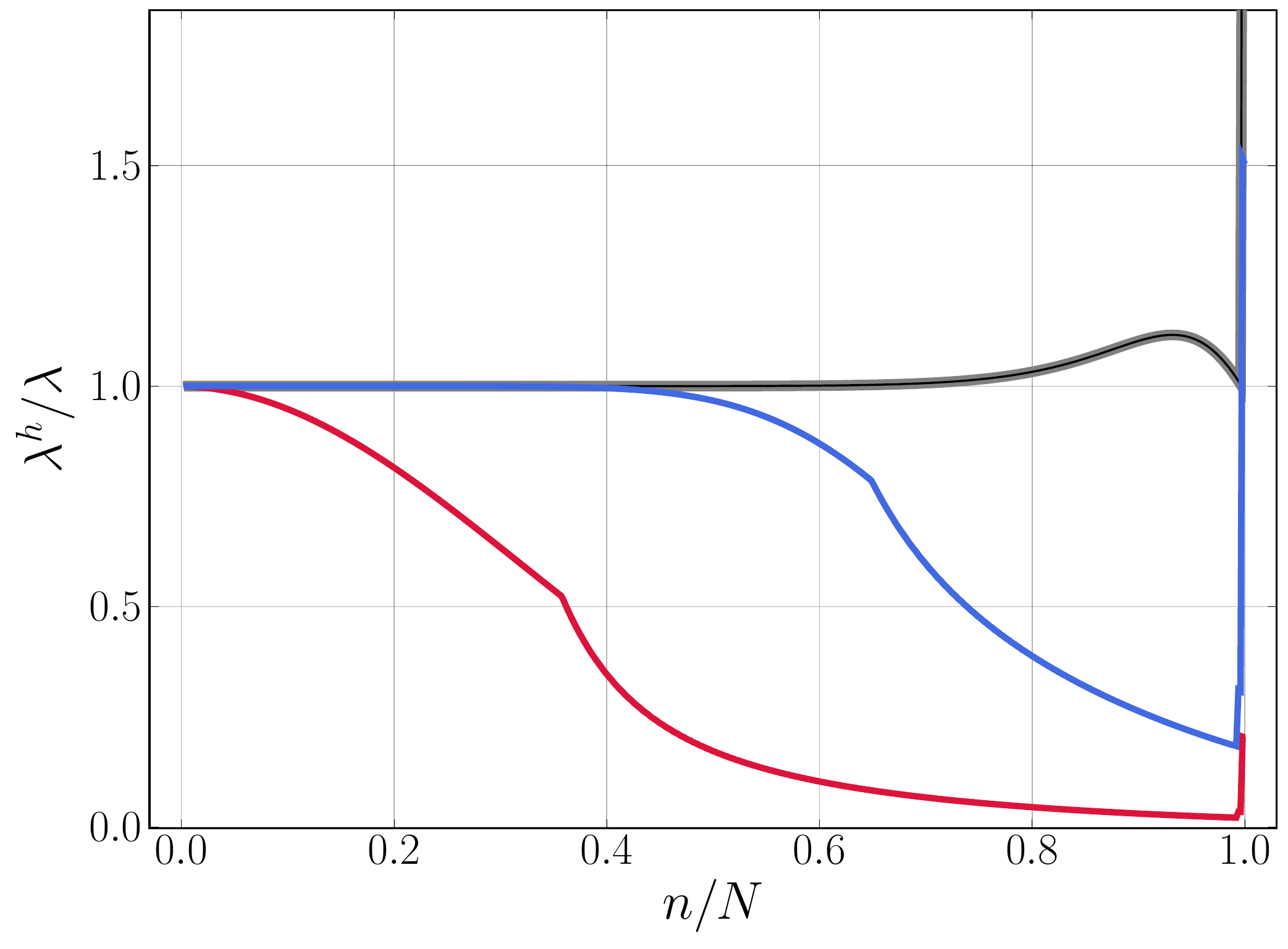} }}
    \vspace{0.2cm}
    \begin{tikzpicture}
    \filldraw[grey1,line width=1pt, solid] (0,0) -- (0.6,0);
    \filldraw[grey1,line width=1pt] (0.6,0) node[right]{\footnotesize Galerkin, consistent mass};
    \filldraw[red1,line width=1pt, solid] (7,0) -- (7.6,0);
    \filldraw[red1,line width=1pt] (7.6,0) node[right]{\footnotesize Galerkin, row-sum lumped mass};
\end{tikzpicture}

\begin{tikzpicture}
    \filldraw[black,line width=1pt, solid] (0,0) -- (0.6,0);
    \filldraw[black,line width=1pt] (0.6,0) node[right]{\footnotesize Petrov-Galerkin, consistent mass};
    \filldraw[blue1,line width=1pt, solid] (6,0) -- (6.6,0);
    \filldraw[blue1,line width=1pt] (6.6,0) node[right]{\footnotesize Petrov-Galerkin, higher-order accurate lumped mass};
\end{tikzpicture}
    \caption{Normalized eigenvalues of a freely vibrating beam, computed on 500 B\'ezier elements.} \label{fig:spectra_eigenval_beam_no_iter}
\end{figure}
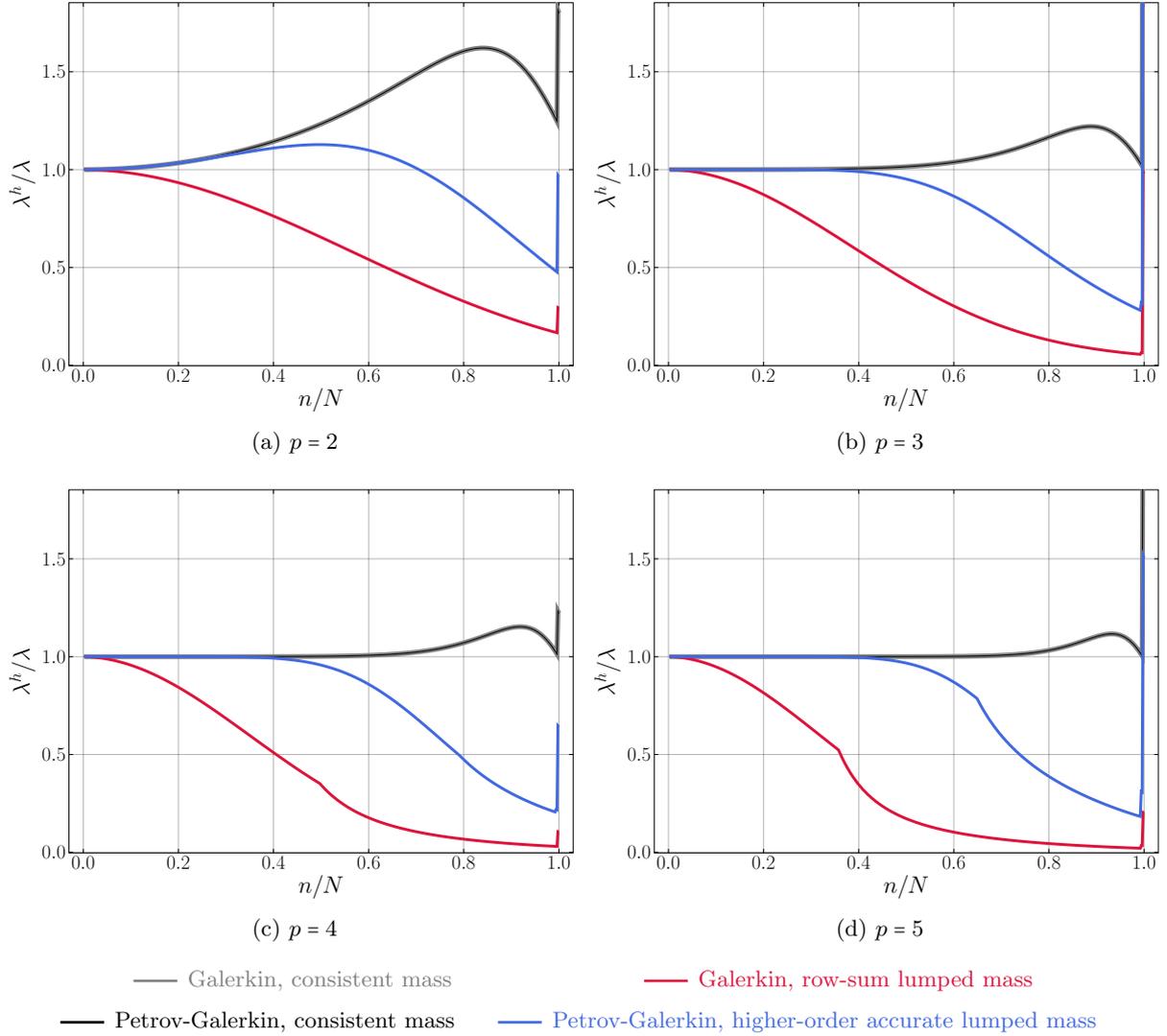

\begin{figure}[h!]
    \centering
    \subfloat[$p=2$]{{\includegraphics[width=0.48\textwidth]{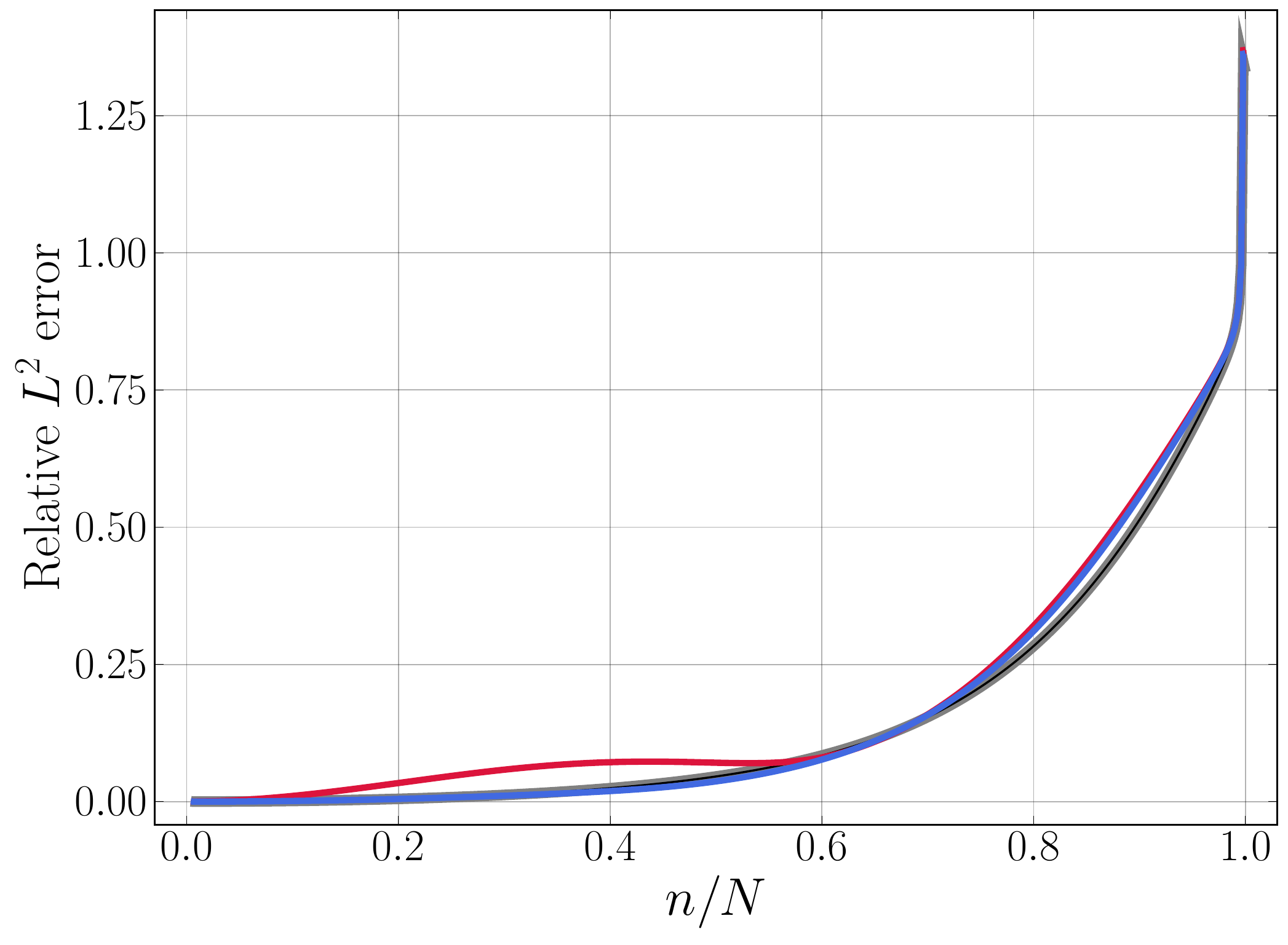} }}
    \subfloat[$p=3$]{{\includegraphics[width=0.48\textwidth]{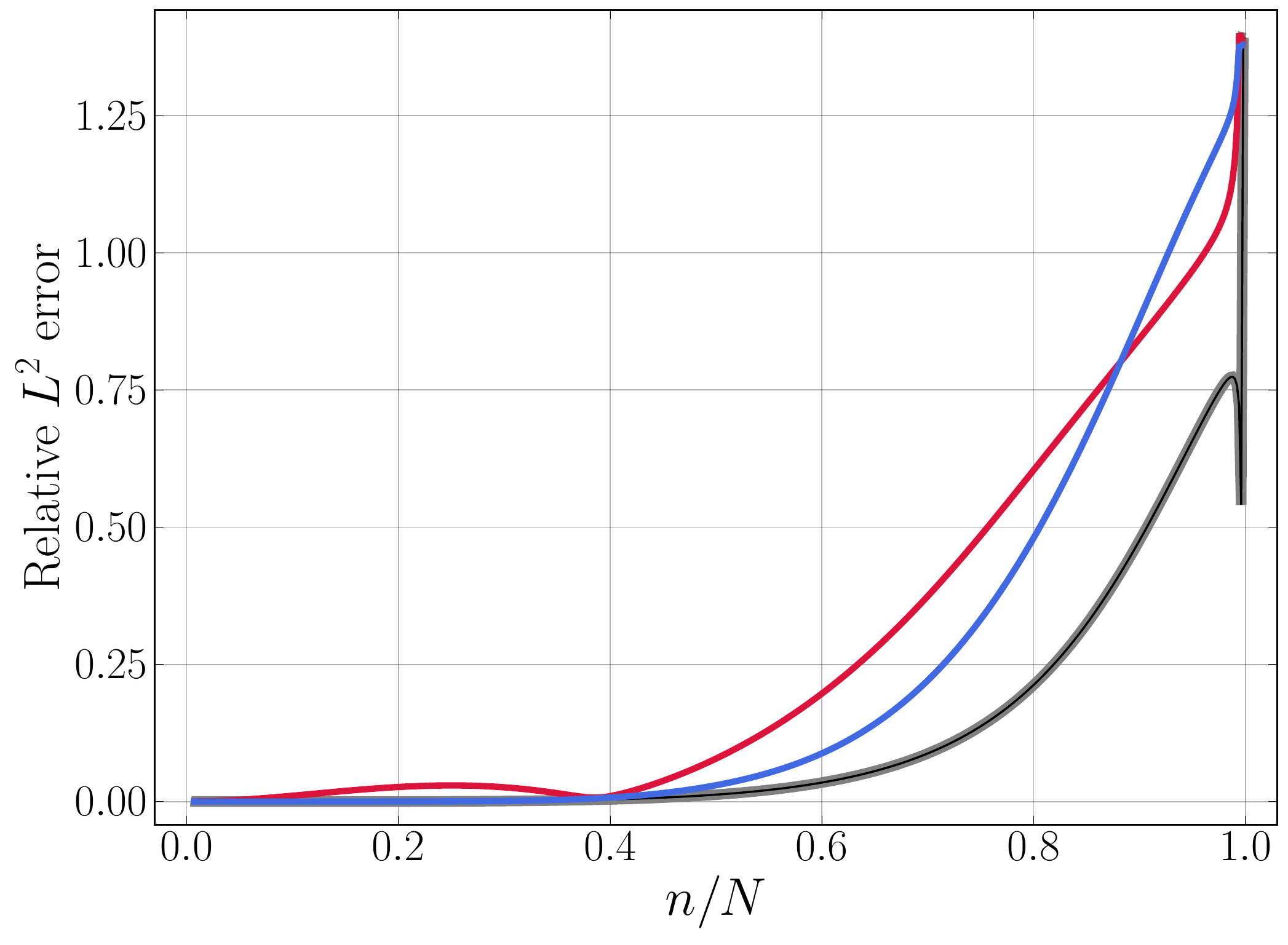} }}

    \subfloat[$p=4$]{{\includegraphics[width=0.48\textwidth]{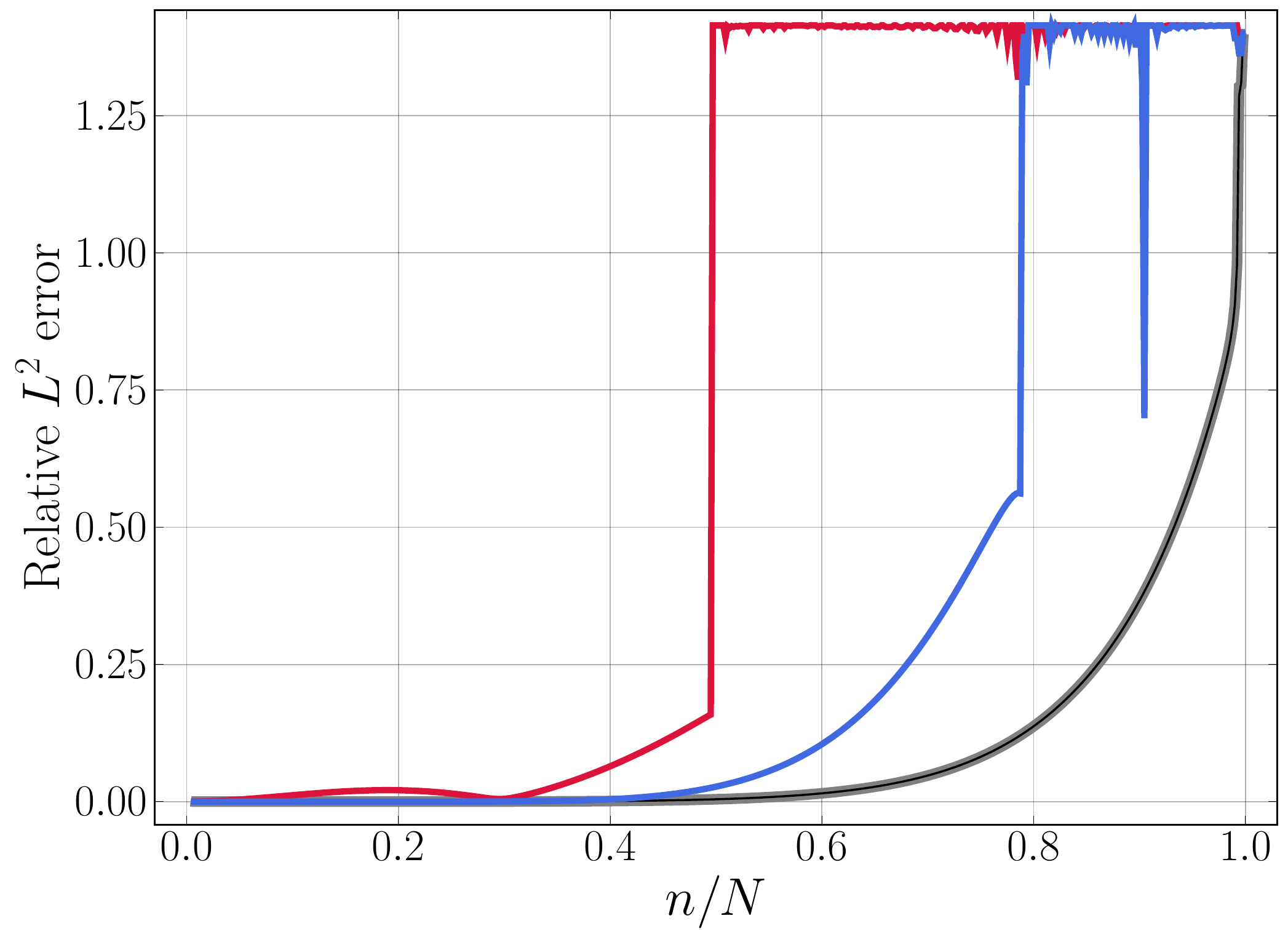} }}
    \subfloat[$p=5$]{{\includegraphics[width=0.48\textwidth]{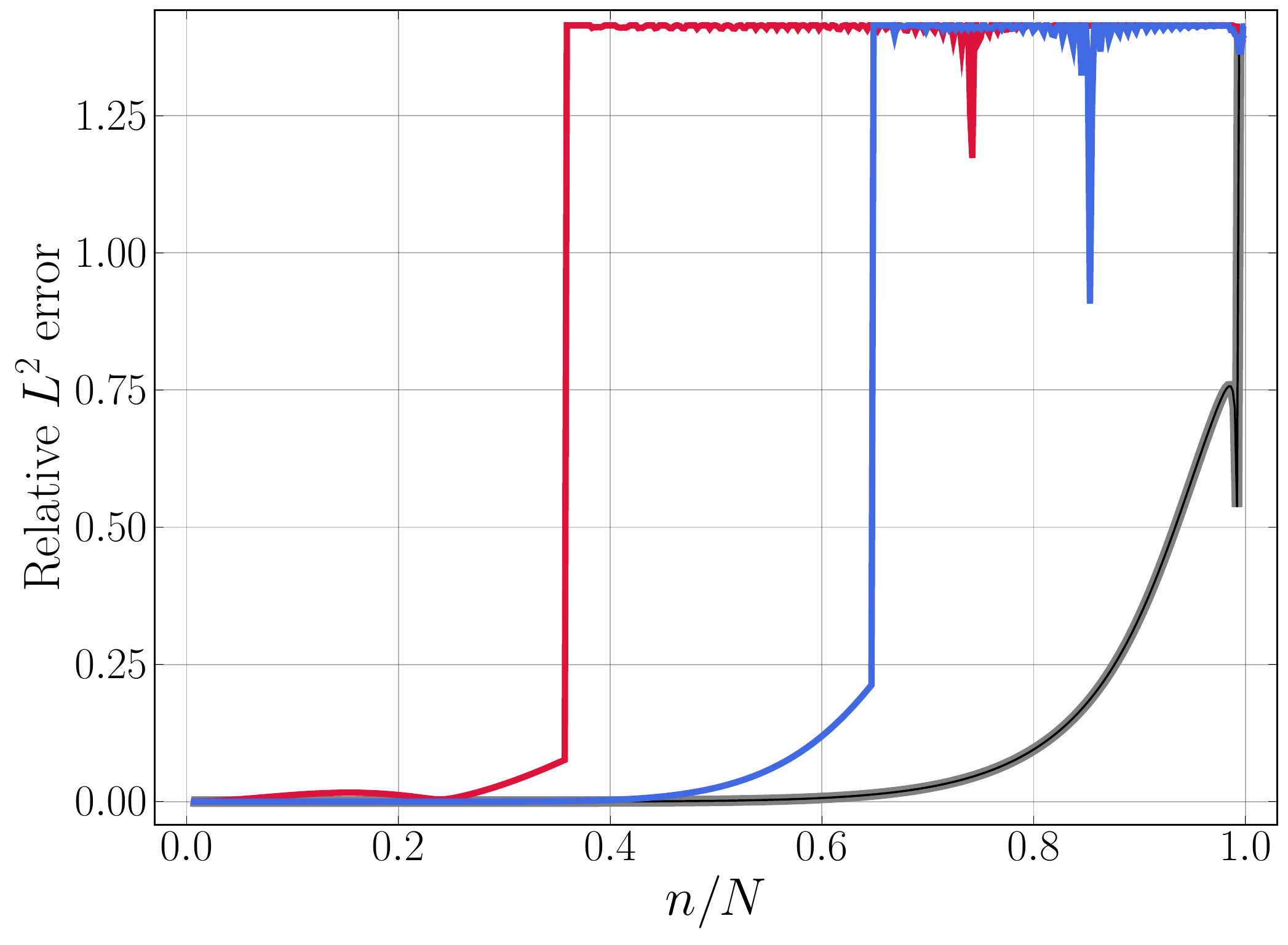} }}
    \vspace{0.2cm}
    \begin{tikzpicture}
    \filldraw[grey1,line width=1pt, solid] (0,0) -- (0.6,0);
    \filldraw[grey1,line width=1pt] (0.6,0) node[right]{\footnotesize Galerkin, consistent mass};
    \filldraw[red1,line width=1pt, solid] (7,0) -- (7.6,0);
    \filldraw[red1,line width=1pt] (7.6,0) node[right]{\footnotesize Galerkin, row-sum lumped mass};
\end{tikzpicture}

\begin{tikzpicture}
    \filldraw[black,line width=1pt, solid] (0,0) -- (0.6,0);
    \filldraw[black,line width=1pt] (0.6,0) node[right]{\footnotesize Petrov-Galerkin, consistent mass};
    \filldraw[blue1,line width=1pt, solid] (6,0) -- (6.6,0);
    \filldraw[blue1,line width=1pt] (6.6,0) node[right]{\footnotesize Petrov-Galerkin, higher-order accurate lumped mass};
\end{tikzpicture}
    \caption{Relative $L^2$ error of the mode shapes of a freely vibrating beam, computed on 500 B\'ezier elements.} \label{fig:spectra_mode_beam_no_iter}
\end{figure}
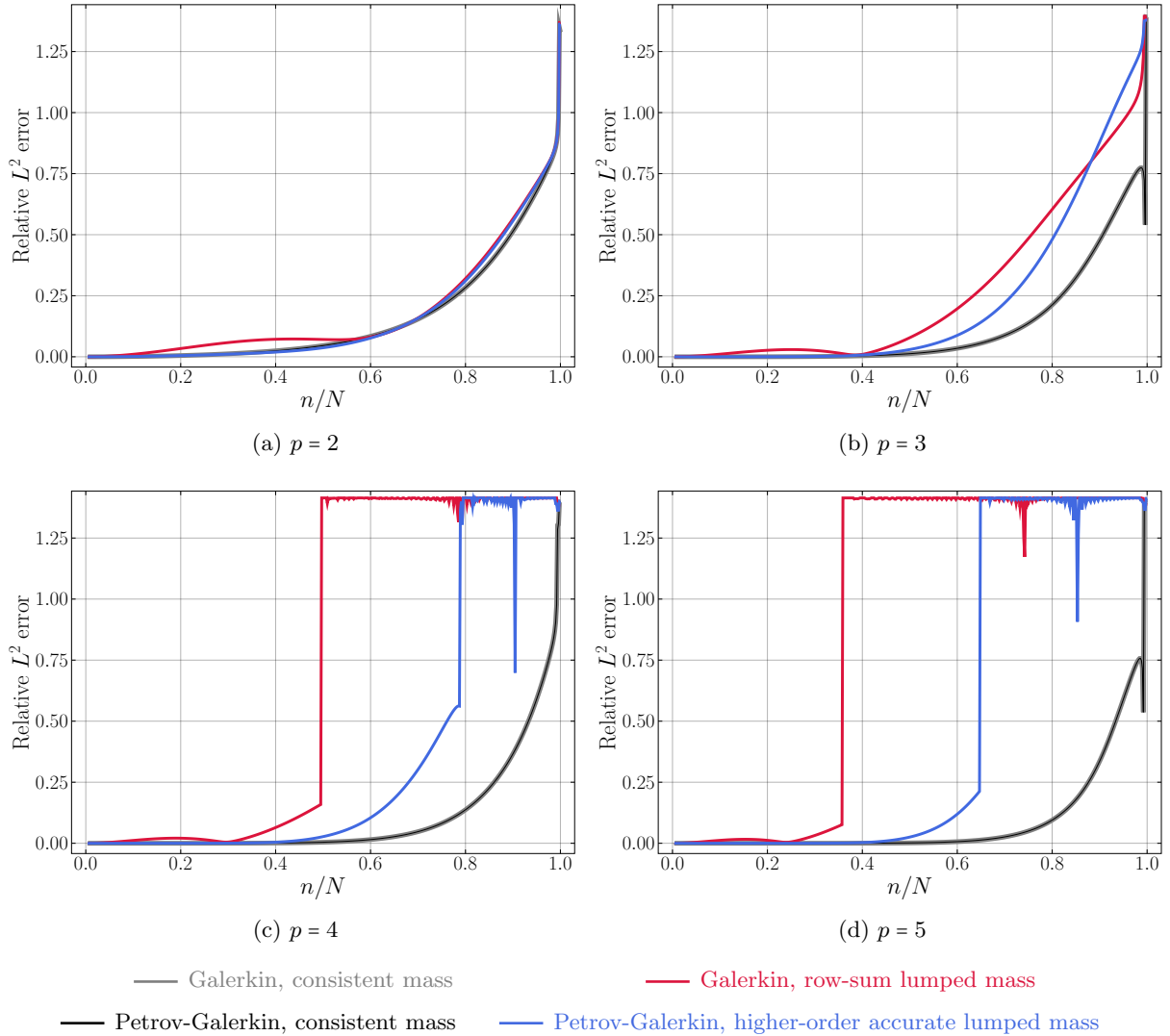

\begin{figure}[h!]
    \centering
    \subfloat[Eigenvalue $\lambda^h_{10}$.]{\includegraphics[width=0.5\textwidth]{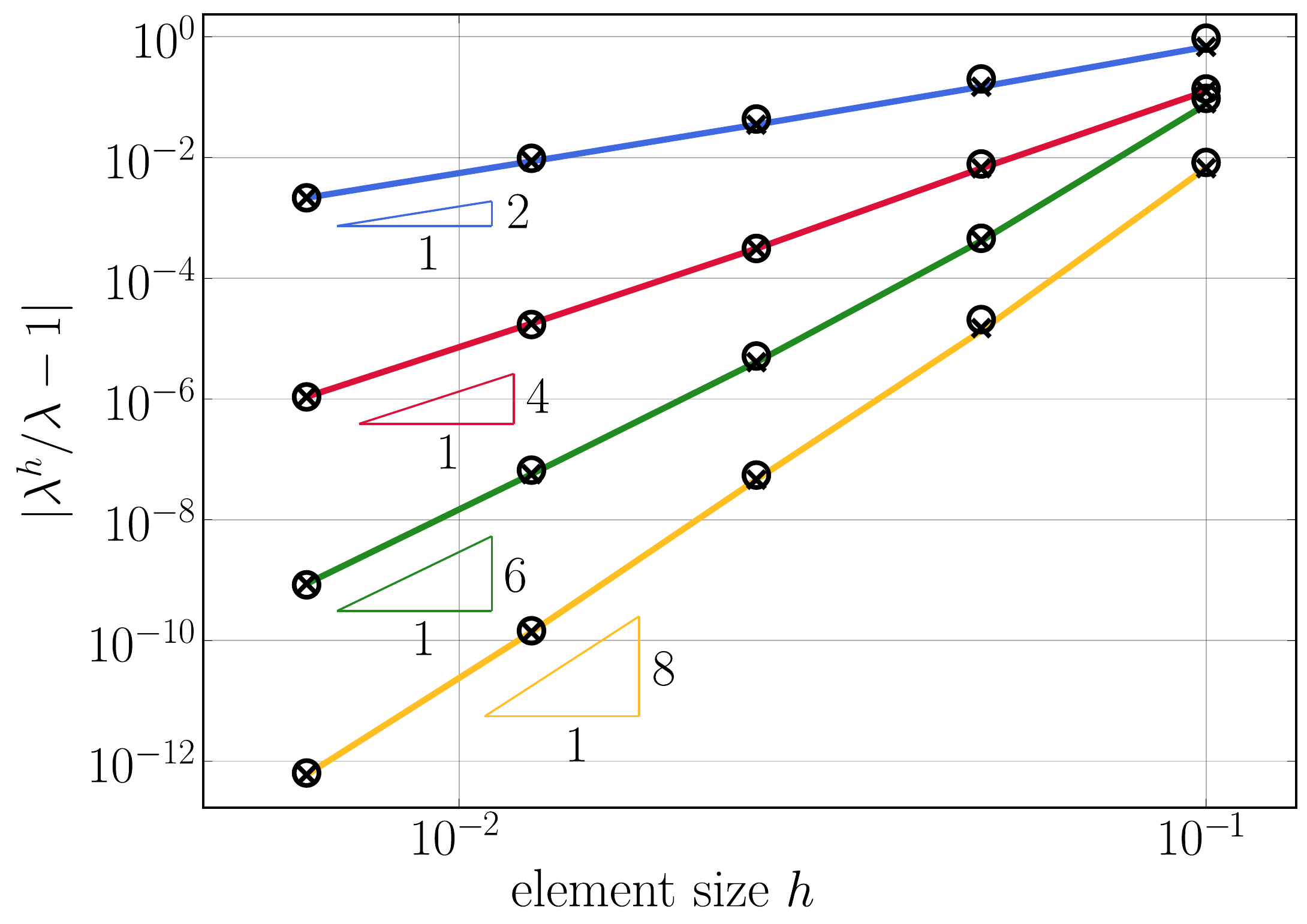} \includegraphics[width=0.5\textwidth]{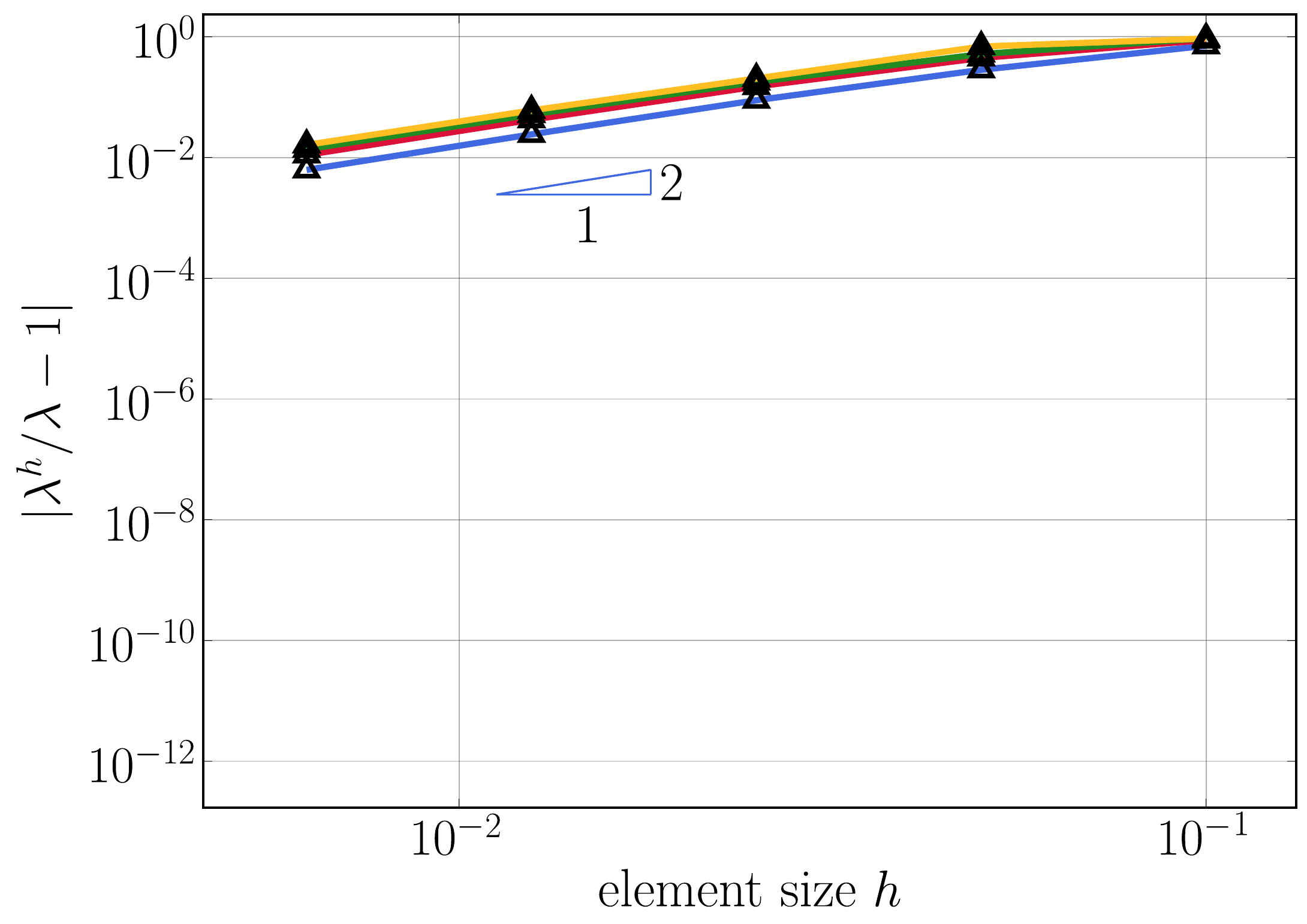} }
    
    \subfloat[Mode shape $\eigenvec^h_{10}$.]{ \includegraphics[width=0.5\textwidth]{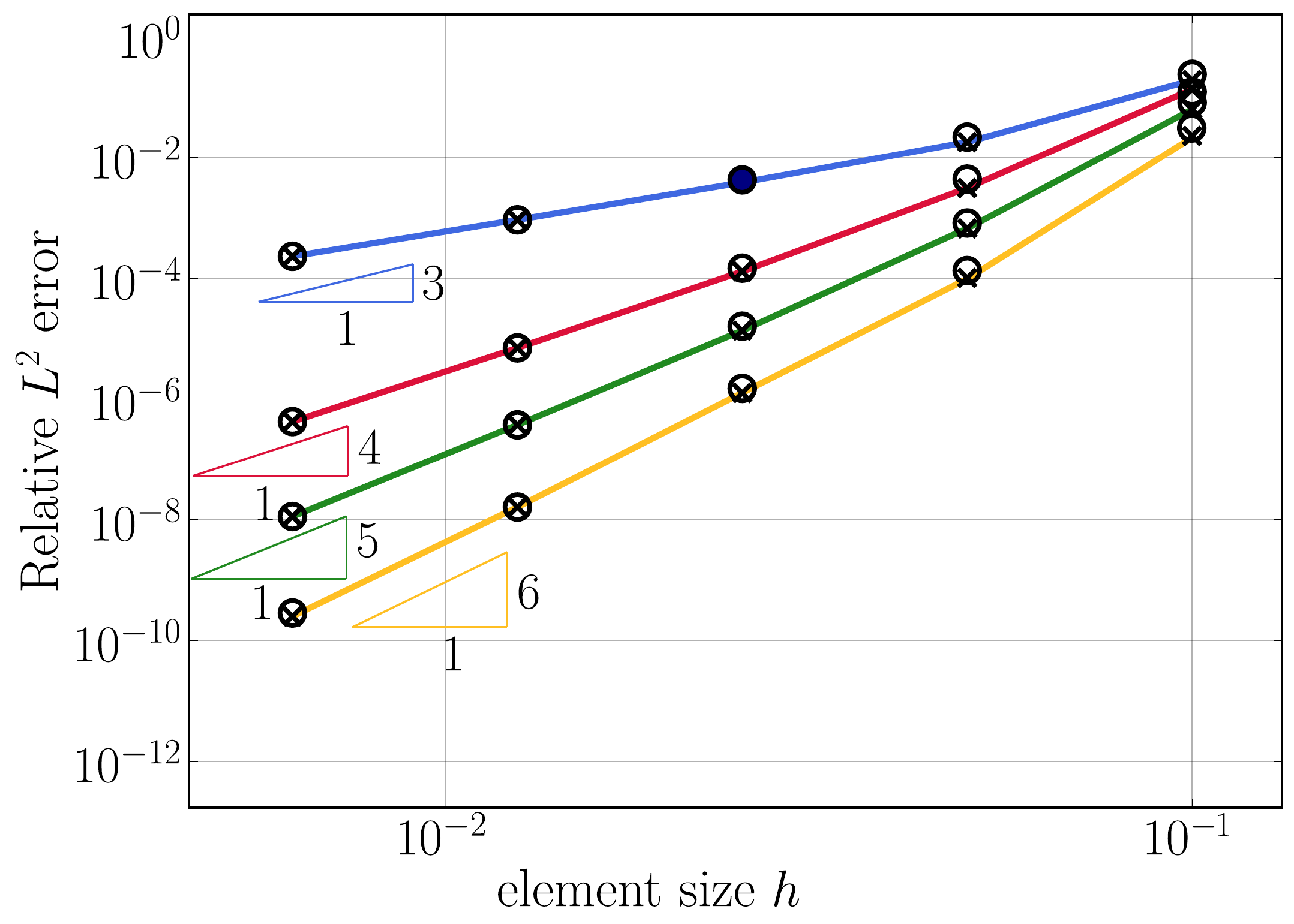} \includegraphics[width=0.5\textwidth]{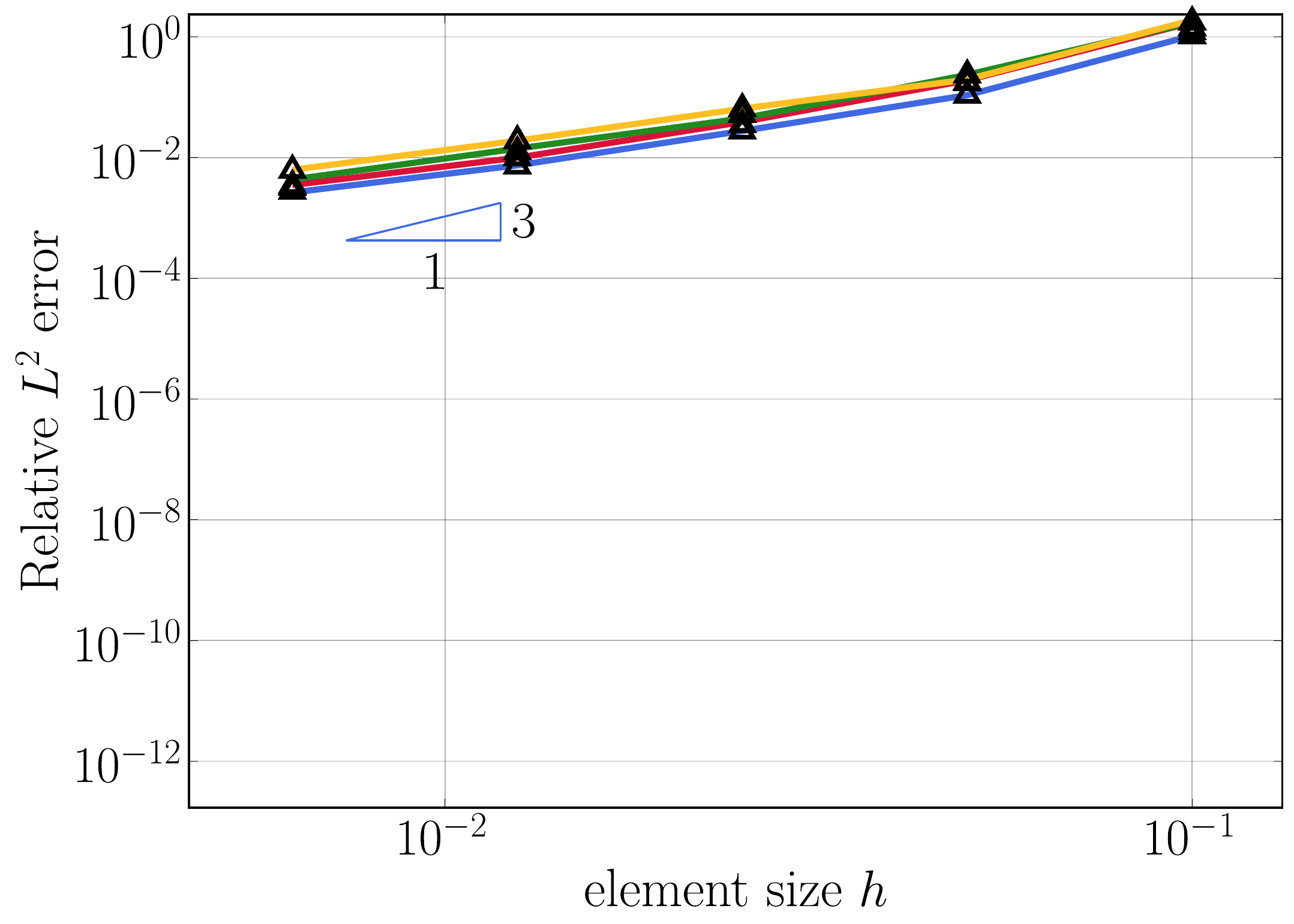} }
    \vspace{0.2cm}
    \begin{tikzpicture}
    \filldraw[blue1,line width=1pt, solid] (0,0) -- (0.6,0);
    \filldraw[blue1,line width=1pt] (0.6,0) node[right]{\footnotesize $p=2$};	
    \filldraw[red1,line width=1pt, solid] (3,0) -- (3.6,0);
    \filldraw[red1,line width=1pt] (3.6,0) node[right]{\footnotesize $p=3$};
    \filldraw[green1,line width=1pt, solid] (6,0) -- (6.6,0);
    \filldraw[green1,line width=1pt] (6.6,0) node[right]{\footnotesize $p=4$};
    \filldraw[yellow1,line width=1pt, solid] (9,0) -- (9.6,0);
    \filldraw[yellow1,line width=1pt] (9.6,0) node[right]{\footnotesize $p=5$};
\end{tikzpicture}
    \begin{tikzpicture}
    \filldraw[black,line width=1pt, solid] (0.0,0) -- (0.6,0);
    \filldraw[black,line width=1pt] (0.0,0) node[right]{\footnotesize $\boldsymbol{\bigtimes}$};
    \filldraw[black,line width=1pt] (0.6,0) node[right]{\footnotesize Galerkin method, consistent mass};
    \filldraw[black,line width=1pt, solid] (7,0) -- (7.6,0);
    \filldraw[black,line width=1pt] (7.0,0) node[right]{\footnotesize $\boldsymbol{\Delta}$};
    \filldraw[black,line width=1pt] (7.6,0) node[right]{\footnotesize Galerkin method, row-sum lumped mass};
\end{tikzpicture}

\begin{tikzpicture}
    \filldraw[black,line width=1pt, solid] (3.0,0) -- (3.2,0);
    \filldraw[black,line width=1pt] (3.3,0) [fill=none] circle (2pt);
    \filldraw[black,line width=1pt, solid] (3.4,0) -- (3.6,0);
    \filldraw[black,line width=1pt] (3.6,0) node[right]{\footnotesize Petrov-Galerkin method, row-sum lumped mass};
\end{tikzpicture}
    \caption{Convergence of the relative $L^2$ error in the tenth eigenvalue and mode shape of a freely vibrating beam, computed on \changed{500 B\'ezier elements}.} \label{fig:spectra_convergence_beam_no_iter}
\end{figure}
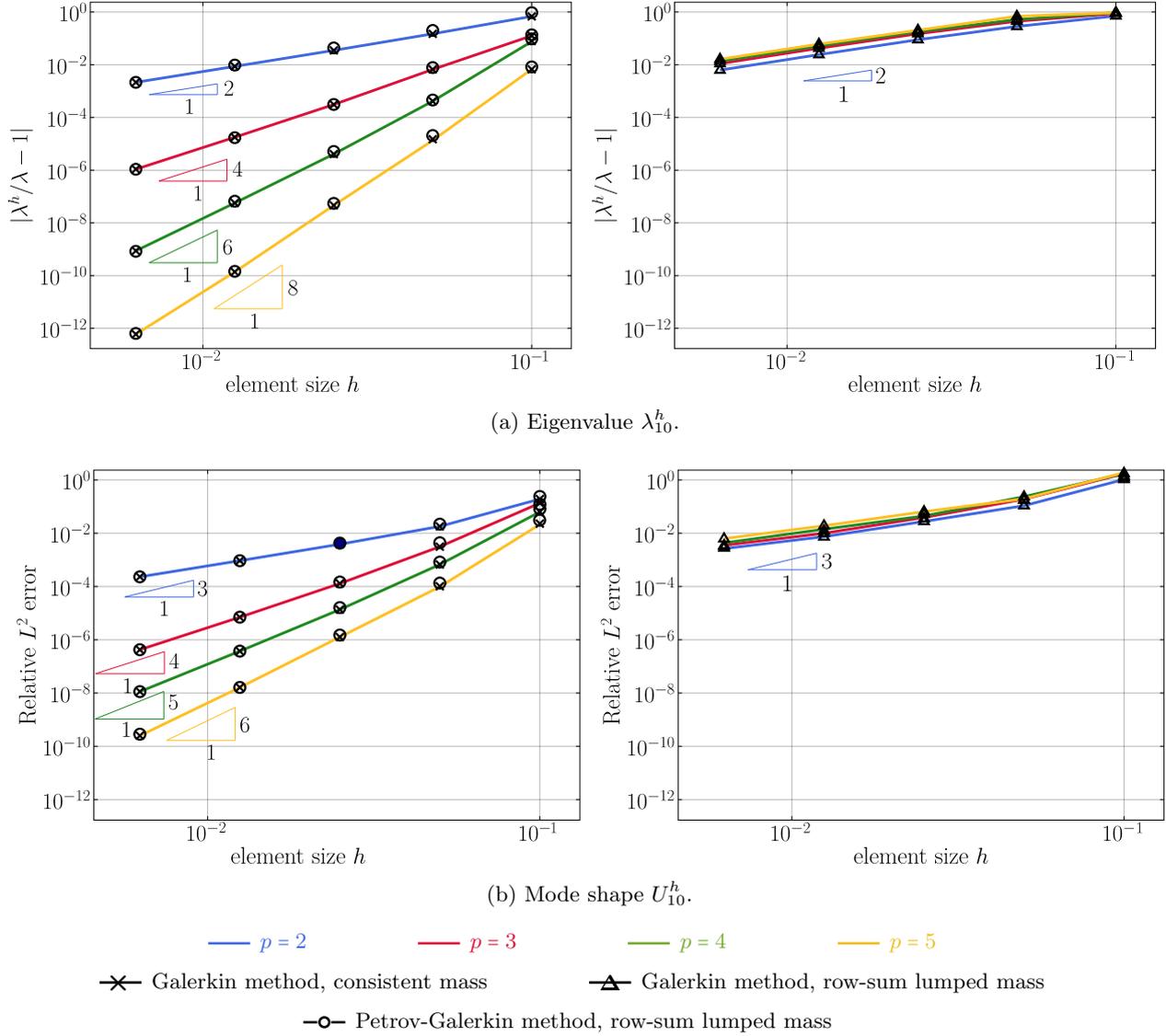

%---------------------------------------------------

\begin{figure}[h!]
    \centering
    \subfloat[$p=2$]{{\includegraphics[width=0.48\textwidth]{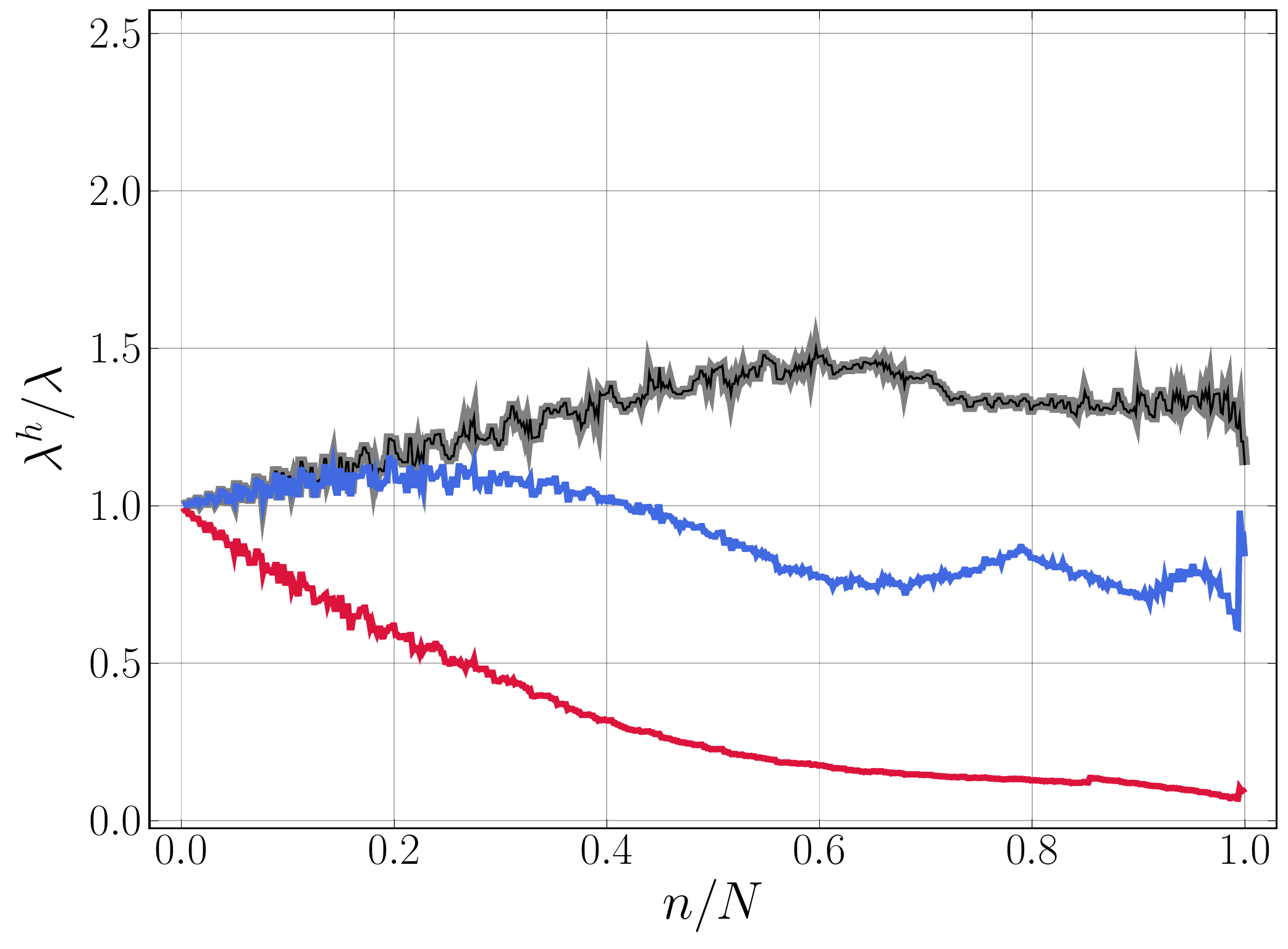} }}
    \subfloat[$p=3$]{{\includegraphics[width=0.48\textwidth]{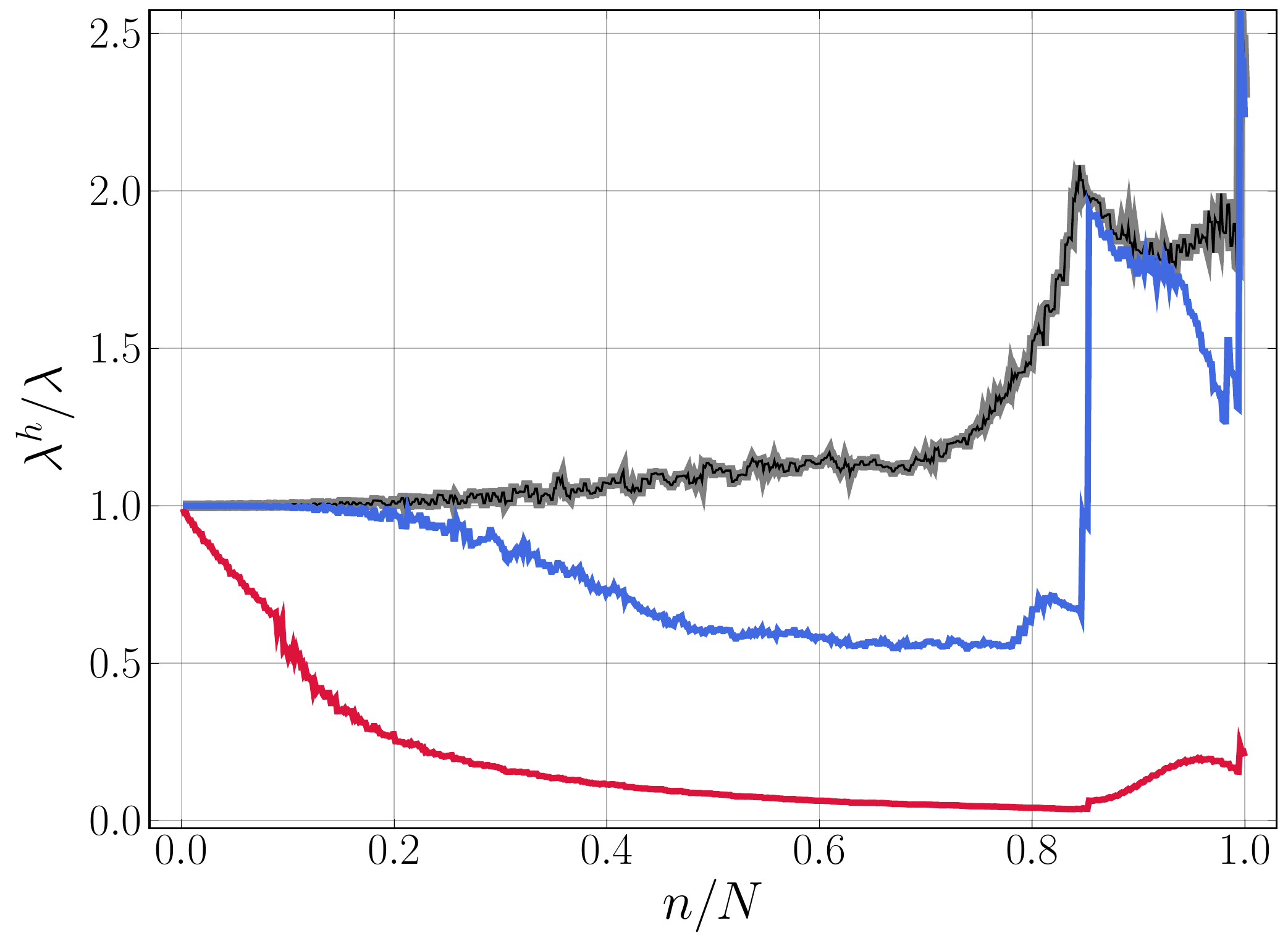} }}

    \subfloat[$p=4$]{{\includegraphics[width=0.48\textwidth]{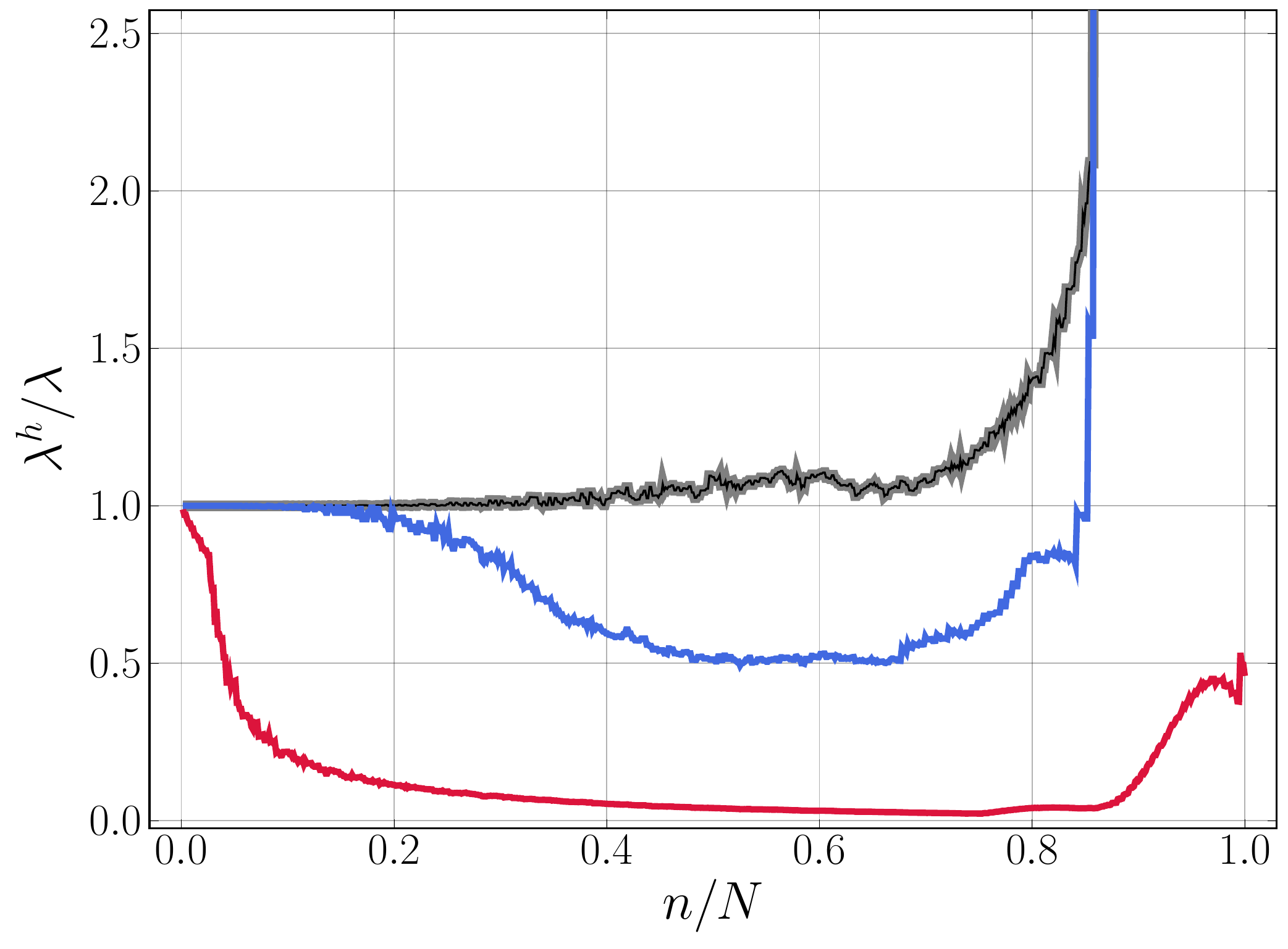} }}
    \subfloat[$p=5$]{{\includegraphics[width=0.48\textwidth]{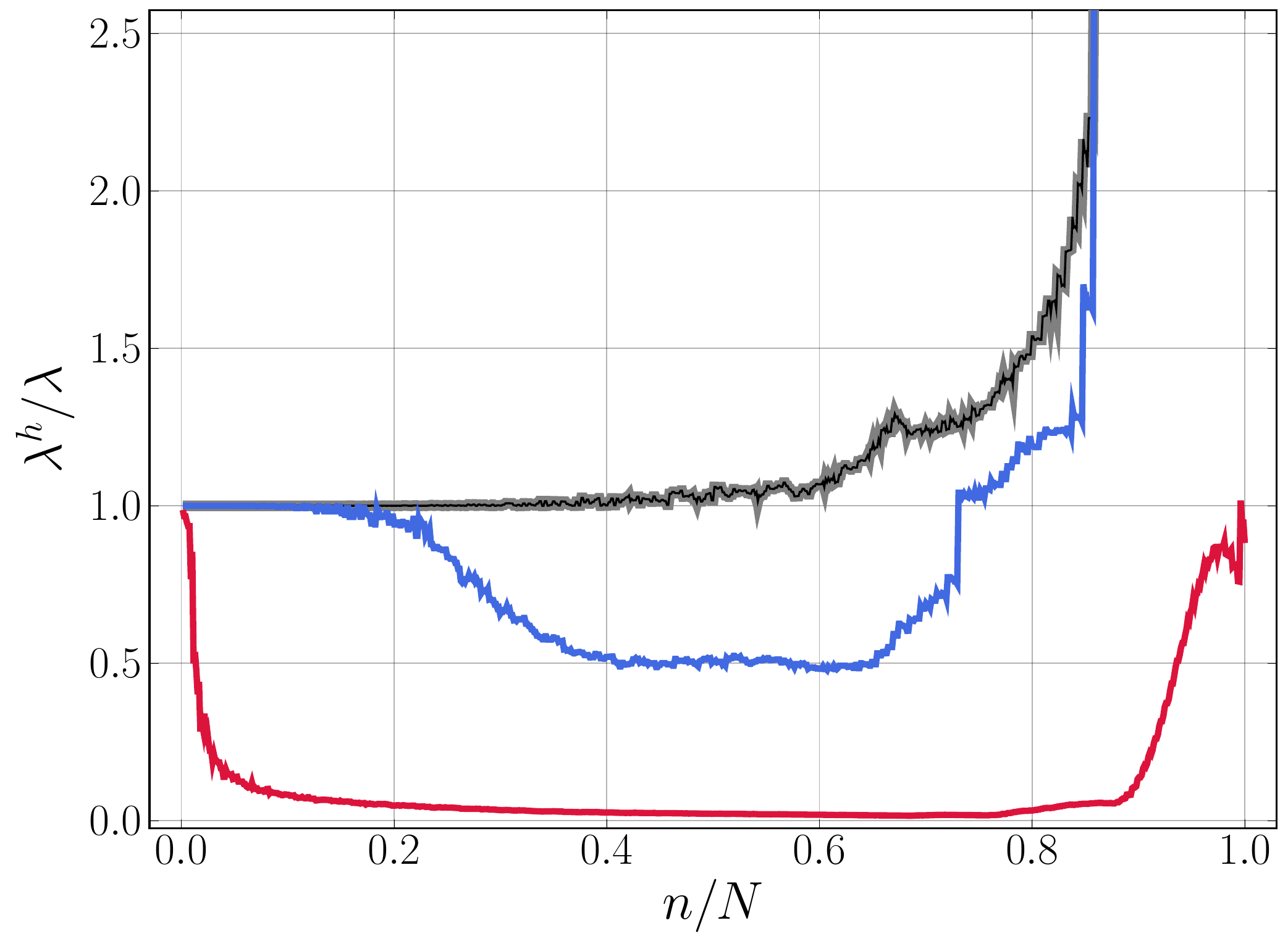} }}
    \vspace{0.2cm}
    \begin{tikzpicture}
    \filldraw[grey1,line width=1pt, solid] (0,0) -- (0.6,0);
    \filldraw[grey1,line width=1pt] (0.6,0) node[right]{\footnotesize Galerkin, consistent mass};
    \filldraw[red1,line width=1pt, solid] (7,0) -- (7.6,0);
    \filldraw[red1,line width=1pt] (7.6,0) node[right]{\footnotesize Galerkin, row-sum lumped mass};
\end{tikzpicture}

\begin{tikzpicture}
    \filldraw[black,line width=1pt, solid] (0,0) -- (0.6,0);
    \filldraw[black,line width=1pt] (0.6,0) node[right]{\footnotesize Petrov-Galerkin, consistent mass};
    \filldraw[blue1,line width=1pt, solid] (6,0) -- (6.6,0);
    \filldraw[blue1,line width=1pt] (6.6,0) node[right]{\footnotesize Petrov-Galerkin, higher-order accurate lumped mass};
\end{tikzpicture}
    \caption{Normalized eigenvalues of a freely vibrating simply supported plate, computed on $25 \times 25$ B\'ezier elements.} \label{fig:spectra_eigenval_plate_no_iter}
\end{figure}
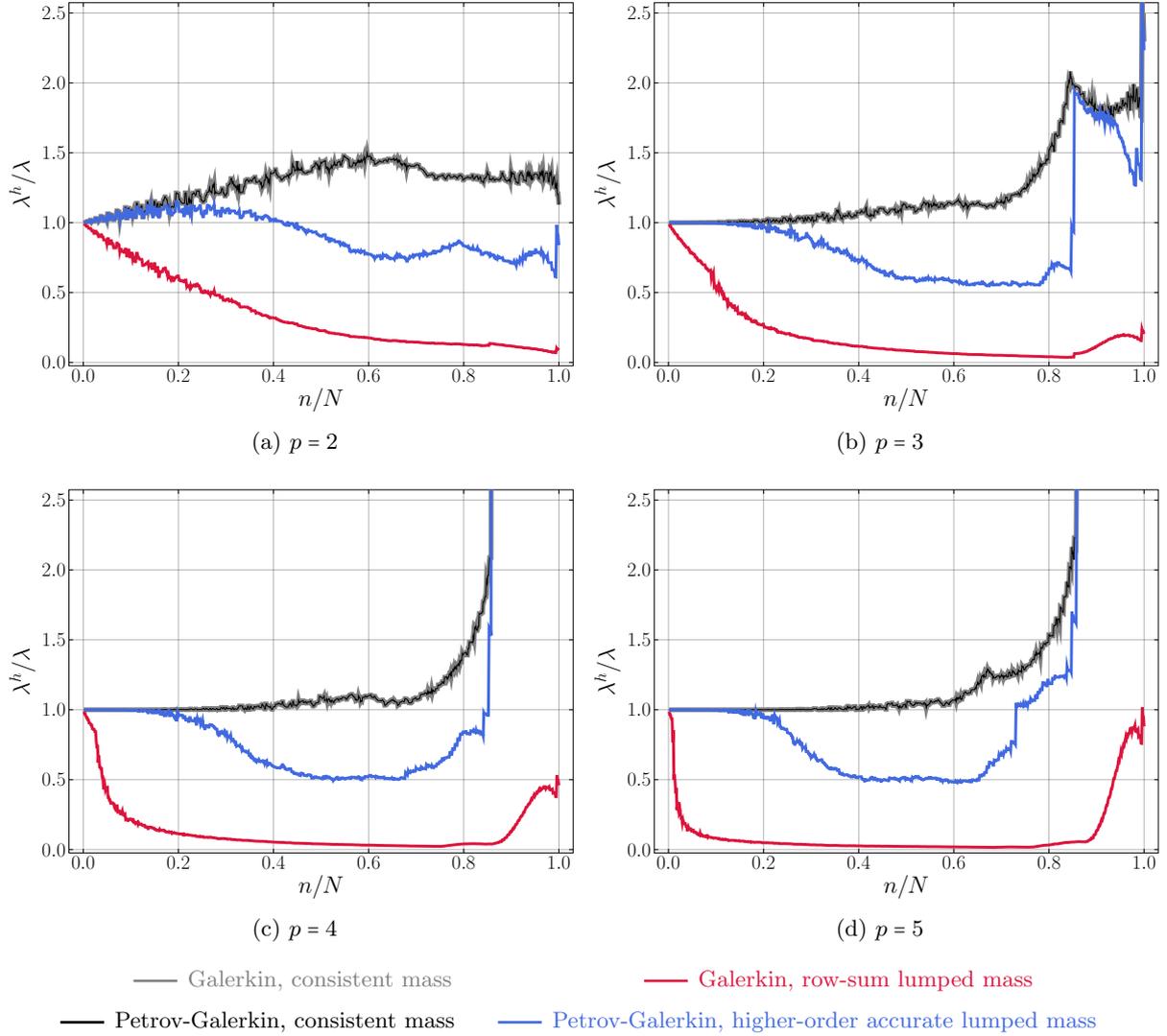

\begin{figure}[h!]
    \centering
   \includegraphics[width=0.49\textwidth]{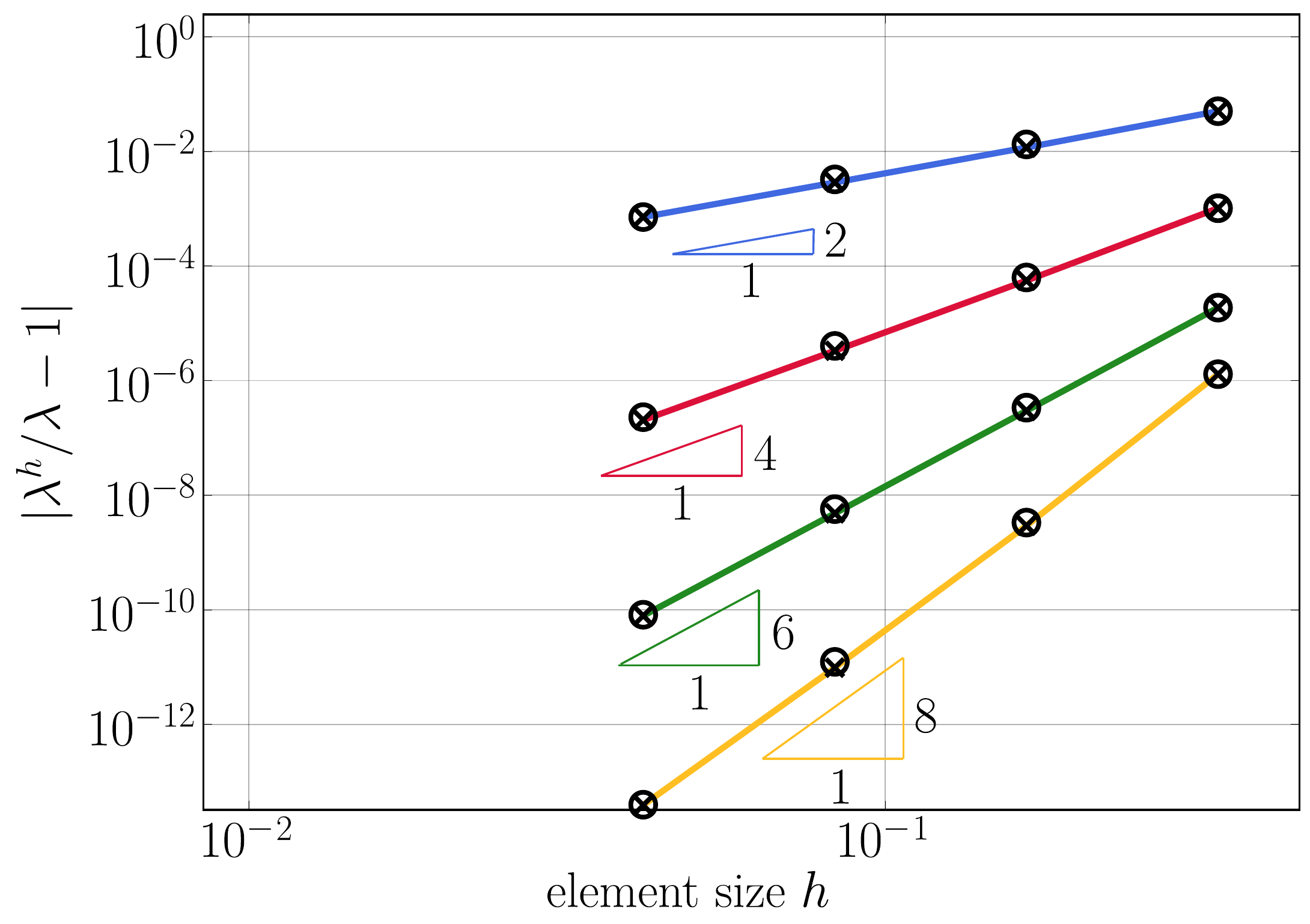} 
   \includegraphics[width=0.49\textwidth]{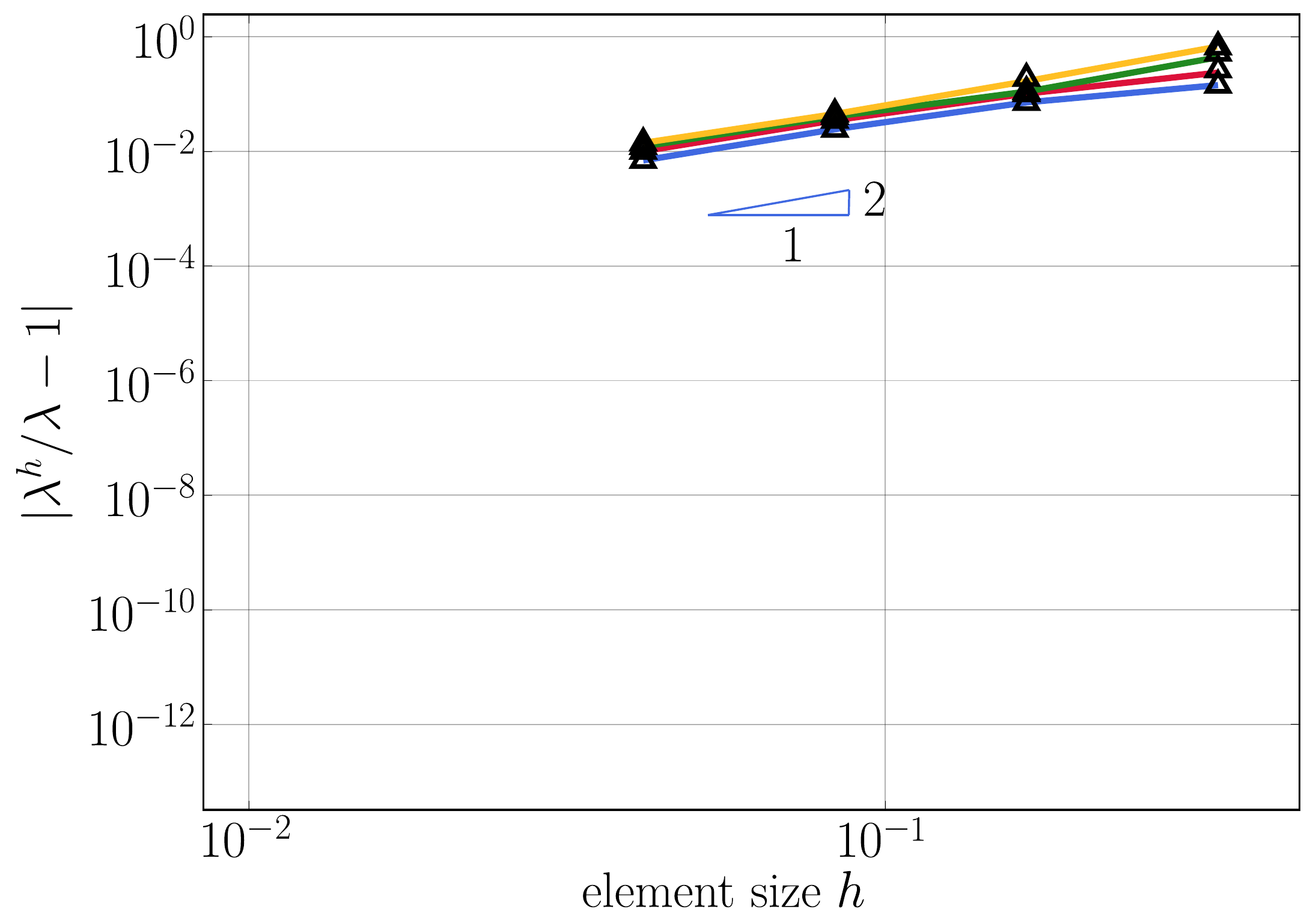}
   
    \vspace{0.2cm}
    \begin{tikzpicture}
    \filldraw[blue1,line width=1pt, solid] (0,0) -- (0.6,0);
    \filldraw[blue1,line width=1pt] (0.6,0) node[right]{\footnotesize $p=2$};	
    \filldraw[red1,line width=1pt, solid] (3,0) -- (3.6,0);
    \filldraw[red1,line width=1pt] (3.6,0) node[right]{\footnotesize $p=3$};
    \filldraw[green1,line width=1pt, solid] (6,0) -- (6.6,0);
    \filldraw[green1,line width=1pt] (6.6,0) node[right]{\footnotesize $p=4$};
    \filldraw[yellow1,line width=1pt, solid] (9,0) -- (9.6,0);
    \filldraw[yellow1,line width=1pt] (9.6,0) node[right]{\footnotesize $p=5$};
\end{tikzpicture}
    \begin{tikzpicture}
    \filldraw[black,line width=1pt, solid] (0.0,0) -- (0.6,0);
    \filldraw[black,line width=1pt] (0.0,0) node[right]{\footnotesize $\boldsymbol{\bigtimes}$};
    \filldraw[black,line width=1pt] (0.6,0) node[right]{\footnotesize Galerkin method, consistent mass};
    \filldraw[black,line width=1pt, solid] (7,0) -- (7.6,0);
    \filldraw[black,line width=1pt] (7.0,0) node[right]{\footnotesize $\boldsymbol{\Delta}$};
    \filldraw[black,line width=1pt] (7.6,0) node[right]{\footnotesize Galerkin method, row-sum lumped mass};
\end{tikzpicture}

\begin{tikzpicture}
    \filldraw[black,line width=1pt, solid] (3.0,0) -- (3.2,0);
    \filldraw[black,line width=1pt] (3.3,0) [fill=none] circle (2pt);
    \filldraw[black,line width=1pt, solid] (3.4,0) -- (3.6,0);
    \filldraw[black,line width=1pt] (3.6,0) node[right]{\footnotesize Petrov-Galerkin method, row-sum lumped mass};
\end{tikzpicture}
    \caption{Convergence of the \changed{relative error} in the first eigenvalue of a freely vibrating simply supported plate.} \label{fig:spectra_convergence_plate_no_iter}
\end{figure}
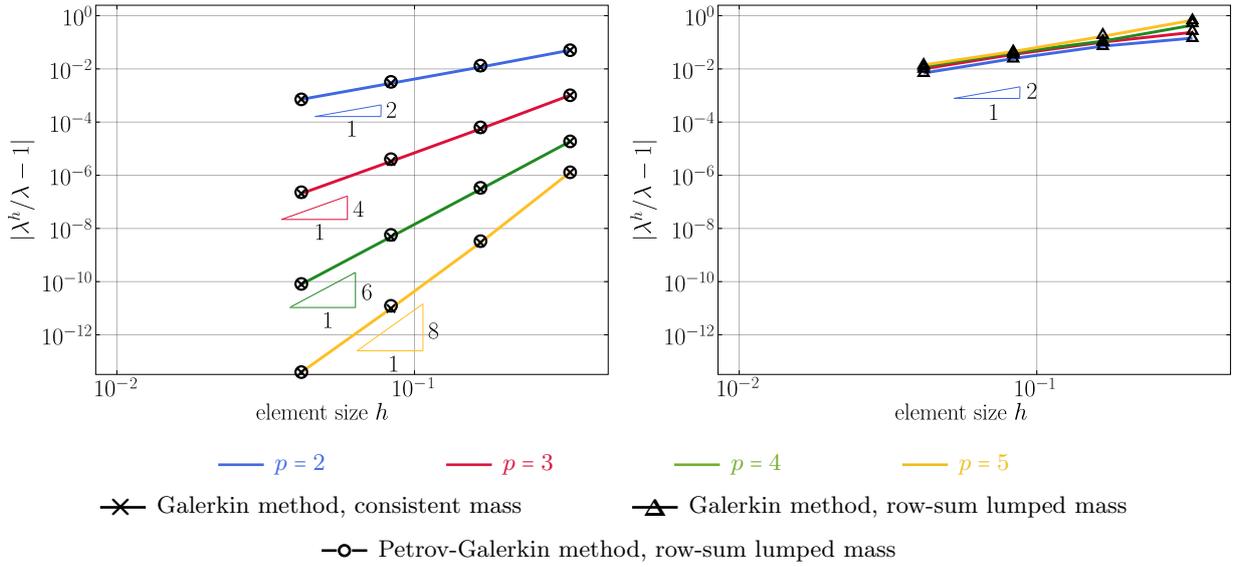

%---------------------------------------------------

\begin{figure}[h!]
    \centering
    \subfloat[$p=2$]{{\includegraphics[width=0.48\textwidth]{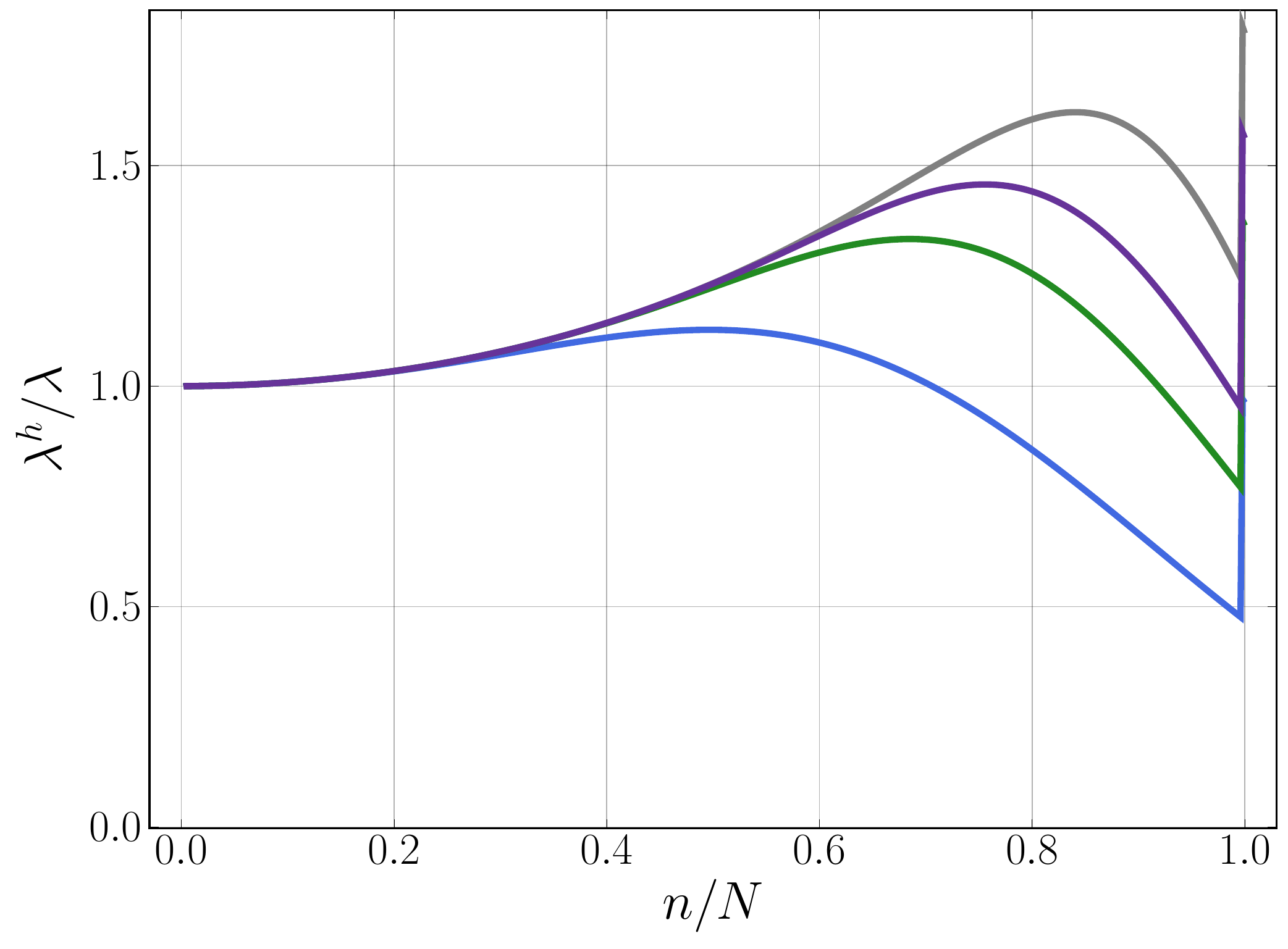} }}
    \subfloat[$p=3$]{{\includegraphics[width=0.48\textwidth]{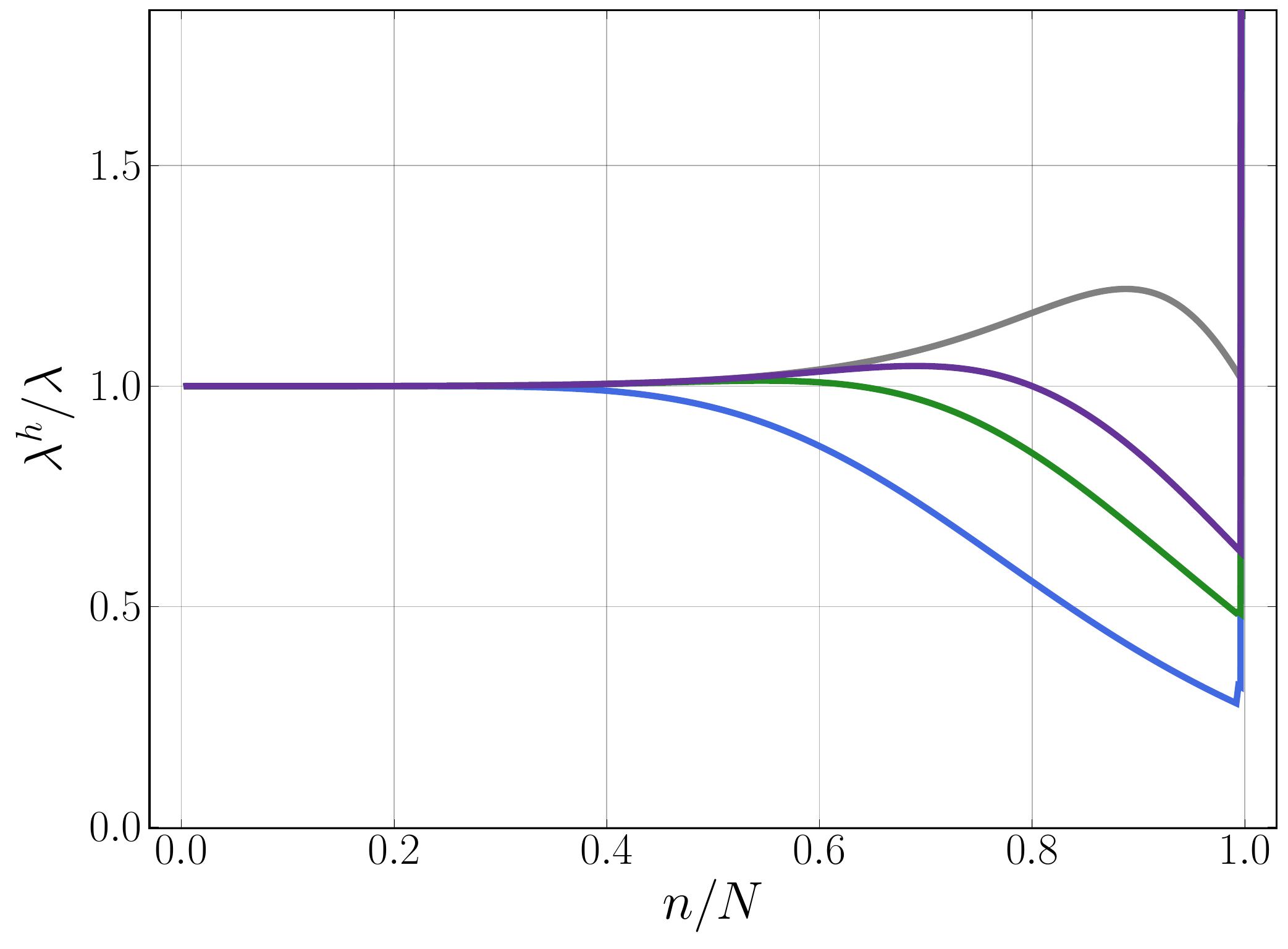} }}

    \subfloat[$p=4$]{{\includegraphics[width=0.48\textwidth]{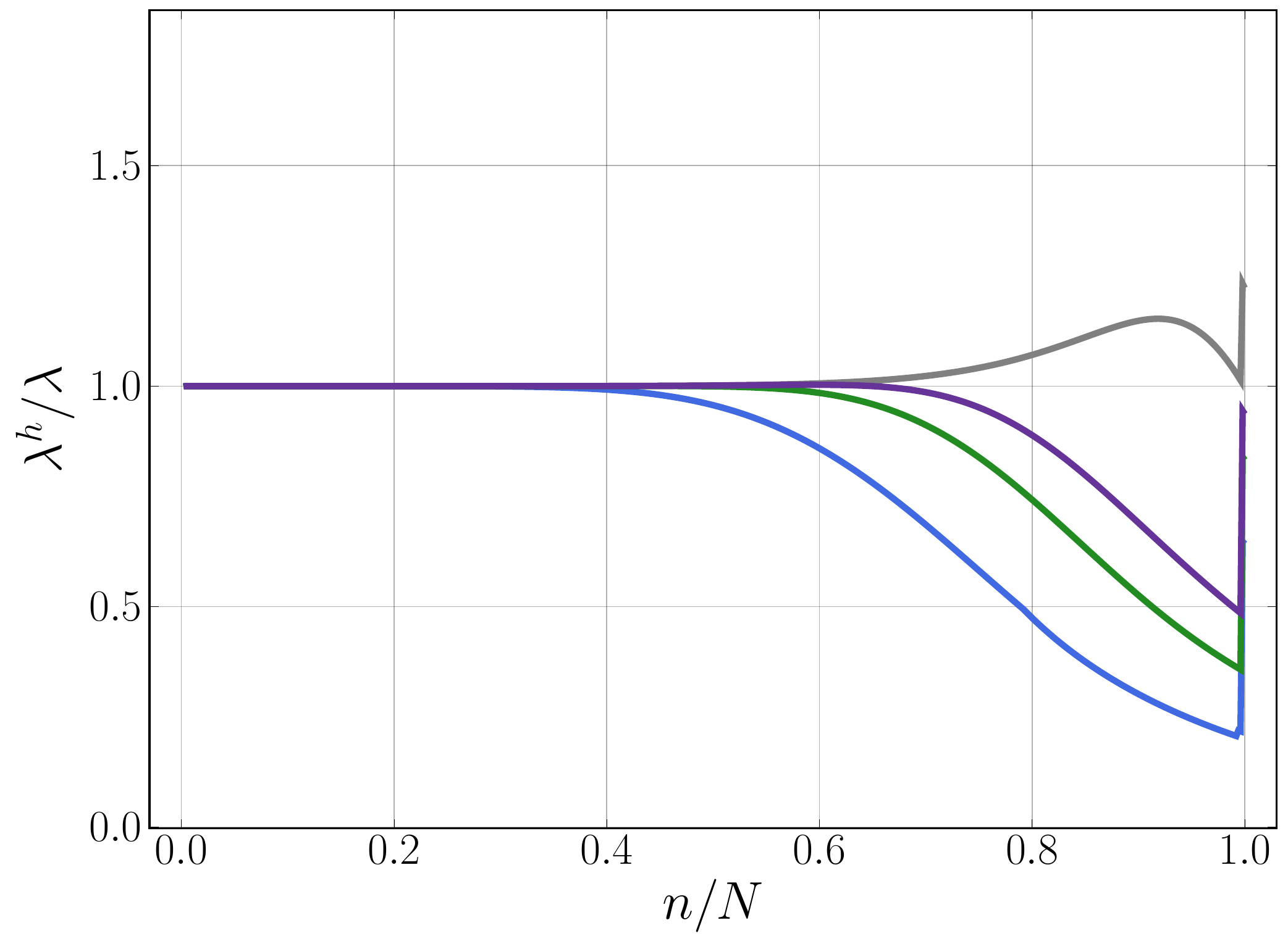} }}
    \subfloat[$p=5$]{{\includegraphics[width=0.48\textwidth]{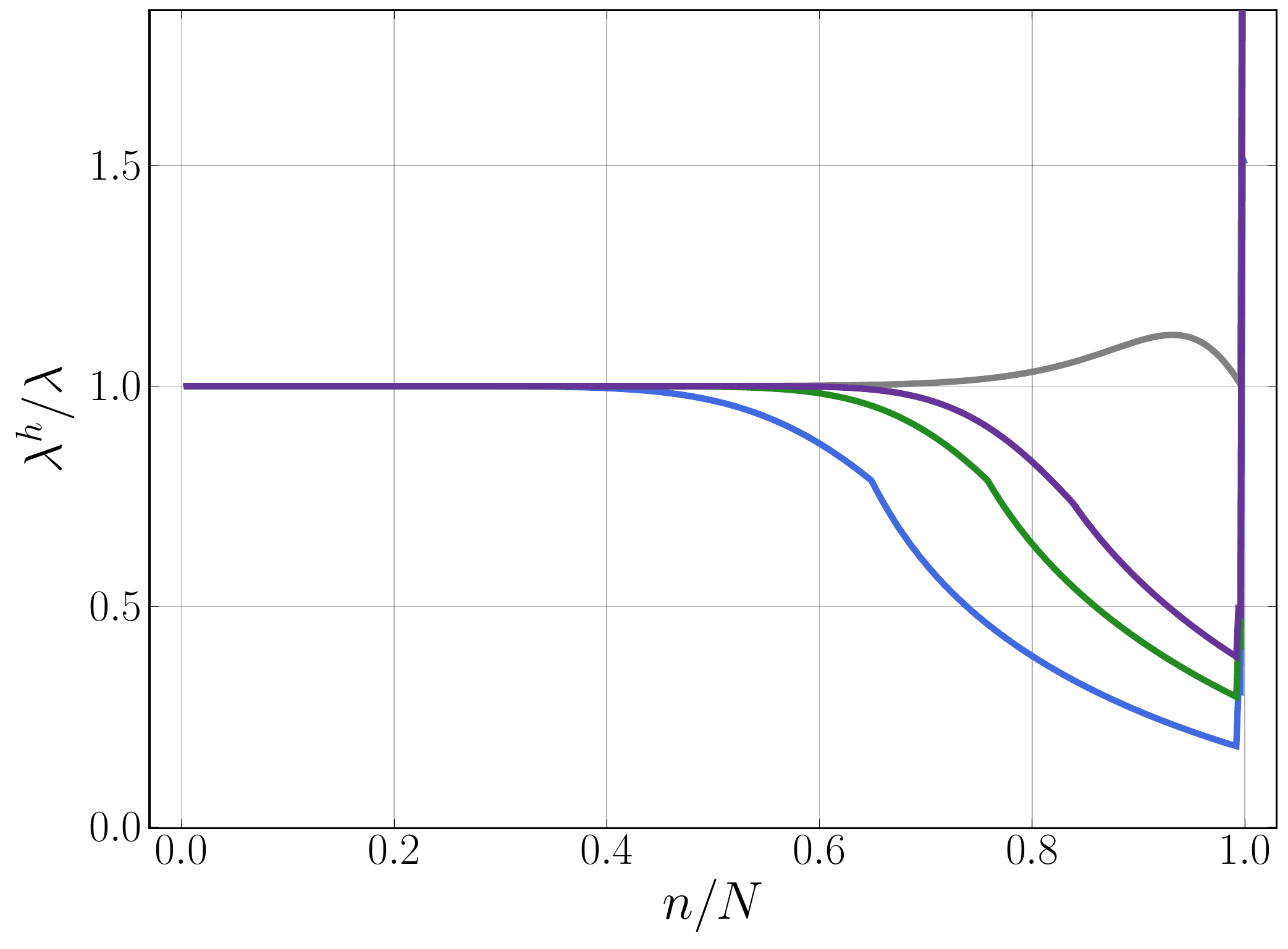} }}
    \vspace{0.2cm}
    \begin{tikzpicture}
    \filldraw[grey1,line width=1pt, solid] (0,0) -- (0.6,0);
    \filldraw[grey1,line width=1pt] (0.6,0) node[right]{\footnotesize Petrov-Galerkin, consistent mass};
    \filldraw[blue1,line width=1pt, solid] (6,0) -- (6.6,0);
    \filldraw[blue1,line width=1pt] (6.6,0) node[right]{\footnotesize $r=0$};
    \filldraw[green1,line width=1pt, solid] (8,0) -- (8.6,0);
    \filldraw[green1,line width=1pt] (8.6,0) node[right]{\footnotesize $r=1$};
    \filldraw[purple1,line width=1pt, solid] (10,0) -- (10.6,0);
    \filldraw[purple1,line width=1pt] (10.6,0) node[right]{\footnotesize $r=2$};
\end{tikzpicture}
    \caption{Normalized eigenvalues of the freely vibrating beam, computed with an iteratively improved approximate dual basis and a different number $r$ of corrector passes.} \label{fig:spectra_eigenval_beam_iter}
\end{figure}
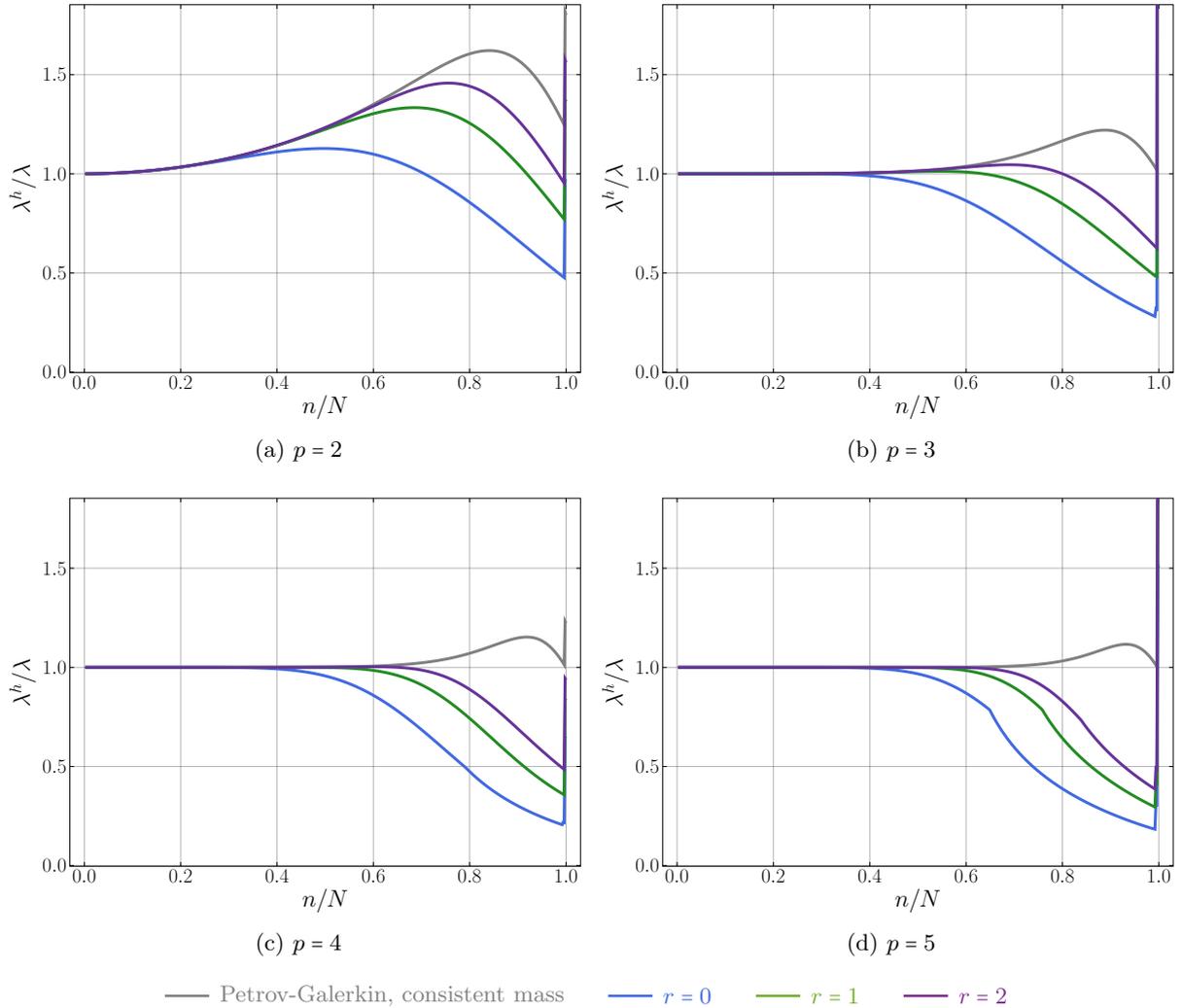

\begin{figure}[h!]
    \centering
    \subfloat[$p=2$]{{\includegraphics[width=0.48\textwidth]{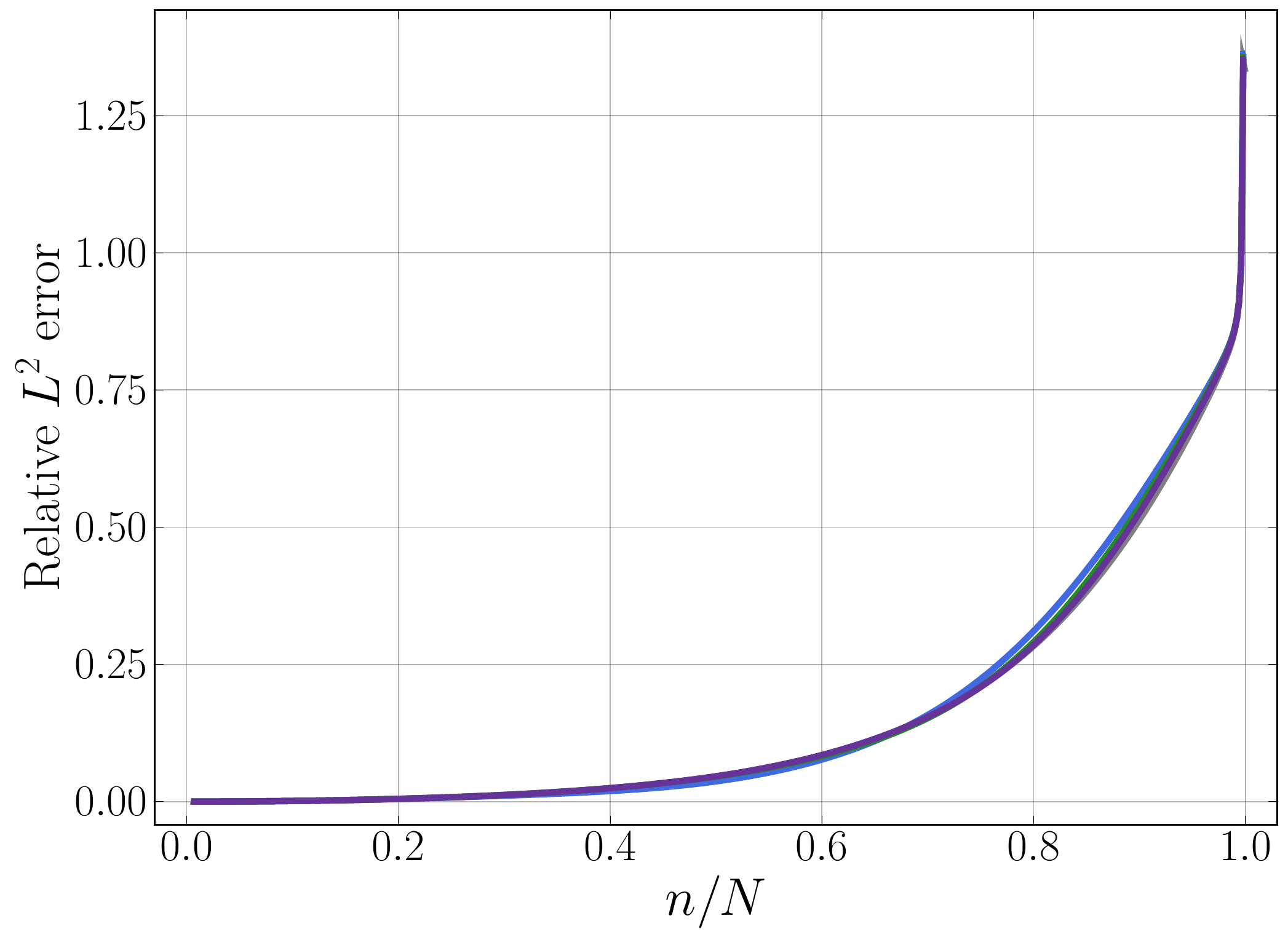} }}
    \subfloat[$p=3$]{{\includegraphics[width=0.48\textwidth]{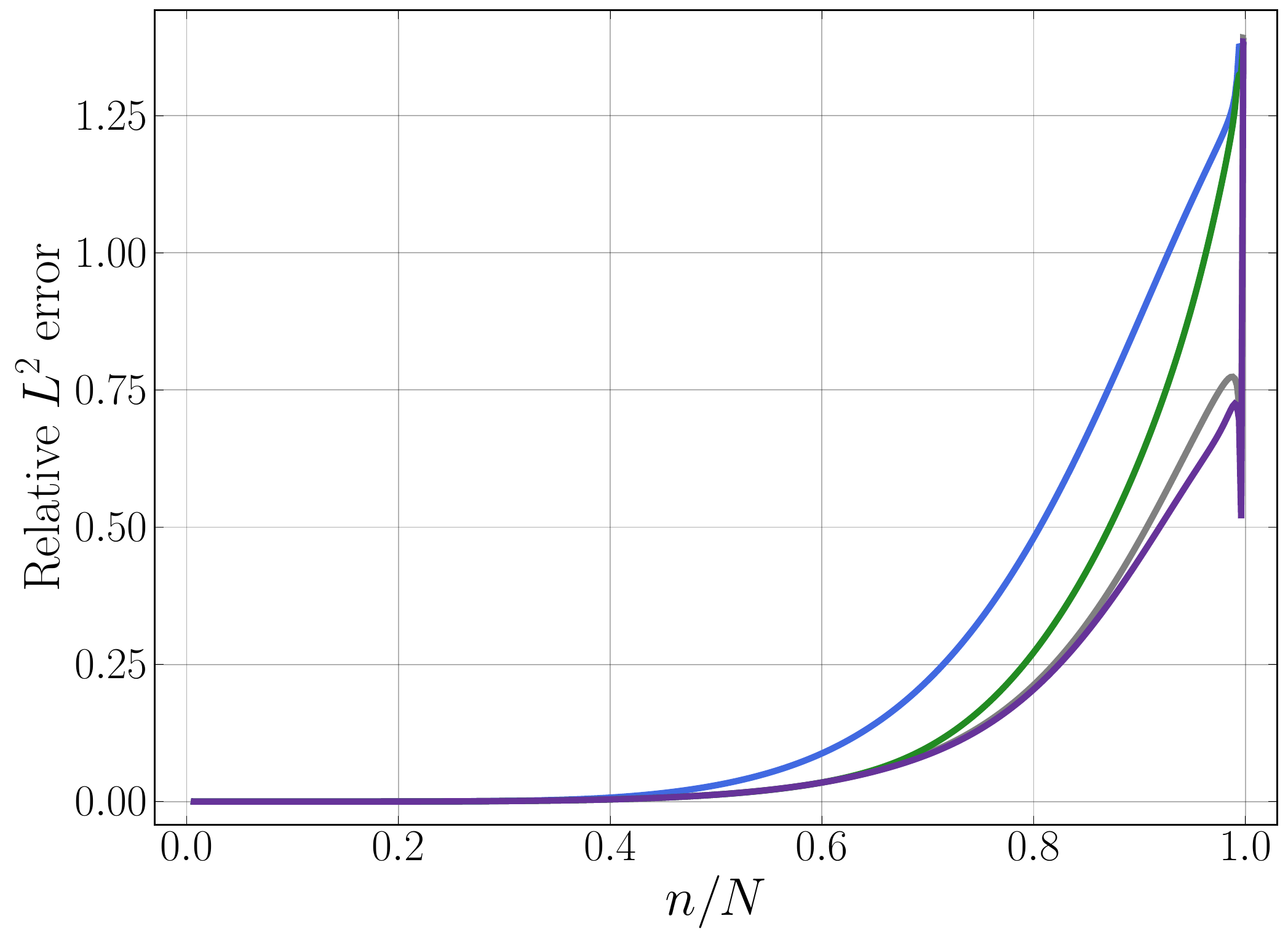} }}

    \subfloat[$p=4$]{{\includegraphics[width=0.48\textwidth]{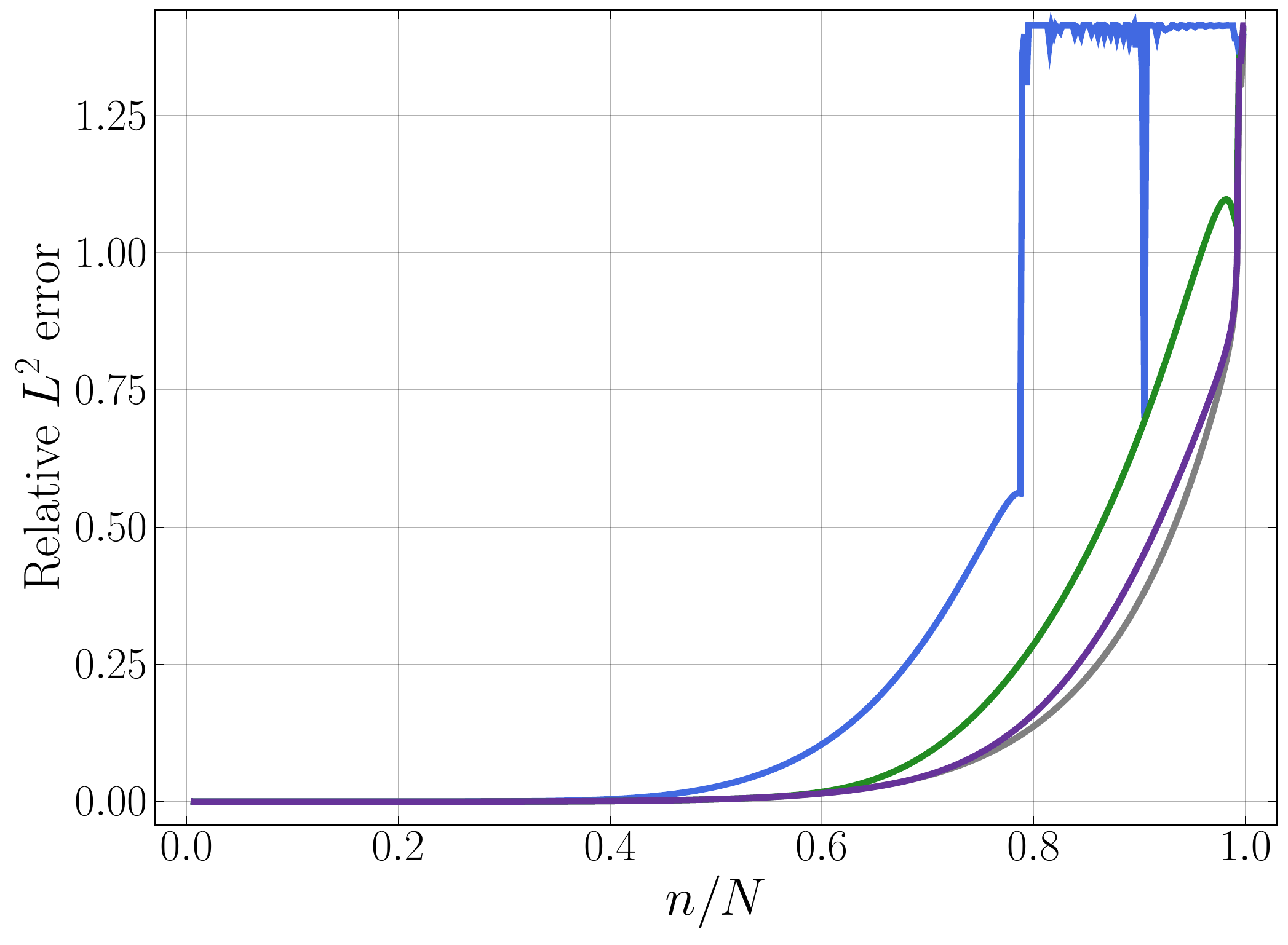} }}
    \subfloat[$p=5$]{{\includegraphics[width=0.48\textwidth]{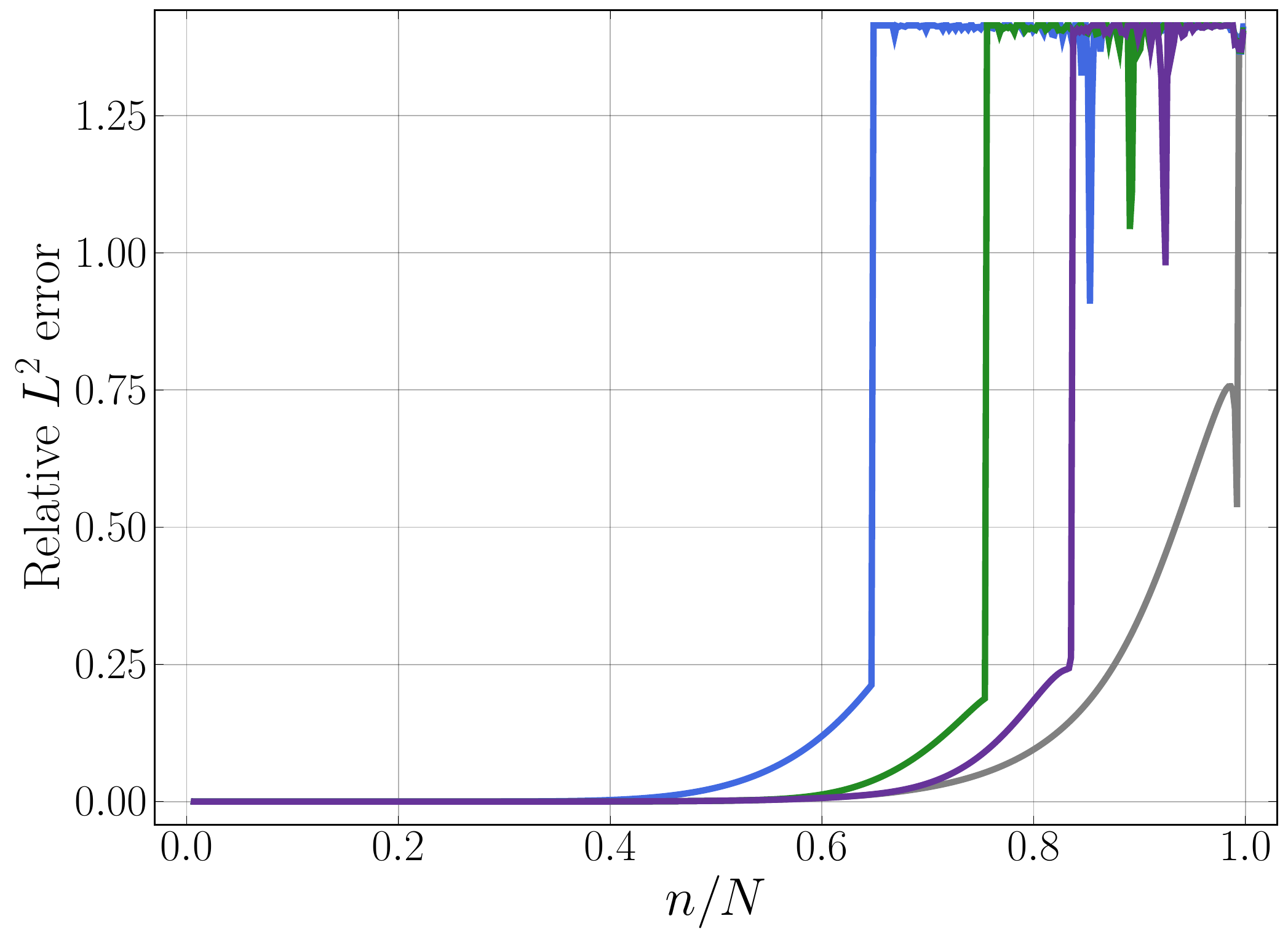} }}
    \vspace{0.2cm}
    \begin{tikzpicture}
    \filldraw[grey1,line width=1pt, solid] (0,0) -- (0.6,0);
    \filldraw[grey1,line width=1pt] (0.6,0) node[right]{\footnotesize Petrov-Galerkin, consistent mass};
    \filldraw[blue1,line width=1pt, solid] (6,0) -- (6.6,0);
    \filldraw[blue1,line width=1pt] (6.6,0) node[right]{\footnotesize $r=0$};
    \filldraw[green1,line width=1pt, solid] (8,0) -- (8.6,0);
    \filldraw[green1,line width=1pt] (8.6,0) node[right]{\footnotesize $r=1$};
    \filldraw[purple1,line width=1pt, solid] (10,0) -- (10.6,0);
    \filldraw[purple1,line width=1pt] (10.6,0) node[right]{\footnotesize $r=2$};
\end{tikzpicture}
    \caption{Relative $L^2$ error in the mode shapes of the freely vibrating beam, computed with an iteratively improved approximate dual basis and a different number $r$ of corrector passes.} \label{fig:spectra_mode_beam_iter}
\end{figure}
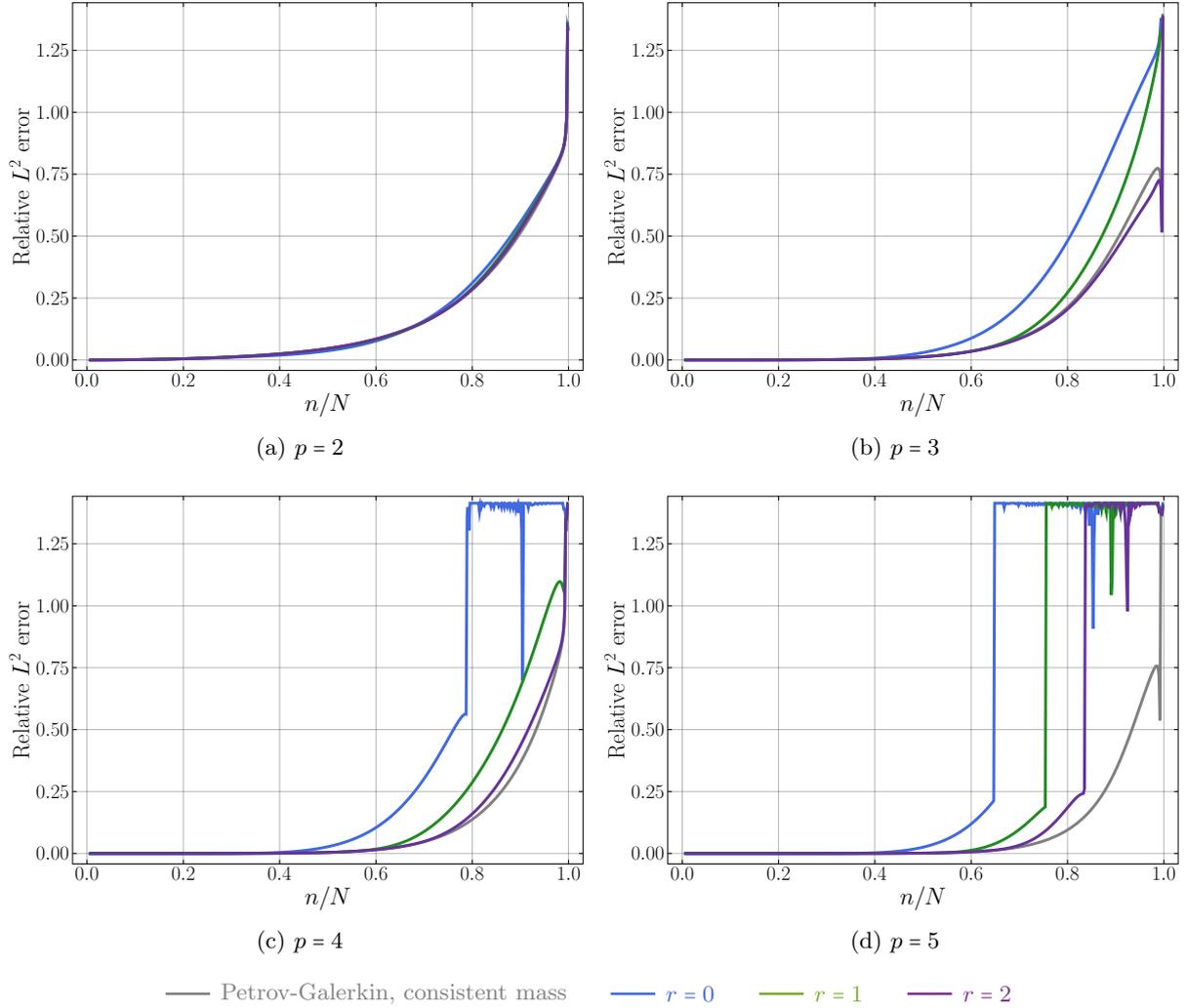

\begin{figure}[h!]
    \centering
    \subfloat[$p=2$]{{\includegraphics[width=0.48\textwidth]{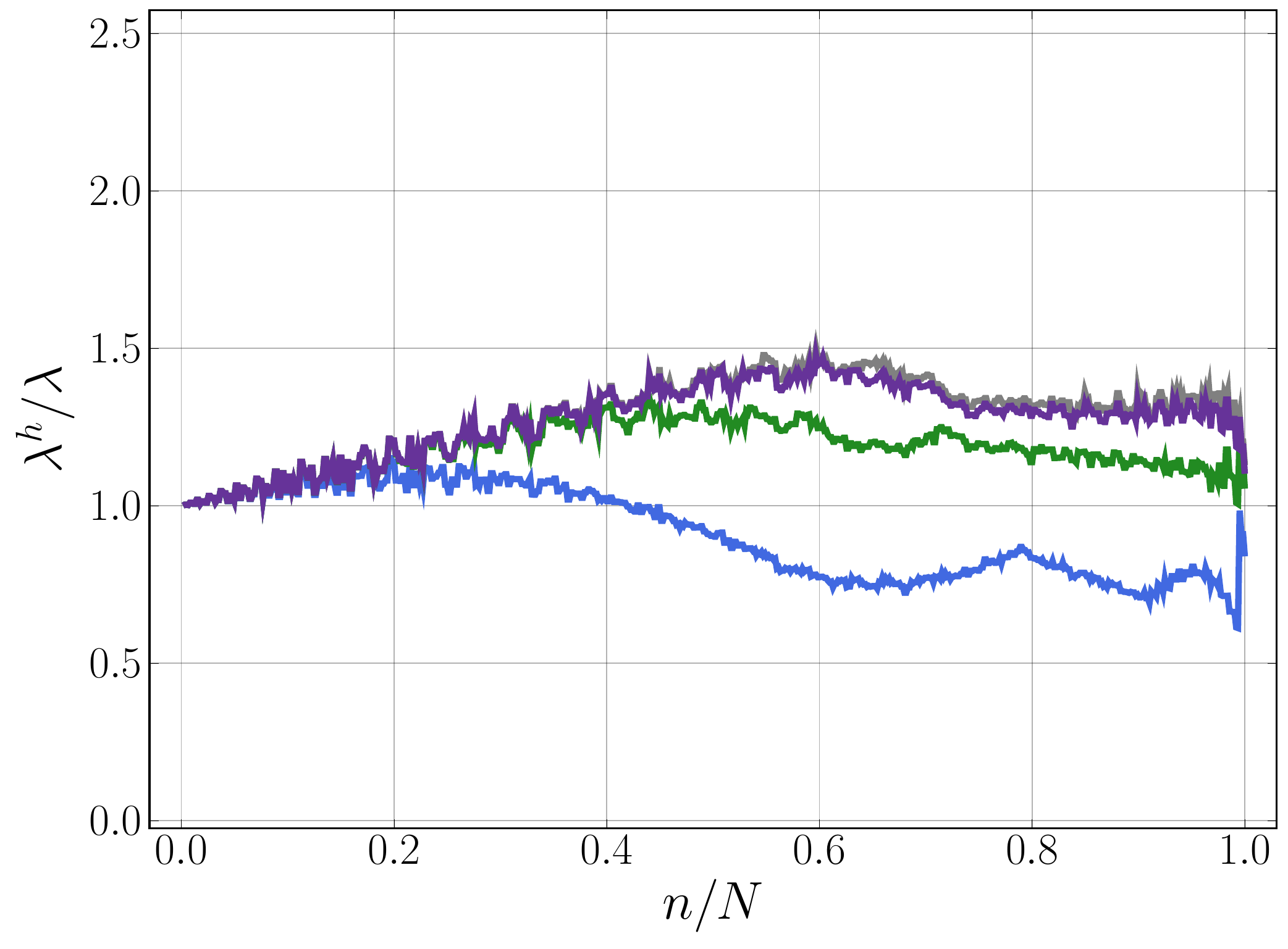} }}
    \subfloat[$p=3$]{{\includegraphics[width=0.48\textwidth]{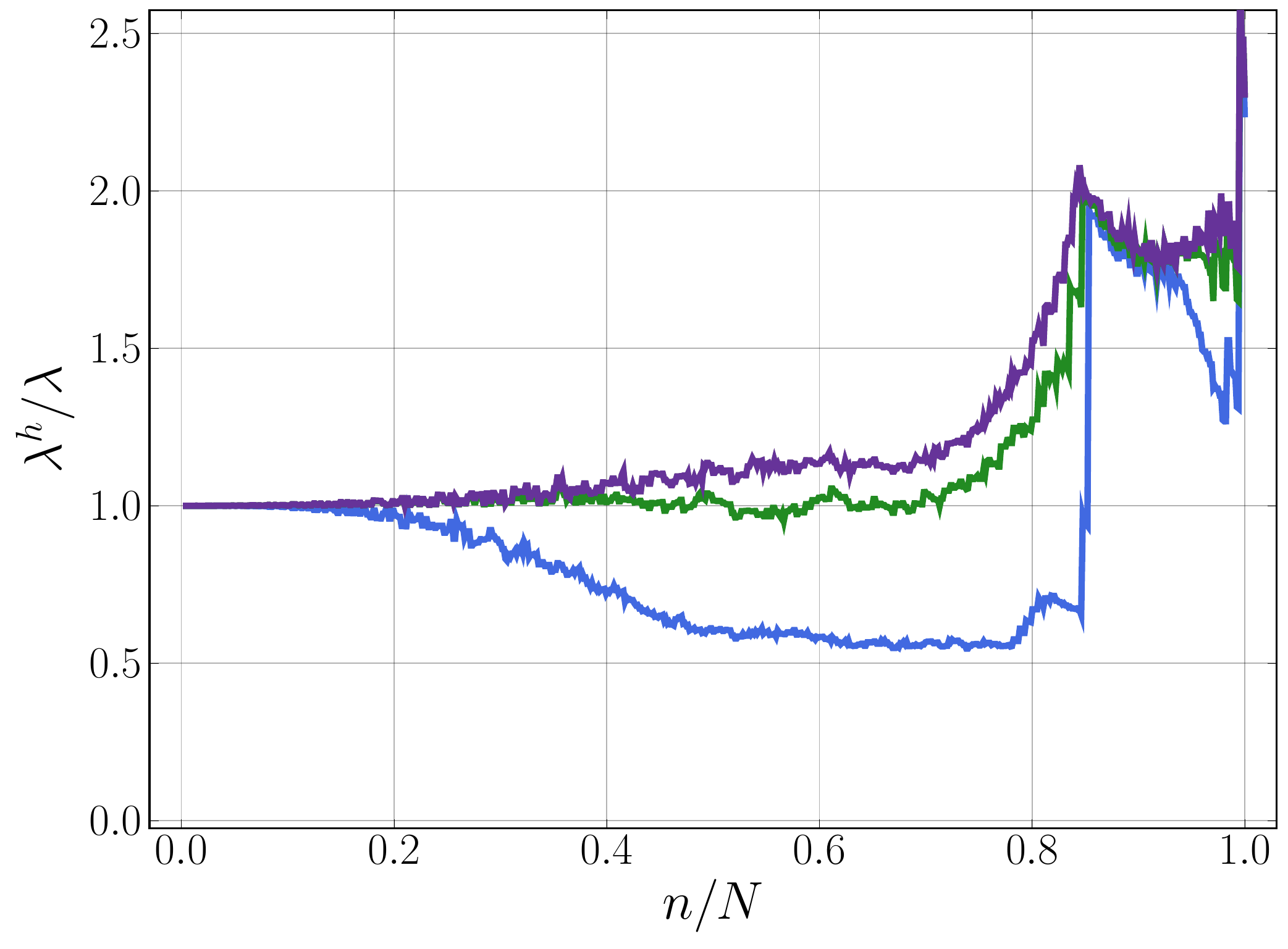} }}

    \subfloat[$p=4$]{{\includegraphics[width=0.48\textwidth]{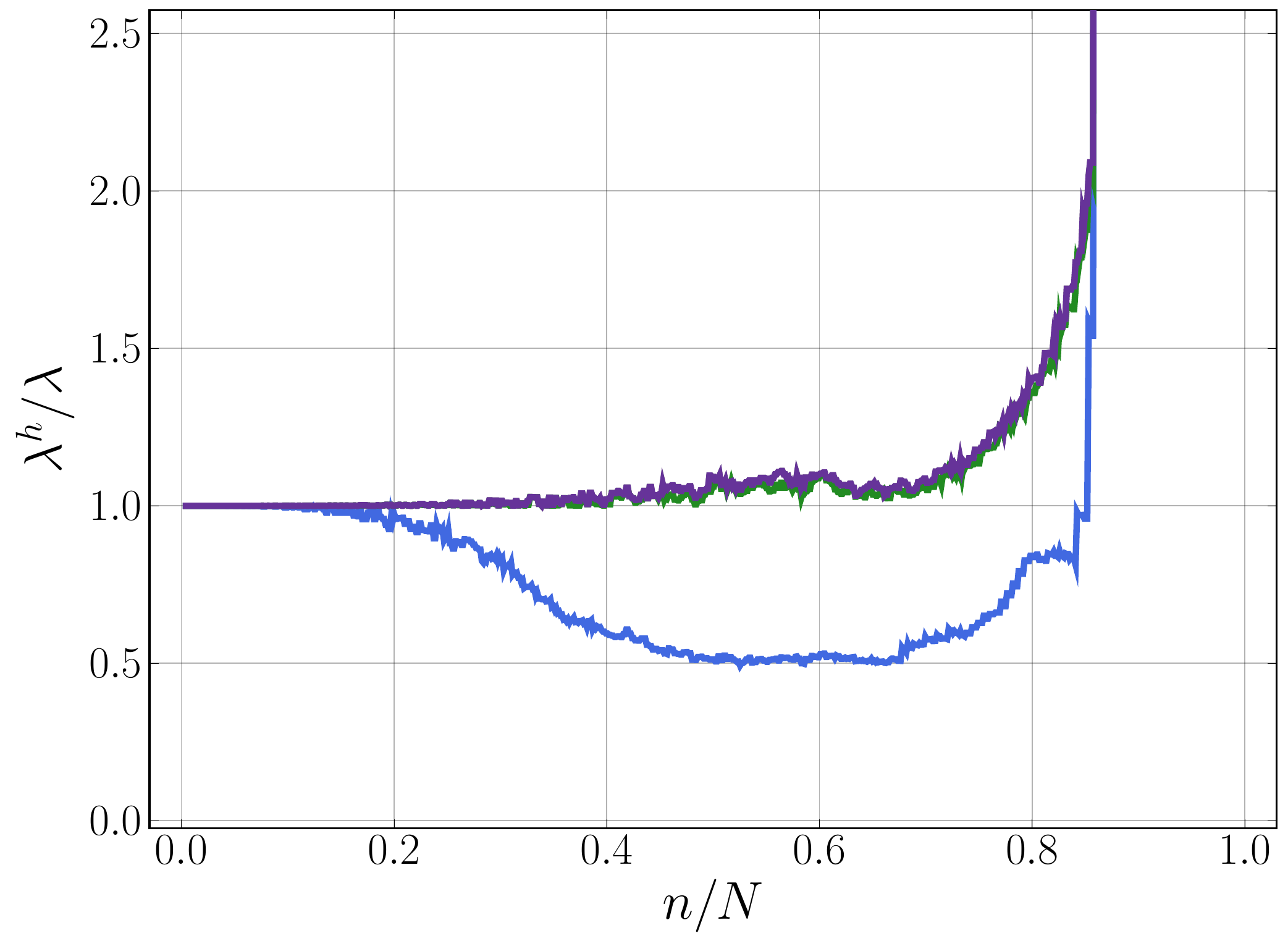} }}
    \subfloat[$p=5$]{{\includegraphics[width=0.48\textwidth]{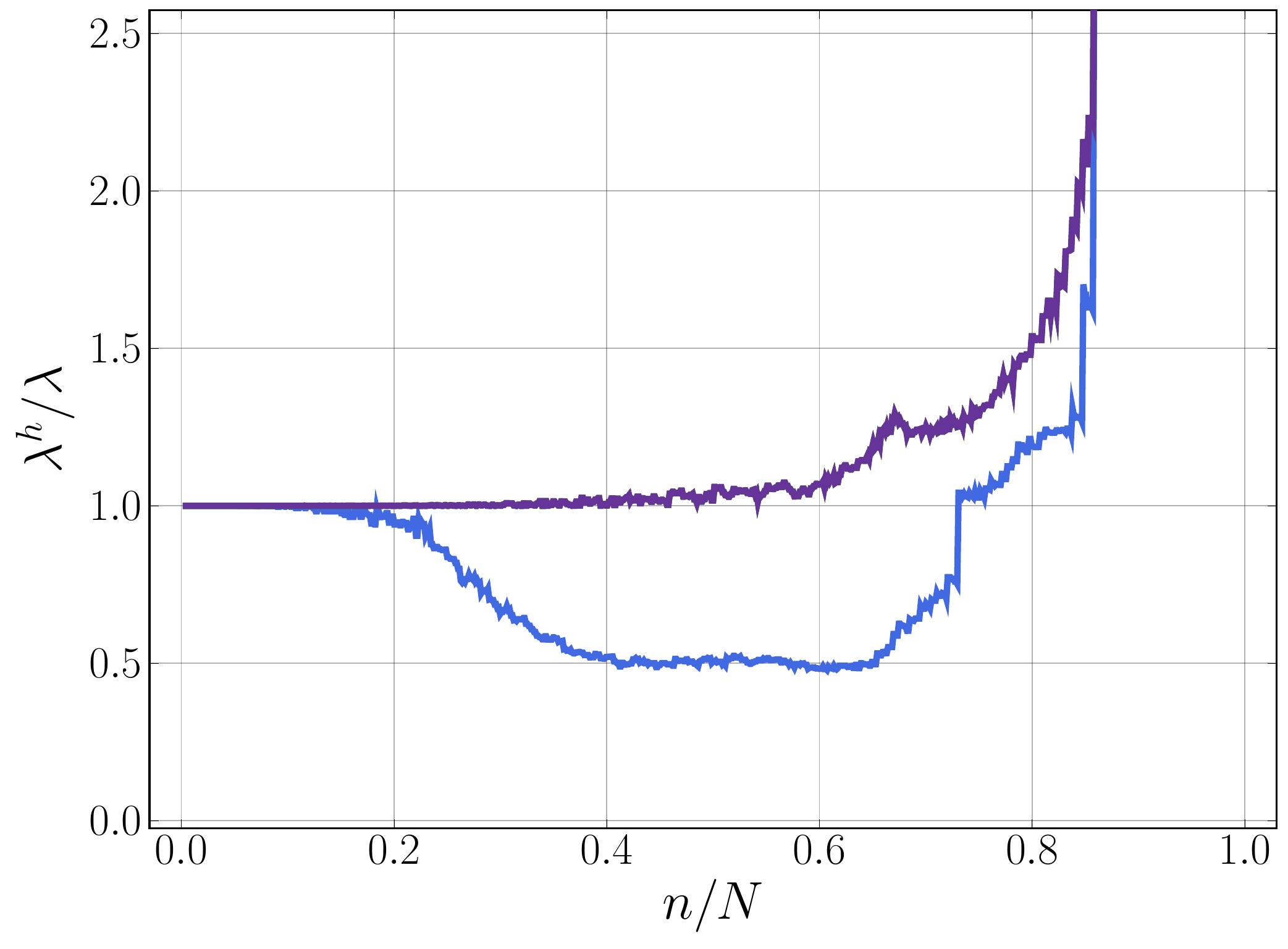} }}
    \vspace{0.2cm}
    \begin{tikzpicture}
    \filldraw[grey1,line width=1pt, solid] (0,0) -- (0.6,0);
    \filldraw[grey1,line width=1pt] (0.6,0) node[right]{\footnotesize Petrov-Galerkin, consistent mass};
    \filldraw[blue1,line width=1pt, solid] (6,0) -- (6.6,0);
    \filldraw[blue1,line width=1pt] (6.6,0) node[right]{\footnotesize $r=0$};
    \filldraw[green1,line width=1pt, solid] (8,0) -- (8.6,0);
    \filldraw[green1,line width=1pt] (8.6,0) node[right]{\footnotesize $r=1$};
    \filldraw[purple1,line width=1pt, solid] (10,0) -- (10.6,0);
    \filldraw[purple1,line width=1pt] (10.6,0) node[right]{\footnotesize $r=2$};
\end{tikzpicture}
    \caption{Normalized eigenvalues of the simply supported plate, computed with an iteratively improved approximate dual basis and a different number $r$ of corrector passes.} \label{fig:spectra_eigenval_plate_iter}
\end{figure}

\subsection{Spectral analysis}

We now use the tool of spectral analysis to elucidate the impact of lumping on the three different schemes. To this end, we recall the 
discrete eigenvalue problem corresponding to the semidiscrete formulation \eqref{dweom}, expressed in matrix form:
\begin{equation} \label{dgep}
    \mat{K} \, \mat{\eigenvec}_\noMode^h \; = \; \lambda_\noMode^h \, \mat{M} \, \mat{\eigenvec}_\noMode^h \, ,
\end{equation}
where $\mat{\eigenvec}_\noMode^h$ denotes the vector of unknown coefficients corresponding to the $\noMode^{\text{th}}$ discrete eigenmode $\eigenvec_\noMode^h$, and $\lambda_\noMode^h$ is the $\noMode^{\text{th}}$ discrete eigenvalue. For further details on spectral analysis and eigenvalue analysis in this context, we refer for instance to our previous papers on the topic \cite{Hiemstra_outlier_2021,Nguyen_outlier_2022,nguyen2022leveraging} and the references therein.

%---------------------------------------------------
\subsubsection{Petrov-Galerkin vs.\ standard Galerkin formulations}

We first consider the free transverse vibration of an unconstrained straight Euler-Bernoulli beam with the following parameters (steel, rectangular cross section): length 1.0~m, width 0.1~m, thickness 0.01~m, Young's modulus $E = 2\,\cdot\,10^{11} \, \text{N/m}^2$, and mass density $\rho=7.84\,\cdot\,10^3\,\text{kg/m}^3$. We can obtain the exact solution in terms of eigenvalues and mode shapes from the continuous problem, given for instance in \cite{Soedel2004}.

We solve the discrete eigenvalue problem \eqref{dgep} by discretizing the beam with 500 B\'ezier elements. We then employ our Petrov-Galerkin formulation discussed above, using B-splines and the corresponding approximate dual functions, and the standard Galerkin formulation, using only B-splines, all for degrees $p=2$ through 5. We then do the analysis of each formulation once with a consistent mass matrix and once with a lumped mass matrix based on the row-sum technique.
Figure~\ref{fig:spectra_eigenval_beam_no_iter} plots the eigenvalues of the discrete problem in ascending order, normalized with respect to the corresponding exact eigenvalues of the continuous problem versus the normalized mode number. Figure~\ref{fig:spectra_mode_beam_no_iter} plots the relative $L^2$ error in the corresponding mode shapes, versus the normalized mode number.

We observe in all plots that the eigenvalues and mode shape errors obtained with the Petrov-Galerkin and Galerkin methods and a consistent mass matrix are identical. The reason behind this observation is that in the current example, due to the constant determinant of the Jacobian matrix, the modified approximate dual functions span the same space as the B-splines, and hence both methods need to produce the same results, when the variational formulations are evaluated and solved consistently. We now focus on the results obtain with mass lumping. In Fig.~\ref{fig:spectra_eigenval_beam_no_iter}, we observe that our Petrov-Galerkin approach with lumped mass matrix leads to significantly better spectral accuracy compared to the standard Galerkin approach with lumped mass matrix. The difference between the results of the two approaches becomes more pronounced with increasing polynomial degree. For $p=5$, for instance, we see that in our Petrov-Galerkin approach, the first 40\% of the eigenvalues are close to the desired ratio of 1.0, whereas in the Galerkin approach, this is achieved by only about 5\%.
Figure~\ref{fig:spectra_mode_beam_no_iter} indicates the same trend for the error in the corresponding mode shapes. We can observe, however, that for $p=2$, all methods, whether consistent or lumped, produce practically the same error, and a clear difference between the two methods with lumped mass matrices starts to be evident only for $p=4$ and $p=5$.

We then investigate the convergence of the error under mesh refinement, focusing on the tenth eigenvalue and eigenmode. We note that the lowest eigenmodes are particularly important for achieving accurate results in space. Figure~\ref{fig:spectra_convergence_beam_no_iter} illustrates the convergence of the relative error of the eigenvalue (upper row) and the convergence of the $L^2$ error in the corresponding mode shape (lower row). We observe that the Petrov-Galerkin method with mass lumping achieves the same optimal convergence as the Galerkin method with consistent mass matrix, while the Galerkin method with mass lumping only achieves second-order convergence for all polynomial degrees.
We note that the optimal convergence rate of the eigenvalue error and the $L^2$ error in the mode is $\mathcal{O}(2(p-1))$ and $\mathcal{O}(p+1)$, respectively \cite{hughes_finite_2003,Cottrell_vibration_2006}. 
These results indicate that unlike in the standard Galerkin method, mass lumping in our Petrov-Galerkin method does not affect optimal convergence of analysis results. 

We now investigate how these results transfer to the two-dimensional case. To this end, we repeat the same analysis for the example of a simply supported Kirchhoff plate with the following parameters (square aluminium sheet): length and width 1.0 m, thickness 0.01 m, Young's modulus $E=7\,\cdot\,10^{10} \, \text{N/m}^2$, and mass density $\rho=2.7\,\cdot\,10^3\,\text{kg/m}^3$. We can obtain the exact solution in terms of eigenvalues and mode shapes from the continuous problem, given for instance in \cite{Soedel2004}. For the discrete problem \eqref{dgep}, we consider meshes with $25 \times 25$ B\'ezier elements of degrees $p=2$ to 5.

Figure~\ref{fig:spectra_eigenval_plate_no_iter} plots the normalized eigenvalues versus the normalized mode number. We note that some outliers at the end of the spectra lie outside the plotted range. Figure~\ref{fig:spectra_convergence_plate_no_iter} plots the convergence of the relative error of the first eigenvalue under uniform mesh refinement. In Fig.~\ref{fig:spectra_eigenval_plate_no_iter}, we can observe again significant accuracy advantages of our Petrov-Galerkin scheme with mass lumping over the standard Galerkin scheme with mass lumping in the spectrum results. Figure~\ref{fig:spectra_eigenval_plate_no_iter} confirms our observation that unlike in the standard Galerkin method, mass lumping does not affect the optimal convergence of our Petrov-Galerkin method.

%---------------------------------------------------
\subsubsection{Iterative improvement of bi-orthogonality}

To further elucidate the accuracy of our Petrov-Galerkin scheme based on approximate dual functions and mass lumping, we use the iterative approach discussed in Section \ref{sec_multicorrector} to gradually improve the bi-orthogonality property of the approximate dual basis. 
To this end, we consider again the two examples of the unconstrained Euler-Bernoulli beam in 1D and the simply supported square plate in 2D. We now employ our Petrov-Galerkin scheme with an iteratively improved mass matrix, where we use one or two corrector passes ($r=1,2$) before mass lumping, and compare the corresponding spectral results to those obtained with the Petrov-Galerkin scheme with a consistent mass matrix and mass lumping without improvement ($r=0$).

For the Euler-Bernoulli beam, the normalized eigenvalues and the relative $L^2$ error in the corresponding mode shapes are plotted versus the normalized mode number in Figs.~\ref{fig:spectra_eigenval_beam_iter} and \ref{fig:spectra_mode_beam_iter}, respectively. We observe that with increasing number of corrector passes, the spectrum curves quickly approach the reference solution obtained with the consistent mass matrix. We see that the iterative improvement is particularly efficient for improving the accuracy of the mode shapes. After two corrector passes, the $L^2$ error curve of the Petrov-Galerkin scheme with mass lumping is practically the same as the one of the reference for polynomial degrees $p=2$ to $4$ throughout the complete spectrum, and a clear difference is only noticeable in the high modes for $p=5$.

Figure~\ref{fig:spectra_eigenval_plate_iter} plots the normalized eigenvalues versus the normalized mode number for the simply supported square plate.  
We observe that in this case, a lumped mass matrix with one or two corrector passes can eliminate the error in the eigenvalue spectrum completely. For $p=5$, the normalized eigenvalue curve obtained with a consistent mass matrix and the lumped mass matrix with only one corrector pass are practically indistinguishable throughout the complete spectrum.
We conclude that - although we have not used the iterative improvement in our simple benchmark calculations here - it could represent a tool to improve spectral accuracy of the Petrov-Galerkin approach in future more challenging applications.

\section{Discussion and outlook}\label{sec_conclusion}

\subsection{Summary and significance of current results}

In this paper, we introduced an isogeometric Petrov-Galerkin formulation that enables higher-order accurate mass lumping. It is based on B-splines to discretize the solution fields and corresponding approximate dual functions with local support to discretize the test functions. 
We demonstrated via convergence studies and spectral analysis for beam and plate models that our Petrov-Galerkin method leads to higher-order accurate solutions in explicit dynamics, also when the mass matrix is lumped. Our results thus confirm that mass lumping as such does not have to be a deal-breaker for higher-order accuracy in the context of isogeometric explicit dynamics calculations, if one adequately exploits the additional opportunities of splines.

We also presented several ideas to resolve related technical issues. Firstly, we showed that the approximate bi-orthogonality property, which is in general destroyed under non-affine geometry mapping, is preserved when we use modified approximate dual functions that are divided by the Jacobian determinant. Secondly, we discussed options for the weak and strong imposition of Dirichlet boundary conditions. For the latter, we suggested to replace the few non-interpolatory approximate dual functions at the Dirichlet boundary by standard B-splines, restoring the interpolation property as a prerequisite for the strong imposition of Dirichlet boundary conditions.
We also presented an algorithm for iteratively improving the approximate bi-orthogonality property of the dual basis. While its application is not necessary for securing higher-order accurate results, in particular given the additional computational cost, it establishes a rigorous link between truly dual and approximate dual basis functions and the accuracy of the associated spectra.

\begin{figure}[h!]%[!htb]
	\centering
    \includegraphics[width=0.65\textwidth]{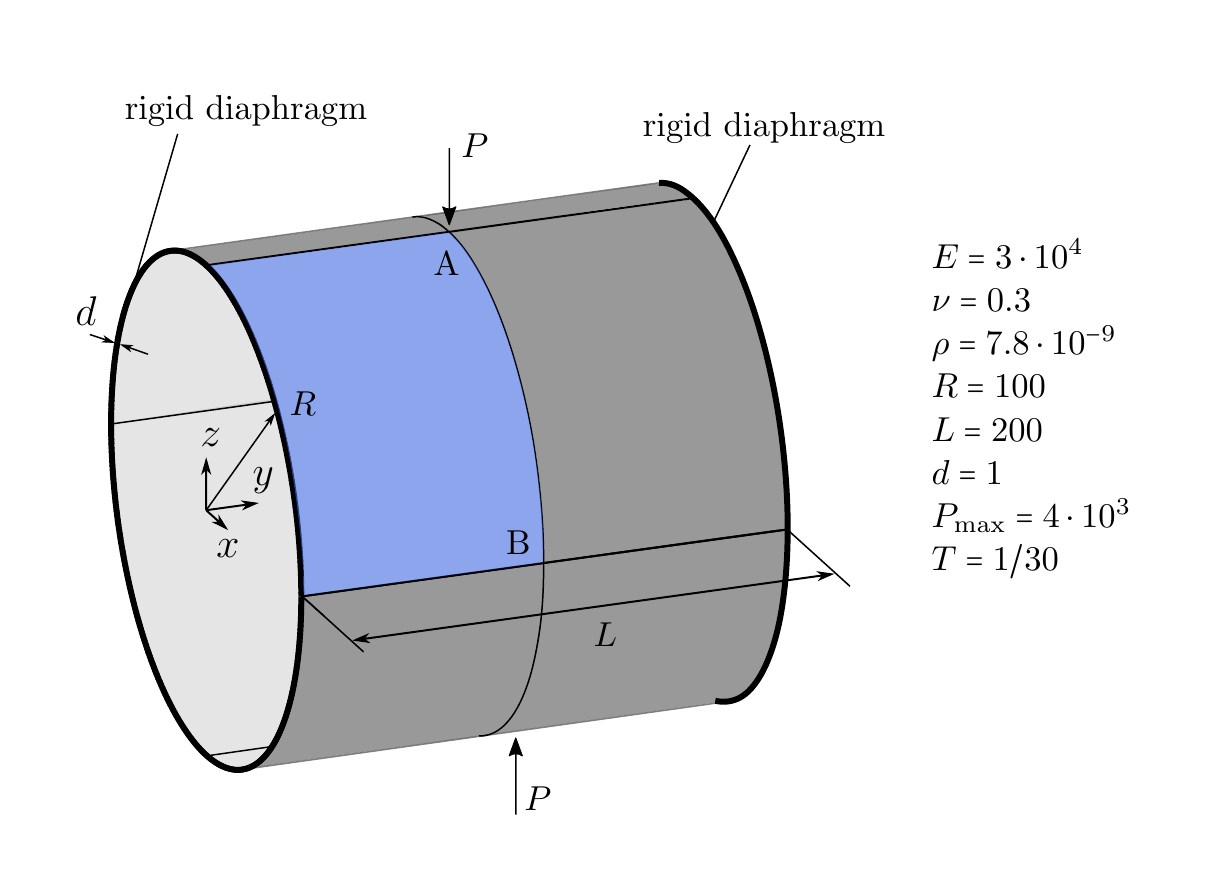}

	\caption{Pinched cylinder constrained by rigid diaphragms.} \label{fig:cylinder_geometry}
\end{figure}

\begin{figure}[h!]
    \centering
    \includegraphics[width=0.99\textwidth]{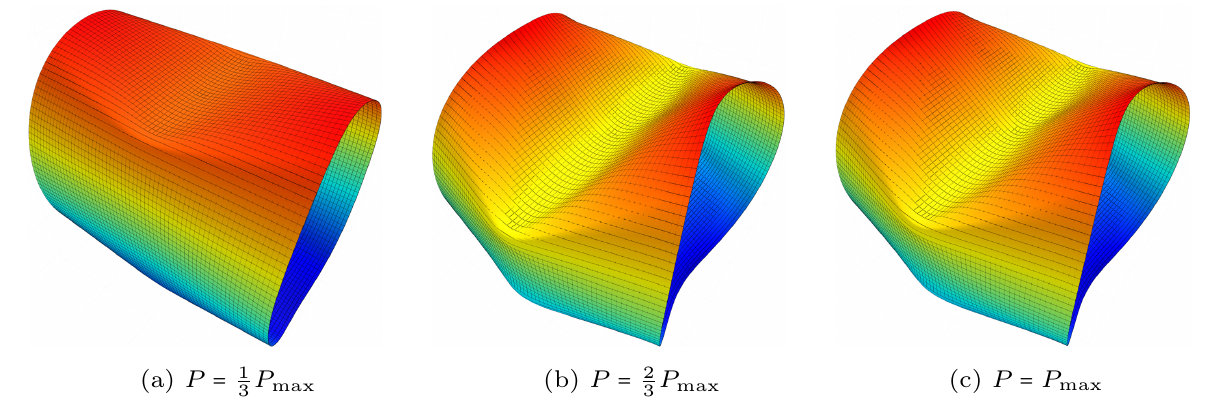}

    \caption{Deformed configurations of the cylinder, computed with our Petrov-Galerkin method with mass lumping.}\label{fig_cylinder-deformed}
\end{figure}

\begin{figure}[t!]
    \centering
    \subfloat[$p=3$, point A.]{{\includegraphics[width=0.432\textwidth]{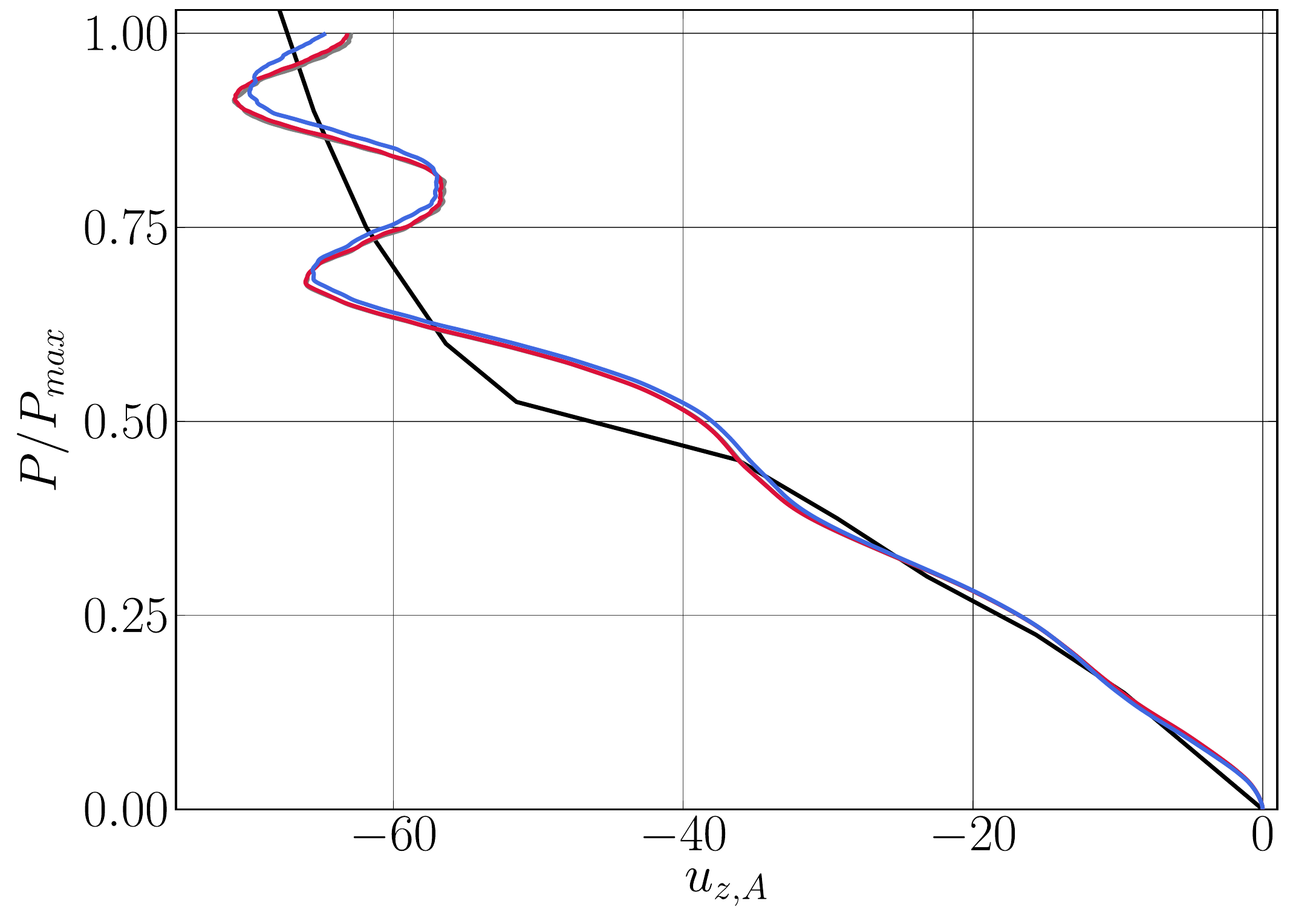} }}
    \subfloat[$p=3$, point B.]{{\includegraphics[width=0.432\textwidth]{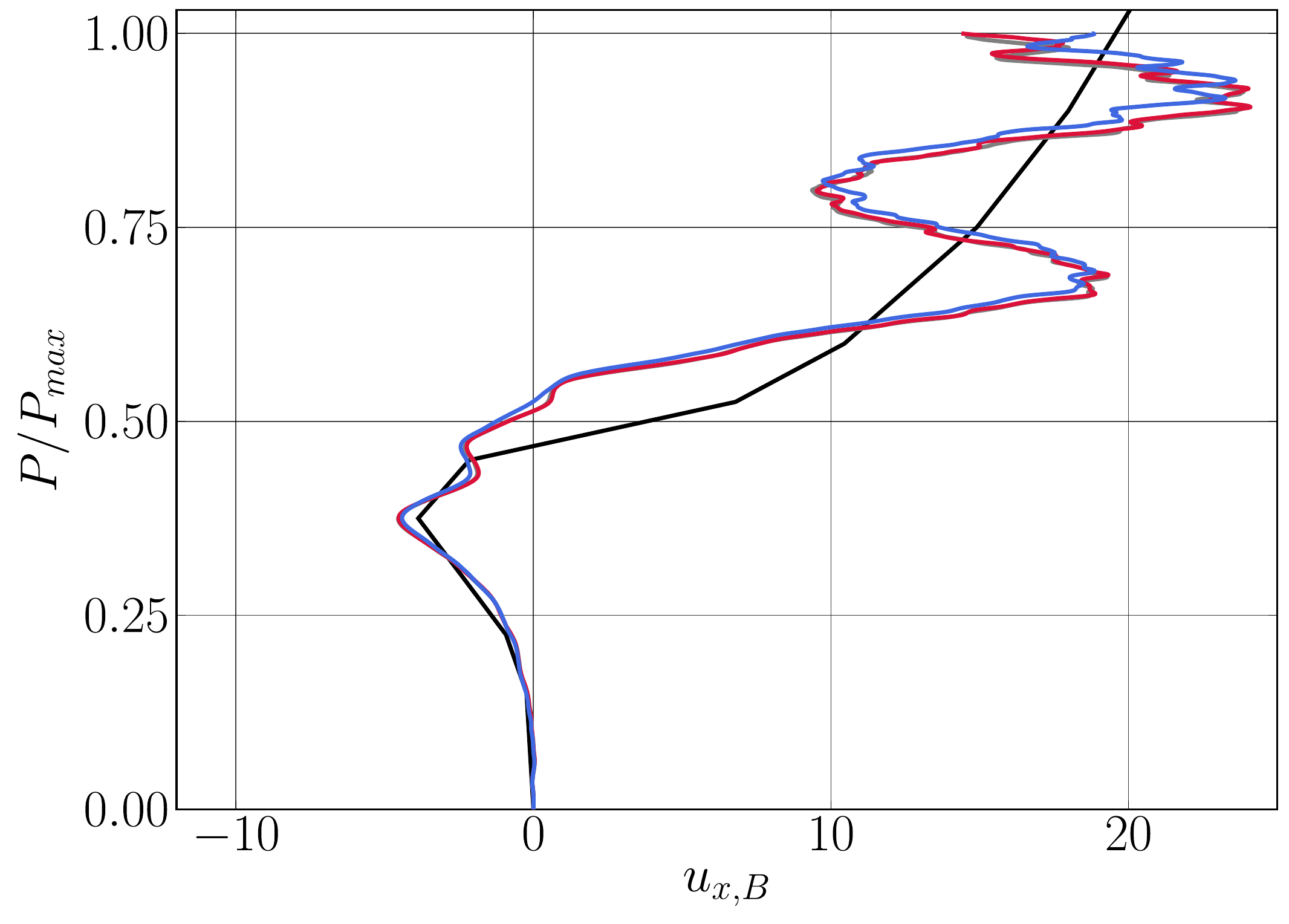} }}

    \subfloat[$p=4$, point A.]{{\includegraphics[width=0.432\textwidth]{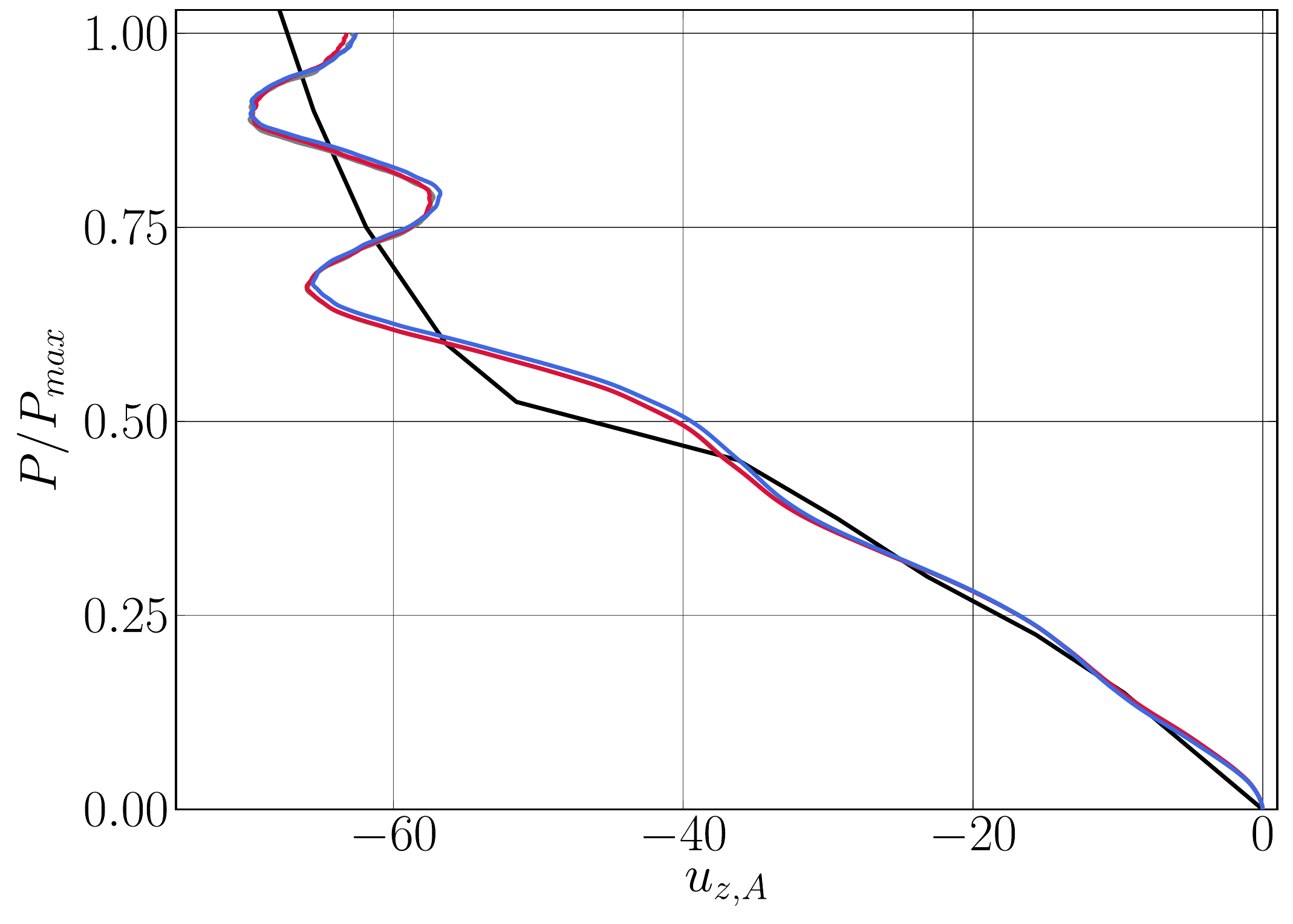} }}
    \subfloat[$p=4$, point B.]{{\includegraphics[width=0.432\textwidth]{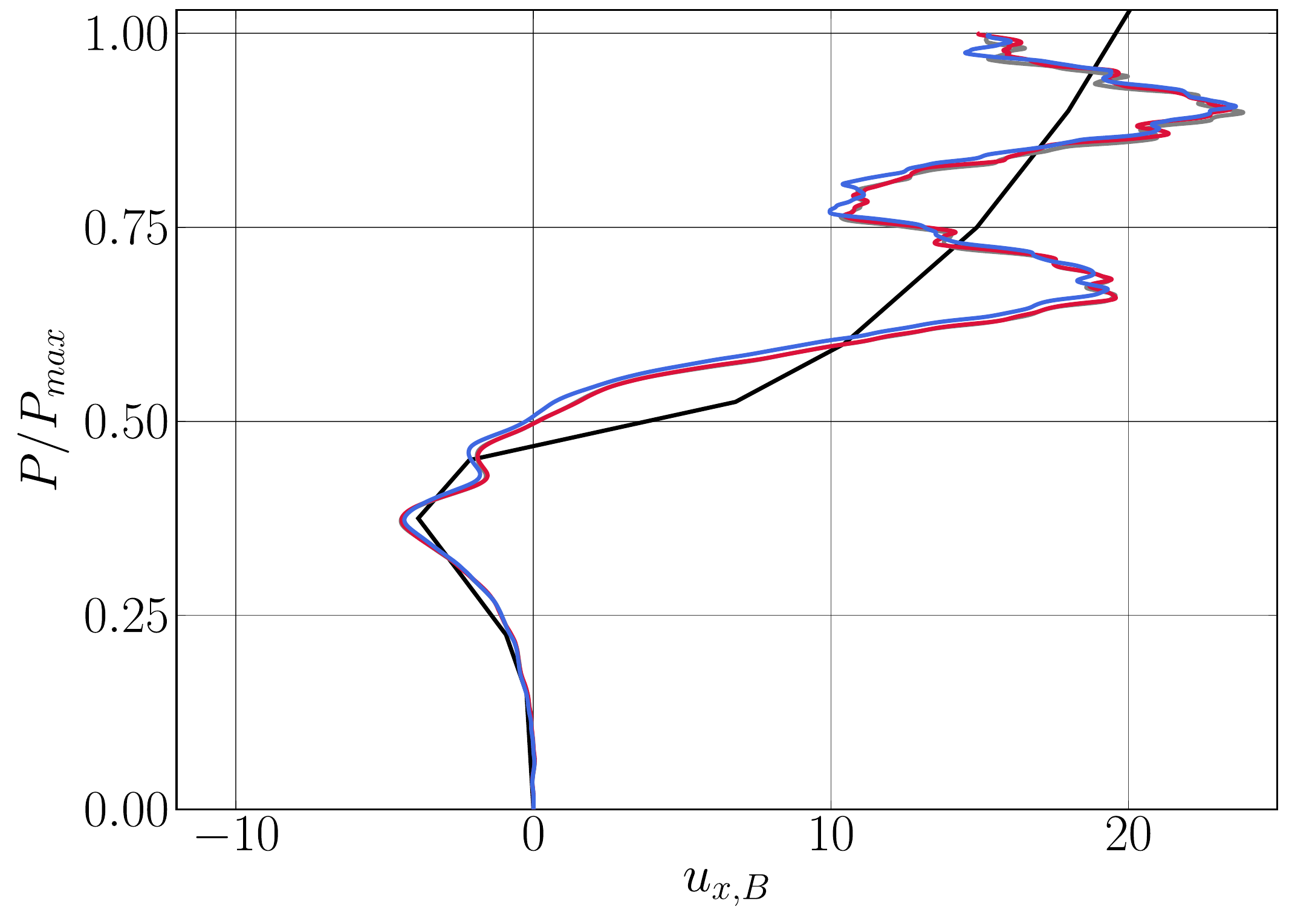} }}

    \subfloat[$p=5$, point A.]{{\includegraphics[width=0.432\textwidth]{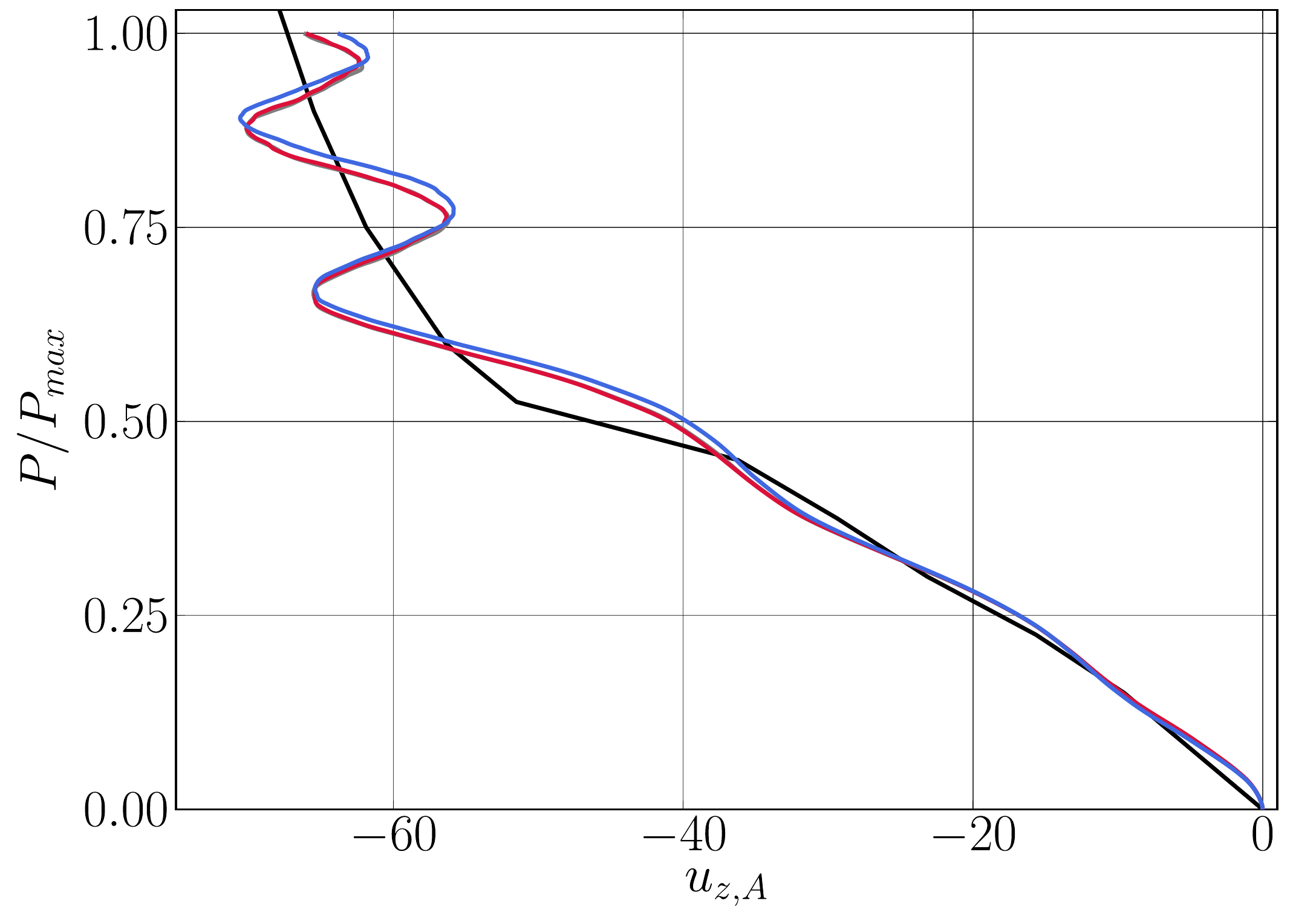} }}
    \subfloat[$p=5$, point B.]{{\includegraphics[width=0.432\textwidth]{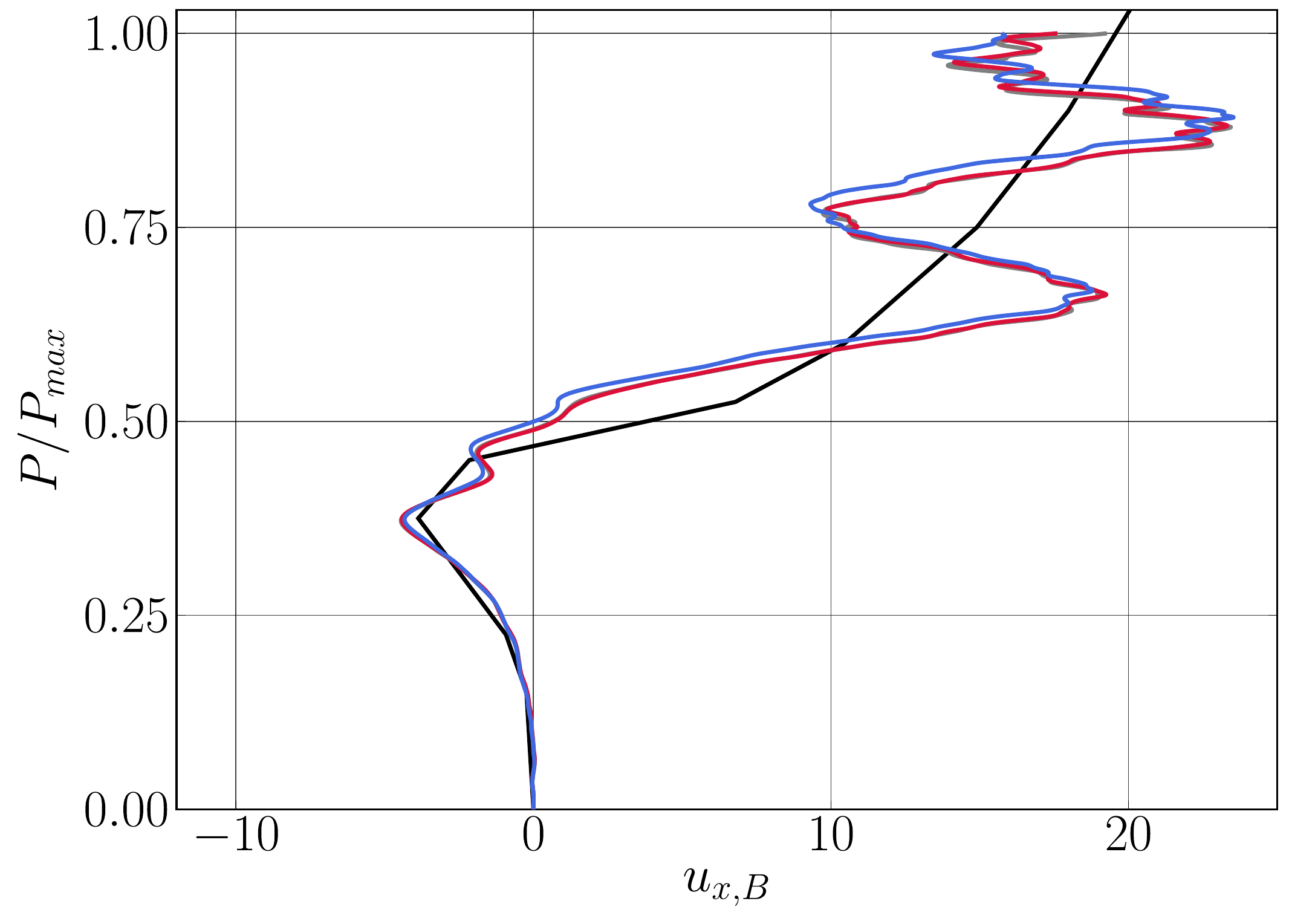} }}
    \vspace{0.2cm}
    \begin{tikzpicture}
    \filldraw[black,line width=1pt, solid] (0,0) -- (0.6,0);
    \filldraw[black,line width=1pt] (0.6,0) node[right]{\footnotesize Static solution \cite{Sze_shell_benchmark_2004}};
    \filldraw[grey1,line width=1pt, solid] (4,0) -- (4.6,0);
    \filldraw[grey1,line width=1pt] (4.6,0) node[right]{\footnotesize Galerkin, consistent mass};
    % \filldraw[green1,line width=1pt, solid] (9,0) -- (9.6,0);
    % \filldraw[green1,line width=1pt] (9.6,0) node[right]{\footnotesize Galerkin, consistent mass, $p=4, 32 \times 32$ elements};
\end{tikzpicture}

\begin{tikzpicture}
    \filldraw[red1,line width=1pt, solid] (2,0) -- (2.6,0);
    \filldraw[red1,line width=1pt] (2.6,0) node[right]{\footnotesize Galerkin, row-sum lumped mass};
    \filldraw[blue1,line width=1pt, solid] (8,0) -- (8.6,0);
    \filldraw[blue1,line width=1pt] (8.6,0) node[right]{\footnotesize Petrov-Galerkin, higher-order accurate lumped mass};
\end{tikzpicture}
    \caption{Load-deflection curves at point A ($u_z$) and at point B ($u_x$) of the pinched cylinder, computed with three different isogeometric schemes on a mesh of $32 \times 32$ B\'ezier elements.}\label{fig_cylinder-uB}
\end{figure}

\subsection{A first look at explicit dynamics of shells}

To get an idea whether its advantages directly transfer to more involved explicit dynamics calculations, we employ our Petrov-Galerkin method with mass lumping to solve the test case of a pinched cylinder, whose set-up is given in Fig.~\ref{fig:cylinder_geometry}. Our implementation follows the Kirchhoff-Love shell model and its isogeometric discretization presented in \cite{Kiendl_shell_2009}, assuming large deformation kinematics and the St.~Venant-Kirchhoff constitutive model. 
In our Petrov-Galerkin approach, we apply B-splines to discretize the displacement solution and modified approximate dual functions to discretize the virtual displacements, while the geometry map is provided by an exact NURBS representation of the cylinder. 
We use the central difference method to integrate in time, with a time step size of $\Delta t = 5 \cdot 10^{-8}$. 
The mass matrix of the Petrov-Galerkin scheme is lumped via the standard row-sum technique, except for the few rows that correspond to coefficients of the boundary B-spline functions.

The point load $P$ is linearly increased from zero to $P_{\text{max}}$ during the time $T=1/30$. Using symmetry, we compute one eighth of the cylinder that we discretize with cubic, quartic and quintic splines on a mesh of $32  \times 32$ B\'ezier elements. This resolution is known in the literature to be adequate for this problem \cite{Leidinger2019}, which exhibits highly localized features in the displacement solution due to the point load and the appearance of wrinkles during deformation. In addition, this resolution guarantees that locking phenomena, in particular membrane locking, are controlled. Figure~\ref{fig_cylinder-deformed} illustrates three snapshots from the deformation history.

To assess the performance of our Petrov-Galerkin method, we compare its displacement response to the results obtained with the standard Galerkin method with a consistent mass matrix and with a row-sum lumped mass matrix. We plot the load-deflection curves at points A and B, that is $u_{z,A}$ and $u_{x,B}$, in Fig.~\ref{fig_cylinder-uB}.
As an additional reference, we also include the static solution (black curves), as the explicit dynamic results oscillate around the static solution after snap-through \cite{Leidinger2019}. We observe that for the given mesh, all three methods lead to practically the same displacement response, irrespective of the polynomial degree considered. We note that we also tried to lower the resolution by repeating the same study on meshes with $16 \times 16$ B\'ezier elements. We found that the three methods still yield displacement results that are overall equivalent, but all deviate from the static solution, and hence are not accurate. 

These results of a more involved test case confirm that our Petrov-Galerkin scheme also works here. But they also indicate that its advantages might not always be as evident in practical applications as in the simple benchmarks shown in Section~\ref{sec:dynamics_plates} above.

\subsection{Future research directions}

From our viewpoint, there are two particularly important technical aspects that remain to be investigated. One is an in-depth performance test of our Petrov-Galerkin method in more involved scenarios, extending our initial results presented above. \reviewerII{In particular, relevant scenarios feature typical challenges in explicit dynamics calculations, including shell elements, large deformations, contact, nonlinear material behavior, and their combinations}. \reviewerI{The other one is a detailed analysis of the computational efficiency, in particular with respect to the computational cost of a competitive matrix-free implementation that fully exploits additional technical opportunities of spline functions such as the exploitation of their tensor-product structure \cite{mika2021matrix}, the use of optimal quadrature rules with minimal number of points \cite{hiemstra2019fast}, and the removal of outliers \cite{Hiemstra_outlier_2021}}.

The results presented in this paper open up further directions for future technical work. One is the capability of directly building interpolatory boundary basis functions and/or the required Dirichlet boundary conditions into the approximate dual space. 
Another one is the combination of locking preventing mechanisms with the approximate dual basis, in order to enable accurate solution fields on coarse meshes. Another challenge, for which no initial idea exists, is to extend the concept of the approximate dual basis as a test function space to trimmed B\'ezier elements.

\section*{Acknowledgments}
The authors gratefully acknowledge financial support from the German Research Foundation (Deutsche Forschungsgemeinschaft) through the DFG Emmy Noether Grant SCH 1249/2-1 and the standard DFG grants SCH 1249/5-1 and EI 1188/3-1.

% \newpage

\appendix

%---------------------------------------------------
\section{Approximate inverse of the Gramian matrix}\label{appendix_approx_inverse}

We summarize the recursive construction of $\invmassMf$ based on \cite{chui_wavelet_2004,Dornisch_dual_basis_2017} in Algorithm \ref{alg:approx_inv0}. 
As discussed in Section \ref{sec:approx_dual}, $\invmassMf$ is an approximate inverse of the Gramian matrix, $\invmassMf \approx \mat{G}^{-1}$.

\begin{algorithm}[h!]
    \textbf{Input}:
        $p$, $N$ and $\knotvect$ \\
    \textbf{Output}: $\invmassMf$
    \begin{algorithmic}[1]
        \State $\mat{D} = \mat{I}_{\ndofs \times \ndofs}$  \Comment{Initialize}
        \State $\invmassMf = \text{compute\textunderscore matrix\textunderscore}\mat{U}(v=0,p,\knotvect,\ndofs)$  \Comment{Initialize. See Algorithm \ref{alg:inv_mass_scaling}}
        \For{$v$ in $1:p$}
        \State $\mat{U} = \text{compute\textunderscore matrix\textunderscore}\mat{U}(v,p,\knotvect,\ndofs)$      \Comment{See Algorithm \ref{alg:inv_mass_scaling}}
        \State $\mat{D} \, \mathrel{*}= \text{compute\textunderscore matrix\textunderscore}\mat{D}(v,p,\knotvect,\ndofs)$    \Comment{See Algorithm \ref{alg:weighted_diff}}
        \State $\invmassMf \, \mathrel{+}= \mat{D} \, \mat{U} \, \mat{D}^T$
        \EndFor
    \caption{Compute the approximate inverse $\invmassMf$ of the Gramian matrix based on \cite{chui_wavelet_2004}}\label{alg:approx_inv0}
    \end{algorithmic}
\end{algorithm}

\begin{algorithm}[h!]
    \textbf{Input}:
        Index $v$ $(v = 0,\ldots,p)$, 
        $p$, $\knotvect$ and $\ndofs$ \\
    \textbf{Output}: $\mat{U}$
    \begin{algorithmic}[1]
        \State $c = \frac{(p+1)\ ! (p-v)\ !}{(p+v+1)\ ! (p+v)\ !}$
        \State $F_v(\xi_{j+1} \ldots \xi_{j+p+v})$ according to Equation (5.1) in \cite{chui_wavelet_2004}  \Comment{See \eqref{eq:Fv} for $v=0,\ldots,5$}
        \For{$j$ in $1:(\ndofs-v)$}
        \State $U_{jj} = c \, \frac{p+v+1}{\xi_{j+p+v+1} - \xi_{j}} \, F_v(\xi_{j+1} \ldots \xi_{j+p+v})$
        \EndFor
    \caption{Compute the diagonal matrix $\mat{U}$ in Algorithm \ref{alg:approx_inv0} \cite{chui_wavelet_2004}}\label{alg:inv_mass_scaling}
    \end{algorithmic}
\end{algorithm}

\begin{algorithm}[h!]
    \textbf{Input}:
    Index $v$ $(v = 0,\ldots,p)$, 
    $p$, $\knotvect$ and $\ndofs$ \\
    \textbf{Output}: $\mat{D}$
    \begin{algorithmic}[1]
        \State $k = p+v$
        \For{$j$ in $1:(\ndofs+p+1-k)$}
        \State $d_{\knotvect,j} = k \, / \, (\xi_{j+k}-\xi_j)$
        \EndFor
        \State $\boldsymbol{\Delta}_{\ndofs+p+1-k} = \begin{bmatrix}
            $1$ & & & \\ $-1$ & $1$ & & \\ & $\ldots$ & $\ldots$ & \\ & & $-1$ & $1$ \\ & & & $-1$ 
        \end{bmatrix}_{(\ndofs+p+1-k) \times (\ndofs+p-k)}$
        \State $\mat{D} = \text{diag}\left(\mat{d}_{\knotvect}\right) \, \boldsymbol{\Delta}_{\ndofs+p+1-k}$
    \caption{Compute matrix $\mat{D}$ in Algorithm \ref{alg:approx_inv0} \cite{chui_wavelet_2004}}\label{alg:weighted_diff}
    \end{algorithmic}
\end{algorithm}

%---------------------------------------------------

In \cite{chui_wavelet_2004}, the authors provide formulas for computing the homogeneous polynomial $F_v(x_1,\ldots,x_r)$ in Algorithm \ref{alg:inv_mass_scaling} with $v=0,\ldots,5$ as follows:
\begin{subequations}\label{eq:Fv}
    \begin{align}
        F_0(x_1,\ldots,x_r) \, = \, & 1.0 \, ,  \\
        F_1(x_1,\ldots,x_r) \, = \, & r^2 \, \sigma_2 \, , \\
        2 F_2(x_1,\ldots,x_r) \, = \, & r^2 \, \left(r^2-3r+3 \right)\, \sigma_2^2 \, - \, r^2\,(r-1)\,\sigma_4 \, ,  \\
        6 F_3(x_1,\ldots,x_r) \, = \, & r^3\,(r-2)\,\left(r^2-7r+15 \right) \, \sigma_2^3 \, - \, 3r^2 \, (r-2) \, \left(r^2-5r+10 \right) \, \sigma_4 \, \sigma_2 \, -  \,  \nonumber \\
         & 2r^2 \, \left(3r^2-15r+20 \right) \, \sigma_3^2 \, + \, 2r^2 \, (r-1) \, (r-2) \, \sigma_6 \, ,  \\
        24 F_4(x_1,\ldots,x_r) \, = \, & r^4 \, \left(r^4 - 18r^3 + 125r^2 - 384r + 441 \right) \, \sigma_2^4 \, - \, \nonumber \\
        & 6r^3 \, \left(r^4 - 16r^3 + 104r^2 - 305r + 336 \right) \, \sigma_4 \, \sigma_2^2 \, + \, \nonumber \\
        & 3r^2 \, \left(r^4 -14r^3 + 95r^2 - 322r + 420 \right) \, \sigma_4^2 \, + \, \nonumber \\
        & 8r^2 \, (r-2) \, (r-3) \, \left(r^2 - 7r + 21 \right) \, \sigma_6 \, \sigma_2 \, - \, \nonumber \\
        & 8r^3 \, (r-3) \, \left(3r^2 - 24r + 56 \right) \, \sigma_3^2 \, \sigma_2 \, + \, 48r^2 \, (r-3) \, \left(r^2 - 7r + 14 \right) \, \sigma_5 \, \sigma_3 \, - \, \nonumber \\
        & 6r^2 \, (r-1) \, (r-2) \, (r-3) \, \sigma_8 \, ,  \\
        120 F_5(x_1,\ldots,x_r) \, = \, & r^5 \, (r-4) \, \left(r^4 - 26r^3 + 261r^2 - 1176r + 2025 \right) \, \sigma_2^5 \, - \, \nonumber \\
        & 10r^4 \, (r-4) \, \left(r^4 - 24r^3 + 230r^2 - 999r + 1674 \right) \, \sigma_4 \, \sigma_2^3 \, + \, \nonumber \\
        & 20r^3 \, (r-4) \, \left(r^4 - 20r^3 + 168r^2 - 645r + 972 \right) \, \sigma_6 \, \sigma_2^2 \, + \, \nonumber \\
        & 15r^3 \, (r-4) \, \left(r^4 - 22r^3 + 211r^2 - 942r + 1620 \right) \, \sigma_4^2 \, \sigma_2 \, - \, \nonumber \\
        & 20r^4 \, \left(3r^4 - 60r^3 + 470r^2 - 1665r + 2232 \right) \, \sigma_3^2 \, \sigma_2^2 \, - \, \nonumber \\
        & 30r^2 \, (r-2) \, (r-3) \, (r-4) \, \left(r^2 - 9r + 36 \right) \, \sigma_8 \, \sigma_2 \, - \, \nonumber \\
        & 20r^2 \, (r-4) \, \left(r^4 - 18r^3 + 173r^2 - 828r + 1512 \right) \, \sigma_6 \, \sigma_4 \, + \, \nonumber \\
        & 240r^3 \, \left(r^4 - 19r^3 + 143r^2 - 493r + 648 \right) \, \sigma_5 \, \sigma_3 \, \sigma_2 \, + \, \nonumber \\
        & 20r^4 \, (r-4) \, \left(3r^2 - 30r + 83 \right) \, \sigma_4 \, \sigma_3^2 \, - \, \nonumber \\
        & 24r^2 \, \left(5r^4 - 90r^3 + 655r^2 - 2250r + 3024 \right) \, \sigma_5^2 \, - \, \nonumber \\
        & 240r^2 \, (r-3) \, (r-4) \, \left(r^2 - 9r + 24 \right) \, \sigma_7 \, \sigma_3 \, + \, \nonumber \\
        & 24r^2 \, (r-1) \, (r-2) \, (r-3) \, (r-4) \, \sigma_{10} \, ,
    \end{align}    
\end{subequations}
where $\sigma_l$ are the centered moments: 
\begin{align}
    \sigma_l = \frac{1}{r} \, \sum_{j=1}^r \, \left(x_j - \bar{x} \right)^l \, \quad \text{with} \, \quad \bar{x} = \frac{1}{r} \, \sum_{j=1}^r \, x_j \, . \nonumber
\end{align}

We demonstrate the computation of $\invmassMf$ in Example \ref{eg:approx_dual} for a space of quadratic $C^1$ B-splines ($p=2$) defined on an open knot vector $\knotvect = [0,0,0,1,2,3,3,3]$.

\begin{example} \label{eg:approx_dual}
    \textup{(Computation of the approximate inverse using Algorithm \ref{alg:approx_inv0} based on \cite{chui_wavelet_2004})}.
    Consider a space of quadratic $C^1$ B-splines ($p=2$) on $\knotvect = \left[0, 0, 0, 1, 2, 3, 3, 3\right]$. The number of B-splines is $\ndofs = 5$.
    The spline function vector $\trialfv = \left[\trialf_1 \; \trialf_2 \; \trialf_3 \; \trialf_4 \; \trialf_5\right]$.

    The matrix $\invmassMf$ computed with Algorithm \ref{alg:approx_inv0} is:
    \begin{align}
        \invmassMf = \mat{U}(v=0) \quad + \quad \mat{D}(v=1) \, & \mat{U}(v=1) \, \mat{D}^T(v=1) \quad + \nonumber \\
        \mat{D}(v=1) \, \mat{D}(v=2) \, & \mat{U}(v=2) \, \mat{D}^T(v=2) \, \mat{D}^T(v=1) \, . \nonumber
    \end{align}
    For $v=0$, we compute the diagonal matrix $\mat{U}$ using Algorithm \ref{alg:inv_mass_scaling} and obtain:
    \begin{align}
        \mat{U}(v=0) = \begin{bmatrix}
            3.0 & & & & \\ & 1.5 & & & \\ & & 1.0 & & \\ & & & 1.5 & \\ & & & & 3.0 
        \end{bmatrix} \, . \nonumber
    \end{align}
    For $v=1$, we compute the values of the homogeneous polynomial $F_1$ using the equivalent formulas in \eqref{eq:Fv} and obtain $F_1 = 2, 6, 6, 2$ for the entries $U_{11}$, $U_{22}$, $U_{33}$ and $U_{44}$ of $\mat{U}$, respectively. The diagonal matrix $\mat{U}$ is then:
    \begin{align}
        \mat{U}(v=1) = \begin{bmatrix}
            1/6 & & & \\ & 3/4 & & \\ & & 3/4 & \\ & & & 1/6
        \end{bmatrix} \, . \nonumber
    \end{align}
    Matrix $\mat{D}$, computed with Algorithm \ref{alg:weighted_diff}, is:
    \begin{align}
        \mat{D}(v=1) = \begin{bmatrix}
            3.0 & & & & \\ & 1.5 & & & \\ & & 1.0 & & \\ & & & 1.5 & \\ & & & & 3.0 
        \end{bmatrix} \, \begin{bmatrix}
            1 & & & \\ -1 & 1 & & \\ & -1 & 1 & \\ & & -1 & 1 \\ & & & -1
        \end{bmatrix} = \begin{bmatrix}
            3.0 & & & \\ -1.5 & 1.5 & & \\ & -1.0 & 1.0 & \\ & & -1.5 & 1.5 \\ & & & -3.0
        \end{bmatrix} \, . \nonumber
    \end{align}
    For $v=2$, analogously, we obtain: 
    \begin{align}
        \mat{U}(v=2) = \begin{bmatrix}
            1/36 & & \\ & 13/144 & \\ & & 1/36
        \end{bmatrix} \, \quad \text{and} \quad
        \mat{D}(v=2) = \begin{bmatrix}
            2.0 & & \\ -4/3 & 4/3 & \\ & -4/3 & 4/3 \\ & & -2.0
        \end{bmatrix} \, . \nonumber
    \end{align}
    The matrix $\invmassMf$ is then: 
    \begin{align}
        \invmassMf = \begin{bmatrix}
            11/2    & -19/12 &  2/9           & \\ 
            -19/12  & 265/72 & -7/6  & 13/36  & \\
            2/9     & -7/6   & 65/27 & -7/6   & 2/9 \\
             0      & 13/36  & -7/6  & 265/72 & -19/12 \\
             0      &   0    &  2/9  & -19/12 &  11/2
        \end{bmatrix} \, . \nonumber
    \end{align}
\end{example}

\bibliographystyle{elsarticle-num}
\bibliography{refs}

\end{document}